\documentclass[preprint,authoryear,3p]{elsarticle}
\usepackage{lipsum}
\usepackage{lineno}
\usepackage{amsmath,amssymb,dsfont}
\usepackage{epsf,graphicx,psfrag}
\usepackage{wrapfig,lipsum,booktabs}
\usepackage{multirow}
\usepackage{longtable,lscape}
\usepackage{cases}
\usepackage{float}
\usepackage{url}
\usepackage{caption}
\usepackage{color}
\usepackage{amsthm}
\usepackage{diagbox}
\usepackage{algpseudocode}
\usepackage[ruled,vlined,titlenotnumbered]{algorithm2e}

\SetCommentSty{mycommfont}
\SetKwInput{KwInput}{Input}                
\SetKwInput{KwOutput}{Output}
\usepackage{makecell}

\usepackage[labelformat=simple]{subfig}

\newcommand\figref[1]{Figure~\ref{#1}}

\usepackage{enumitem}
\DeclareMathOperator{\st}{s.t.}
\DeclareMathOperator*{\argmax}{arg\,max}  
\DeclareMathOperator*{\argmin}{arg\,min}  
\usepackage[thinc]{esdiff}
\usepackage{fancyhdr}
\usepackage[utf8]{inputenc}
\usepackage[T1]{fontenc}
\usepackage{nomencl}
\makenomenclature

\renewcommand{\nompreamble}{In this paper, we use the following notations:}
\usepackage{etoolbox}
\renewcommand\nomgroup[1]{%
 \item[\bfseries
  \ifstrequal{#1}{A}{Lower Level}{%
  \ifstrequal{#1}{B}{Upper Level}{%
  \ifstrequal{#1}{C}{Other Notations}{}}}%
]}

\newtheorem{lemma}{Lemma}
\newtheorem{corollary}{Corollary}

\theoremstyle {plain}
\theoremstyle{definition}
\newtheorem{definition}{Definition}[section]

\newtheorem*{remark}{Remark}

\usepackage{tikz}
\usetikzlibrary{shapes,arrows}

\newenvironment{frcseries}{\fontfamily{frc}\selectfont}{}

\numberwithin{equation}{section}

\journal{Transportation Research Part C}

\begin{document}

\begin{frontmatter} 


\noindent
\textcolor{blue}{
Published in: Transportation Research Part C 118 (2020) 102710.}

\noindent
\textcolor{blue}{
Please cite this Paper as: Di, X., Chen, X., Talley, E., 2020. Liability design for autonomous vehicles and human-driven vehicles: A hierarchical game-theoretic approach. Transportation Research Part C: Emerging Technologies 118, 102710. doi:https://doi.org/10.1016/j.trc.2020.102710}
\title{Liability Design for Autonomous Vehicles and Human-Driven Vehicles: A Hierarchical Game-Theoretic Approach}

\date{\today}


\author[cu1,dsi]{Xuan Di\corref{cor1}}
\ead{sharon.di@columbia.edu}
\author[cu1]{Xu Chen}
\author[cu2,dsi]{Eric Talley}


\cortext[cor1]{Corresponding author. Tel.: +1 212 853 0435;}
\address[cu1]{Department of Civil Engineering and Engineering Mechanics, Columbia University}
\address[dsi]{Data Science Institute, Columbia University}
\address[cu2]{Law School, Columbia University}

\begin{abstract}
Autonomous vehicles (AVs) are inevitably entering our lives with potential benefits for improved traffic safety, mobility, and accessibility. However, AVs' benefits also introduce a serious potential challenge, in the form of complex interactions with human-driven vehicles (HVs). The emergence of AVs introduces uncertainty in the behavior of human actors and in the impact of the AV manufacturer on autonomous driving design. This paper thus aims to investigate how AVs affect road safety and to design socially optimal liability rules in comparative negligence for AVs and human drivers. 
A unified game is developed, including a Nash game between human drivers, a Stackelberg game between the AV manufacturer and HVs, and a Stackelberg game between the lawmaker and other users. We also establish the existence and uniqueness of the equilibrium of the game. The game is then simulated with numerical examples to investigate the emergence of human drivers' moral hazard, the AV manufacturer's role in traffic safety, and the lawmaker's role in liability design. Our findings demonstrate that human drivers could develop moral hazard if they perceive their road environment has become safer and an optimal liability rule design is crucial to improve social welfare with advanced transportation technologies. More generally, the game-theoretic model developed in this paper provides an analytical tool to assist policy-makers in AV policymaking and hopefully mitigate uncertainty in the existing regulation landscape about AV technologies. 
\end{abstract}


\end{frontmatter}

\section{Introduction}
\label{sec:intro} 

\subsection{Motivation}

\par Already an acknowledged transportation ``game changer", autonomous vehicles (AVs) are projected to arrive on public roads over the course of the next decade and disrupt the landscape of transportation ecosystems \citep{fag2015av}. Despite AVs' potential benefits for improved traffic safety, mobility, emissions, and accessibility \citep{gu2020lstm,huang2020game,huang2020stable,huang2019stable,kroger2019does,chen2017quantifying,wadud2016help,talebpour2016influence,fag2015av}, the first-of-its-kind traffic fatality in Tempe, Arizona \citep{tempe2018arizo,vergeUber} involving a self-driving automobile also elicited tremendous attention among the public and policy makers about who should be liable in the interaction between autonomous vehicles and human drivers, cyclists and pedestrians. 
There are several possibilities, ranging from the human operator to prevailing conditions to the behavior of the pedestrian.
From a legal perspective, these factual challenges are all but inevitable: Under Arizona law (and that of many other states), legal liability for accidents between automobiles and pedestrians typically involves a complex calculus of ``comparative fault” assessments for each of the aforementioned groups. The involvement of an autonomous vehicle can complicate matters further by adding other parties to the mix, such as the manufacturers of hardware and programmers of software. Insurance coverage distorts matters further by including third party insurers to the mix.
According to a letter issued by the Yavapai County Attorney \citep{uberdoc}, 
``there is no basis for criminal liability for the Uber corporation", and the accident is now under further investigation by National Transportation Safety Board (NTSB) \citep{ntsbdoc}. 
An article in Bloomberg \citep{bloomUber} indicated that no cause has been assigned to this crash based on a recently released NTSB report. 
Such uncertainty in accident laws in the presence of AVs urges a prudently designed legal and policy system in place.  
In this paper, we seek to understand the research question: 
\emph{In an accident involving AVs, what an efficient liability rule is to apportion loss between road users so that the total social cost is minimized}? 
A liability rule is ``efficient" when both the injurer and the victim executes optimal care levels that minimizes a total social cost \citep{jain2002efficient}.
Negligence-based liability is shown to be both necessary and sufficient to ensure an efficient liability rule \citep{jain2002efficient}, thus will be the focus of this paper. 

As the ecosystem of motorized transportation transitions from human- to autonomously-driven vehicles, many transportation and legal experts anticipate that the underlying role of law and regulation will play a pivotal role in mediating the myriad local interactions within that ecosystem \citep{marchau2019editorial}. 
Autonomous vehicle policies cover a broad spectrum, ranging from cybersecurity, privacy, to vehicle licensing and land use. Accordingly, recent years have seen a growing body of literature on identifying and addressing challenges brought forth by AVs in regulatory regimes \citep{shladover2017regulatory}, urban planning policies \citep{fraedrich2019autonomous}, and social interaction protocol design \citep{straub2019takes}. 
However, one of the biggest challenges is that, 
unlike manufacturers of conventional human-driven vehicles (HVs), AV manufacturers can directly influence traffic by programming driving algorithms, making manufacturers an indispensable actor to be modeled in the new ecosystem. 
The changing composition of human and autonomous vehicles makes pressing the need to apportion liability risk among accident victims and the businesses who manufacture, design and market AVs \citep{marchant2012coming,Geistfeld2017road,smith2017automated}. 
However, research that quantifies the economic efficiency of liability rules for AVs is still in its nascent stage \citep{chatterjee2013evolutionary,chatterjee2016understanding,eric2019law,shavell2019accident}. \cite{shavell2019accident} argued for what is effectively a Pigouvian tax on accidents in environments where AVs completely saturate the traffic ecosystem (e.g., at some future date when AVs have fully penetrated and dominate the market). Because this scenario is unlikely to unfold in the near term (or perhaps ever, when one considers the presence of pedestrians, cyclists, and other human traffic actors), we concentrate instead on mixed AV-HV environment as a more immediately interesting case. Like us, \cite{chatterjee2013evolutionary,chatterjee2016understanding,eric2019law} also explored these mixed environments, analyzing how varying the legal standard associated with negligence and contributory / comparative negligence can distort human drivers' interaction with AVs. Moreover, \cite{eric2019law} analyzed how various liability rules affect care levels of human drivers and AV manufacturers. Both studies do not assume that the AV is itself a strategic “actor”. Our analysis is distinct from these other efforts in at least two respects.  First, we posit a legal rule for inferring causation, which in turn is fed into a general class of comparative negligence rules.  And second, we do model the AV as having the capacity to take precautions, where its technology is the product of investments made by the manufacturer.

\subsection{Literature Review}

Tort law constitutes one of the oldest ``common law" institutions for regulating behavior, deriving its authority from longstanding precedent that has for centuries imposed on actors a legal duty to exercise appropriate precautions when engaging in activities that are potentially harm-creating.  Should a harm occur in such a context that is caused by the actor's fault (or breach of that duty), she can be held liable in damages to foreseeable victims who suffer from damages \citep{anderson2014autonomous}.  In cases where potential victims were also active participants capable of taking precautions, a subsidiary doctrine of ``contributory" or ``comparative" negligence also plays a role, either eliminating or limiting (respectively) recoverable damages even when an injuring party is at fault when the injured party was negligent in taking precautions. Jurisdictions tend to vary on what degree of fault is sufficient to trigger liability, ranging from strict liability on one end of the spectrum to no-fault on the other, with various versions of ``negligence" occupying the middle ground \citep{eric2019law}. During its centuries-long existence, accident law has been forced to evolve many times, often on the heels of significant technological shocks to human interaction that involve significant risks. In the transportation sector, the advent of the automobile proved particularly significant, in large part because the automobile introduced the salient challenge of how to apportion responsibility between a human driver causing harm (who may or may not have exercised care) and the vehicle itself (manufactured by a third party). Approximately a century ago, courts began routinely to allow injured parties to bring lawsuits not only against the injuring driver(s), but also against the manufacturer of either / both vehicles, alleging that they were produced in a fashion that caused them to be dangerously unsafe \citep{cardozo1996macpherson,hyltonl2013}.  Although it is often said to be a form of ``strict liability," products liability law itself also tends to vary in the extent of fault required, both to trigger potential liability and to measure comparative / contributory negligence \citep{heve2014res}. 
A now well-known literature in the economic analysis of law has studied how the various combinations of tort law and products liability are likely to perform in the context of human driven interactions \citep{shavell2007liability,shavell2009foundations}. An immediate implication of this literature is that driver interactions often involve joint investments in risk reduction -- a context that is challenging to regulate efficiently, since drivers bear direct costs of their own care taking but may not enjoy the full economic benefits. 

Development of liability rules requires one to first model the interactions between AVs and other road users. 
There has been a relatively long history of using game-theoretic approaches to model the interactions between road users and predict the behavior of these road users,  
when changes in road environment arise from new traffic management countermeasures \citep{yang2000comp,lim2005net} or technologies \citep{assum1999risk,risa1992moral}. 
Interested readers can refer to \cite{elvik2013review} for a comprehensive review of these studies.   

A majority of existing studies have focused on designing AVs' driving algorithms in various scenarios to ensure traffic efficiency, safety, or ethics \citep{reece1993computational,yoo2013stackelberg,kim2014game,yu2018human,wang2015game,Talebpour2015,wang2016cooperative,gong2016constrained,gong2018cooperative,li2018game,huang2020game,huang2020stable,huang2020stable,smith2019ethics}, 
but ignoring human drivers' behavioral adaptation to AVs as humans are exposed to more and more traffic encounters with AVs. 
Human drivers may have a weaker incentive to exercise ``due care" when faced with AVs. 
Since human drivers perceive AVs as
intelligent agents with the ability to adapt to more aggressive and potentially dangerous human driving behavior, the so-called ``moral hazard" effect may lower human driver's caution. 
It is a well-studied phenomenon in economics \citep{pedersen2001game,pedersen2003moral} 
and is also observed in traffic contexts \citep{boy1987rule,risa1992moral,chatterjee2013evolutionary,chatterjee2016understanding,millard2016pedestrians}. 
In a rear-end crash scenario, \cite{chatterjee2013evolutionary} demonstrated that human drivers tended to have long reaction times and short car-following headways as the proportion of AVs on the road increases, there-by offsetting, to a certain degree, the expected benefit of AVs in reducing the risk of rear-end crashes. In another study, \cite{millard2016pedestrians} showed that a pedestrian may cross an intersection recklessly, even if it is the AV's right-of-way, because of the general perception that AVs, equipped with smart sensing technologies, should be able to take evasive actions in a timely manner. 

\subsection{Contributions of This Paper}

The most relevant paper to our work is \cite{eric2019law}, which explored how accident law should adapt to the emergence of AVs using the multilateral precaution framework from the economics of tort law. Accordingly, no fault, strict liability, and a family of negligence-based rules, i.e., fault-based products liability, are all candidates for efficient legal rules. However, it did not explore the full range of strategic interactions between the marginal cost of manufacturer precautions and the possibility that the AV itself can take precautions on the road.

This paper aims to develop a game theory model to capture the strategic interactions among different agents, namely road users comprised of AVs and HVs, the AV manufacturer, and the lawmaker,   
that coexist in the transportation ecosystem. 
To eliminate human drivers' moral hazard and ensure an optimal design of autonomous driving algorithms, design of liability rules will be discussed. 
The contributions of this paper are:
\begin{enumerate}
	\item 
	Building on a good understanding of both AVs and HVs equilibrium behaviors in the developed game, we would like to explore human drivers' \emph{moral hazard} incurred by the presence of AVs. 
	
	\item 
	We aim to model how the AV manufacturer selects safety specifications for AVs as the market becomes larger. 
	Accordingly, the role of the AV manufacturer on traffic safety is explored.   
	
	\item 
	
	In addition to modeling traffic rules, our proposed traffic flow model will allow us to simulate a variety of permutations of conventional tort law and products liability law, varying the degree of fault required to trigger a liability obligation / defense. 
	Accordingly, a sequence of sensitivity analysis is performed to investigate how the transportation system performance may change when the related parameters vary. 
	
\end{enumerate}

The remainder of this paper is organized as follows. 
In Section \ref{sec:pre}, we first introduce the necessary knowledge needed to formulate games.    
In Section \ref{sec:hiergame}, a hierarchical game is formulated considering the game between road users (HVs and AVs), one between the AV manufacturer and HVs, and one between the lawmaker and other users.  
In Section \ref{sec:property}, the mathematical properties of the game are explored, including the existence and uniqueness of the equilibrium, and an algorithm to solve the equilibrium is devised.  
The game and the algorithm are tested on numerical results in Section \ref{sec:numerical}. 
Finally, conclusions and future work are presented in Section \ref{sec:conclusions}.

\section{Preliminaries}
\label{sec:pre}

\nomenclature[A, 01]{$c_{A}$}{care level of autonomous vehicles, which is predetermined by the AV manufacturer's sensors.}
\nomenclature[A, 03]{$c_{A_i}^{(AA)}$}{care level of autonomous vehicles in the $AA$ scenario where $i=1,2$ and $c_{A_i}^{AA}=c_A$.}
\nomenclature[A, 03]{$c_{A}^{(AH)}$}{care level of autonomous vehicles in the $AH$ scenario where $c_{A}^{AH}=c_A$.}
\nomenclature[A, 04]{$c_H^{(AH)}$}{care level of human drivers in the $AH$ scenario.}
\nomenclature[A, 05]{$c_{H_i}^{(HH)}$}{care level of human drivers in the $HH$ scenario where $i=1,2$.}
\nomenclature[A, 06]{$\hat{c_A}$}{upper bound of AVs' care level. }
\nomenclature[A, 07]{${\cal C}_{c_A}$}{feasible set for AVs' care level.}
\nomenclature[A, 06]{$\bar{c_H}$}{upper bound of HVs' care level. }
\nomenclature[A, 07]{${\cal C}_{c_H}$}{feasible set for HVs' care level.}
\nomenclature[A, 08]{$s_A^{(AH)}$}{share function of an AV when a crash happens in the $AH$ scenario.}
\nomenclature[A, 09]{$s_H^{(AH)}$}{share function of a HV when a crash happens in the $AH$ scenario.}
\nomenclature[A, 10]{$s_{H_i}^{(HH)}$}{share function of HV player $H_i$ when a crash happens in the $HH$ scenario where $i=1,2$.}
\nomenclature[A, 11]{$s_{A_i}^{(AA)}$}{share function of AV player $A_i$ when a crash happens in the $AA$ scenario where $i=1,2$.}
\nomenclature[A, 12]{$S_H$}{cost of executing a care level for a human driver, also called ``precaution cost".}
\nomenclature[A, 13]{$C_H^{(AH)}$}{cost of HVs in the $AH$ scenario.}
\nomenclature[A, 14]{$C_{H_i}^{(HH)}$}{cost of HVs in the $HH$ scenario where $i=1,2$.}
\nomenclature[A, 15]{$S_A$}{cost of executing a care level for an AV, also called ``precaution cost" or ``sensor cost", representing the cost of sensor production for the AV manufacturer.}
\nomenclature[A, 16]{$C_A$}{total cost for the AV manufacturer including sensor's cost and crash loss caused by AVs.}
\nomenclature[A, 17]{$\alpha$}{parameter of sensor cost function $S$.}
\nomenclature[A, 18]{$\beta$}{parameter of care level cost function $u$.}
\nomenclature[A, 19]{$w_a^{sen}$}{trade-off coefficient of sensors in the AV manufacturer's cost.}
\nomenclature[A, 19]{$w_a^{loss}$}{trade-off coefficient of the $AA$ scenario loss in the AV manufacturer's cost.}
\nomenclature[A, 20]{$w_h$}{trade-off coefficient in human drivers' cost.}
\nomenclature[B, 01]{$\sigma_H$}{care level standard of HVs, measuring the relative importance of HVs' contributions to car crashes.}
\nomenclature[B, 02]{$\sigma_A$}{care level standard of AVs, measuring the relative importance of AVs' contributions to car crashes.}
\nomenclature[B, 03]{$k$}{care level standard ratio, representing the lawmaker's punishments on road users where $k=\frac{\sigma_H}{\sigma_A}$.}
\nomenclature[B, 05]{${\cal C}_k$}{feasible set for the lawmaker's decision $k$.}
\nomenclature[B, 07]{$SC$}{social cost for a lawmaker, representing the sum of total crash loss and cost of precautions.}
\nomenclature[B, 07]{$TC$}{total cost of road users' care levels.}
\nomenclature[B, 08]{$w_l$}{trade-off coefficient in the lawmaker's payoff .}
\nomenclature[C,01]{$p$}{penetration rate(market share) of AVs, representing the proportion of AVs in the market.}
\nomenclature[C,02]{$P$}{crash probability of a car accident which is determined by road users' care level.}
\nomenclature[C,03]{$T$}{crash severity of a car accident which is determined by road users' care level.}
\nomenclature[C,03]{$L$}{crash loss of a car accident where $L=P\cdot T$.}
\nomenclature[C,04]{$TL$}{total crash loss on the road which is the sum of crash loss in all car scenarios.}
\nomenclature[C,04]{$R$}{crash rate for each car scenario.}
\nomenclature[C,04]{$TR$}{total crash rate in the market.}
\nomenclature[C,05]{$a$}{parameter of crash probability function $P$}
\nomenclature[C,06]{$h$}{parameter of crash probability function $P$}
\nomenclature[C,07]{$M$}{parameter of crash severity function $P$}
\nomenclature[C,08]{$s$}{parameter of crash severity function $P$}
\nomenclature[C,09]{$t$}{parameter of crash severity function $P$}

\printnomenclature

\subsection{Crashes for Three Types of Vehicle Encounters}
\label{sec:crashforthree}

On the road, there are three types of vehicular encounters. They are:  
\begin{enumerate}
	\item an HV encounters the other HV (i.e., $HH$ scenario). The two players in this scenario are denoted as $H_1^{(HH)}$ and $H_2^{(HH)}$, respectively;
	\item an AV encounters an HV (i.e., $AH$ scenario). The two players in this scenario are denoted as $A^{(AH)}$ and $H^{(AH)}$, respectively;
	\item an AV encounters the other AV (i.e., $AA$ scenario). The two players in this scenario are denoted as $A_1^{(AA)}$ and $A_2^{(AA)}$, respectively.  
\end{enumerate}

We first assume a designated market penetration rate of AVs and the probability of three types of vehicle encounters can be computed. 
Define 
$p$ as the AVs' market penetration rate. 
Accordingly, 
the probabilities of three encounters are 
$(1-p)^2$ for the $HH$ scenario, 
$2p(1-p)$ for the $AH$ scenario, 
and $p^2$ for the $AA$ scenario, respectively.  
The sum of three probabilities equals one. 

Based on different values of $p$, three markets can be defined, shown in Table \ref{table:market}. 
The market is a pure HV market when $p=0$. In a pure HV market, there is only one type of vehicle encounter, which is the $HH$ scenario. 
The market is a pure AV market when $p=1$. In a pure AV market, there is only one type of vehicle encounter, which is the $AA$ scenario. 
The market is a mixed AV-HV market when $0 < p < 1$. In a mixed AV-HV market, there are three types of vehicle encounters, which are the $HH$, $AA$, and $AA$ scenarios.

\begin{table}[H]
	\centering
	\begin{tabular}{c|c|c|c}
		\hline   \diagbox{Vehicle Encounter}{Market}& \makecell{\small{Pure HV Market} \\($p=0$)} & \makecell{\small{Mixed AV-HV Market} \\($0<p<1$)} & \makecell{\small{Pure AV Market}\\ ($p=1$)} \\
		\hline
		 HV-HV & \checkmark & \checkmark &  \\
		\hline
		 AV-HV & & \checkmark &  \\
		\hline
		 AV-AV & & \checkmark & \checkmark \\
		\hline
	\end{tabular}
	\caption{\small{Three markets with vehicular encounter scenarios}}
	\label{table:market}
\end{table} 

\begin{remark}
	In this paper, we are primarily focused on Level-5 vehicle automation, i.e., autonomous vehicles. Accordingly, three discrete encounter scenarios are included (HV-HV, AV-HV, AV-AV). 
	We acknowledge that in-between degree of autonomous driving already exists in reality, considering Level-2 or 3 automated vehicles. 
	Also, HVs manufacturers play a crucial role in the incorporation of semi-autonomous driving functions into HVs, leading to a spectrum of mixed automation. 
	However, modeling these scenarios requires a deep understanding of human factor and human-machine interaction, which goes beyond the scope of this paper. 	
\end{remark}


Within one encounter, an accident may or may not happen. 
The occurrence of an accident depends on the care level of two vehicles. Care level is an abstract quantity to represent how attentive one driver is while driving:   
A higher care level implies a more cautious attitude to road safety. 
For human drivers, it can be specified as the difference between reaction time and following headway in the context of rear-end crashes \citep{chatterjee2013evolutionary,chatterjee2016understanding}. 
For AVs, their care level, determined by the AV manufacturer in the production process, represents a safety specification level \citep{shavell2019accident}. 
Such a specification is determined by both hardware and software of sensors with a combined performance of sensing, computational (tracking, perception, and recognition), and actuating.
Accordingly, in this paper, we will use sensor quality to represent AVs' safety level. 
The more expensive sensors are deployed, the higher care level AVs take, and the safer these AVs are.     
We use $c_{H_i}^{(HH)},\ i=1,2$ to represent HVs' care level in the $HH$ scenario, $c_{A_i}^{(AA)},\ i=1,2$ to represent AVs' care level in the $AA$ scenario and $c_{A}^{(AH)},\ c_{H}^{(AH)}$ to represent HVs' and AVs' care level in the $AH$ scenario. 

The higher care level one vehicle chooses, the less likely an accident happens. 
We thus define a crash probability function between vehicles $i$ and $j$ as $P(c_i,c_j)$, which is monotonically decreasing with respect to care levels \citep{pedersen2003moral}.  
In other words, 
$\frac{\partial P(c_i,c_j)}{\partial c_i}<0,\ \frac{\partial P(c_i,c_j)}{\partial c_j}<0$. 
Crash probability functions for each scenario are $P(c_{A_1}^{(AA)},c_{A_2}^{(AA)})$, $P(c_{H_1}^{(HH)},c_{H_2}^{(HH)})$, and $P(c_A^{(AH)},c_H^{(AH)})$, respectively. 
	In addition, we define a crash severity function between vehicles $i$ and $j$ as $T(c_i,c_j)$, which is monotonically decreasing with respect to care levels \citep{pedersen2003moral,chatterjee2016understanding}. In other words, $\frac{\partial T(c_i,c_j)}{\partial c_i}<0,\ \frac{\partial T(c_i,c_j)}{\partial c_j}<0$. Crash severity functions for each scenario are $T(c_{A_1}^{(AA)},c_{A_2}^{(AA)})$, $T(c_{H_1}^{(HH)},c_{H_2}^{(HH)})$, and $T(c_A^{(AH)},c_H^{(AH)})$, respectively. 
	Mathematically, crash loss of each scenario is denoted by $L(c_{A_1}^{(AA)},c_{A_2}^{(AA)})=P(c_{A_1}^{(AA)},c_{A_2}^{(AA)}) \cdot T(c_{A_1}^{(AA)},c_{A_2}^{(AA)})$, $L(c_{H_1}^{(HH)},c_{H_2}^{(HH)})=P(c_{H_1}^{(HH)},c_{H_2}^{(HH)}) \cdot T(c_{H_1}^{(HH)},c_{H_2}^{(HH)})$ and $L(c_A^{(AH)},c_H^{(AH)})=P(c_A^{(AH)},c_H^{(AH)}) \cdot T(c_A^{(AH)},c_H^{(AH)})$.


Based on encounter probability and crash probability, we define crash rate for each scenario. Crash rate shows the proportion of cars involved in crashes, equalling the product of encounter probability and crash probability. Accordingly, the crash rates are $R(c_{H_1}^{(HH)},c_{H_2}^{(HH)})=(1-p)^2 \cdot P(c_{H_1}^{(HH)},c_{H_2}^{(HH)})$ for the $HH$ scenario, $R(c_{A_1}^{(AA)},c_{A_2}^{(AA)})=p^2 \cdot P(c_{A_1}^{(AA)},c_{A_2}^{(AA)})$ for the $AA$ scenario and $R(c_A^{(AH)},c_H^{(AH)})=2p(1-p) \cdot P(c_A^{(AH)},c_H^{(AH)})$ for the $AH$ scenario. Total crash rate equals the sum of crash rates of all scenarios:
\begin{linenomath*}
\begin{equation}
  TR(c_{H_1}^{(HH)},c_{H_2}^{(HH)},c_{A_1}^{(AA)},c_{A_2}^{(AA)},c_A^{(AH)},c_H^{(AH)})=R(c_{H_1}^{(HH)},c_{H_2}^{(HH)})+R(c_{A_1}^{(AA)},c_{A_2}^{(AA)})+R(c_A^{(AH)},c_H^{(AH)}).
\end{equation}
\end{linenomath*}

Note that although AVs are likely to have greater powers in perception with a higher standard of care than human drivers, car crashes may still occur in the $AA$ scenario \citep{shavell2019accident}.

\subsection{Share Function} 


A liability rule is a mechanism to apportion direct costs associated with an accident between involved drivers. 
As we mentioned in the previous section, \cite{eric2019law} and \cite{shavell2019accident} discussed the possibility of using no fault, strict liability, and negligence-based liability policies for AVs. 
In this paper, we will use negligence-based liability, comparative negligence in particular, for both driver liability and product liability to regulate human's driving behavior and AVs' driving algorithms. 

Negligence-based liability, including contributory and comparative negligence \citep{gary1978comp,parisi2004cause,comp2012con}, are primarily designed to incentivize one to exercise ``due care", which is a predefined care level threshold. 
Contributory negligence stipulates that the negligent driver pays for the entire crash loss. 
If both drivers are negligent, they have the same liability. 
Comparative negligence, on the other hand, divides crash loss according to drivers' relative contribution to a crash. 
In other words, one's negligence can be quantified by the difference between her care level and a predefined standard, 
which is a desirable care level one should execute while driving on the road. 
A causation function -- ``share function" is used to measure one's loss share based on her comparative negligence level to the care level standard. 
Different causation functions have been proposed in the literature \citep{parisi2004cause,singh2007com,chatterjee2016understanding,eric2019law}. 
For example, one's causation contribution depended on both care level and activity level in \cite{parisi2004cause,singh2007com}. 
Assuming a constant driving activity level, \cite{chatterjee2016understanding,eric2019law} defined one's causation contribution as the ratio of her relative care level (to the due care) to the sum of two players' relative care levels. 
However, these functions become problematic when a crash happens even if care levels of both parties are higher than the due care. 
For example,  
\cite{chatterjee2016understanding} implicitly assumed that no crash happened if care levels of both parties were higher than the due care, and thus no share loss was defined;     
If the share function in \cite{eric2019law} is used, a higher loss is assigned to the player whose relative care level to the due care is higher, which is undesirable. 
Accordingly, we propose the following share function to address the above shortcomings: 
\begin{linenomath*}
	\begin{equation}
	s_{i}(c_{i}^{'},c_{j}^{'},\sigma_{i}^{'},\sigma_{j}^{'})=\frac{e^{\sigma_i^{'}-c_i^{'}}}{e^{\sigma_i^{'}-c_i^{'}}+e^{\sigma_{j}^{'}-c_j^{'}}}, 
	\end{equation}
\end{linenomath*} 
where $s_{i}$ is the share function of road user $i$.  
$c_{i}^{'},\ c_{j}^{'}$ are care level of road users $i$ and $j$ in a crash, respectively. 
$\sigma_{i}^{'},\ \sigma_{j}^{'}$ are care level standards for road users $i$ and $j$, respectively. 
\begin{remark}
	The care level standard $\sigma$ is the decision variable of the lawmaker. The lawmaker changes the liability rule by specifying the care level standards for two road users. 	
	In other words, $\sigma$ is exogenous to the two games between HVs, and between HVs and the AV manufacturer, but endogenous to the game between the lawmaker and other road users.  When we define the share function here, we treat $\sigma$ as an exogenous parameter. 
\end{remark}

To simplify it, we convert the exponential form into a linear fractional form by substituting  $e^{c_i^{'}}$ with $c_i$, $e^{c_j^{'}}$ with $c_j$, $e^{\sigma_i^{'}}$ with $\sigma_i$, and $e^{\sigma_j^{'}}$ with $\sigma_j$, 
where $c_{i},\ c_{j}$ are care level of road user $i$ and $j$ in a crash, and $\sigma_{i},\ \sigma_{j}$ are care level standards for road user $i$ and $j$, respectively. 
The share function is then reformulated as:
\begin{linenomath*}
	\begin{equation}
	s_{i}(c_{i},c_{j},\sigma_{i},\sigma_{j})=\frac{\sigma_{i}c_i^{-1}}{\sigma_{i}c_i^{-1}+\sigma_{j}c_j^{-1}},
	\end{equation}
\end{linenomath*} 

Define the care level standards for HVs and AVs as $\sigma_{H}$ and $\sigma_{A}$, respectively. 
The share function for the $HH$ scenario is:  
\begin{linenomath*}
	\begin{equation}
	s_{H_i}^{(HH)} (c_{H_1}^{(HH)},c_{H_2}^{(HH)},\sigma_{H}) =\frac{\sigma_{H} (c_{H_i}^{(HH)})^{-1}}{\sigma_{H} (c_{H_1}^{(HH)})^{-1}+\sigma_{H} (c_{H_2}^{(HH)})^{-1}},\ i=1,2, 
	\end{equation}
\end{linenomath*}
where $s_{H_i}^{(HH)}$ is the share function and $c_{H_i}^{(HH)}$ is the care level of HV player $H_i$.

The share function for the $AH$ scenario is:  
\begin{equation}
\begin{aligned}
	s_{A}^{(AH)}(c_A^{(AH)},c_H^{(AH)},\sigma_A,\sigma_H)=\frac{\sigma_{A} (c_{A}^{(AH)})^{-1}}{\sigma_{H} (c_{H}^{(AH)})^{-1}+\sigma_{A} (c_{A}^{(AH)})^{-1}},\\ s_{H}^{(AH)}(c_A^{(AH)},c_H^{(AH)},\sigma_A,\sigma_H)=\frac{\sigma_{H} (c_{H}^{(AH)})^{-1}}{\sigma_{H} (c_{H}^{(AH)})^{-1}+\sigma_{A} (c_{A}^{(AH)})^{-1}},
\end{aligned}
\end{equation}  
where $s_{A}^{(AH)},s_{H}^{(AH)}$ are share functions and $c_{A}^{(AH)},c_{H}^{(AH)}$ are care levels of AV player $A$ and HV player $H$. 

The share function for the $AA$ scenario is:  
\begin{linenomath*}
	\begin{equation}
	s_{A_i}^{(AA)}(c_{A_1}^{(AA)},c_{A_2}^{(AA)},\sigma_A)=\frac{\sigma_{A} (c_{A_i}^{(AA)})^{-1}}{\sigma_{A} (c_{A_1}^{(AA)})^{-1}+\sigma_{A} (c_{A_2}^{(AA)})^{-1}},\ i=1,2,
	\end{equation}
\end{linenomath*} 
where $s_{A_i}^{(AA)}$ is the share function and $c_{A_i}^{(AA)}$ is the care level of AV player $A_i$.

\subsection{Crash Loss Share} 
\label{sec:crashlossshare}

A share function apportions an expected crash loss of an accident between two involved parties. 
Denote the average crash loss of a car accident as $T$. Then the expected crash loss for an accident is the product of crash probability and average loss. 
Accordingly, one's crash loss share is the expected crash loss of an accident times its share. 

\noindent In the $HH$ scenario, the crash loss share for player $H_i^{(HH)}$ is: 
$L(c_{H_1}^{(HH)},c_{H_2}^{(HH)}) \cdot s_{H_i}^{(HH)}(c_{H_1}^{(HH)},c_{H_2}^{(HH)},\sigma_H)$. 

\noindent In the $AH$ scenario, the crash loss share for player $H^{(AH)}$ is: $L(c_{A}^{(AH)},c_{H}^{(AH)})\cdot s_{H}^{(AH)}(c_{H}^{(AH)},c_{A}^{(AH)},\sigma_A,\sigma_H)$  and that for player $A^{(AH)}$ is: $L(c_{A}^{(AH)},c_{H}^{(AH)})\cdot s_{A}^{(AH)}(c_{H}^{(AH)},c_{A}^{(AH)},\sigma_A,\sigma_H)$. 

\noindent In the $AA$ scenario, the crash loss share for player $A_i^{(AA)}$ is: $L(c_{A_1}^{(AA)},c_{A_2}^{(AA)})\cdot s_{A_i}^{(AA)}(c_{A_1}^{(AA)},c_{A_2}^{(AA)},\sigma_A)$.


To summarize, Table \ref{table:scenario} shows crash probability, crash rate, and crash loss share for each vehicle encounter scenario. 
\begin{table}[H]
	\centering
	\begin{tabular}{c|c|c|c|c}
		\hline Scenario  & Player &Crash loss & Crash rate &Crash loss share\\
		\hline \tiny{$HH$} & \tiny{$H_1^{(HH)}$,$H_2^{(HH)}$} & \tiny{$L(c_{H_1}^{(HH)},c_{H_2}^{(HH)})$} & \tiny{$(1-p)^2  P(c_{H_1}^{(HH)},c_{H_2}^{(HH)})$} &\tiny{$L(c_{H_1}^{(HH)},c_{H_2}^{(HH)})\cdot s_{H_i}^{(HH)},\ i=1,2$.} \\
		\hline \tiny{$AH$} & \tiny{$A^{(AH)}$,$H^{(AH)}$} &\tiny{$L(c_{A}^{(AH)},c_{H}^{(AH)})$} & \tiny{$2p(1-p) P(c_{A}^{(AH)},c_{H}^{(AH)})$} &\makecell{\tiny{$L(c_{A}^{(AH)},c_{H}^{(AH)})\cdot s_{H}^{(AH)}$,}\\ \tiny{$L(c_{A}^{(AH)},c_{H}^{(AH)}) \cdot s_{A}^{(AH)}$.}} \\
		\hline \tiny{$AA$} & \tiny{$A_1^{(AA)}$,$A_2^{(AA)}$} &
		\tiny{$L(c_{A_1}^{(AA)},c_{A_2}^{(AA)})$} & \tiny{$p^2 P(c_{A_1}^{(AA)},c_{A_2}^{(AA)})$}&\tiny{$L(c_{A_1}^{(AA)},c_{A_2}^{(AA)}) \cdot s_{A_i}^{(AA)},\ i=1,2$.}  \\
		\hline
	\end{tabular}
	\caption{\small{Quantities for three vehicular encounter scenarios}}
	\label{table:scenario}
\end{table} 

\section{Game Formulations} 
\label{sec:hiergame}

In this section, we will state the problem 
and then introduce its mathematical formulations.  
Assumptions we will use for the game-theoretic framework are first presented: 
\begin{enumerate} 
	\item There is only one AV manufacturer that produces one type of sensor. Thus all AVs on the road use the same type of sensors. 
	Mathematically, $c_A^{(AH)}= c_{A_1}^{(AA)} = c_{A_2}^{(AA)} \equiv c_A$. 
	Accordingly, when there is an accident happening between two AVs, the AV manufacturer endures the entire crash loss. 
	
	\item The more expensive the sensors are, the less likely a crash happens; 
	and vice versa. 	
	Define the sensor cost as $S_A(c_A)$, which is an increasing function with respect to $c_A$. 
	Mathematically, $\frac{dS_A(c_A)}{dc_A} > 0$. 
	\item To achieve the same care level $c$, the cost of precautions for AVs (i.e., sensor cost) should be lower than that for HVs. Mathematically,  
	$\frac{dS_H(c_H)}{dc_H}|_{c_H=c}>\frac{dS_A(c_A)}{dc_A}|_{c_A=c} >0 $,  
	where $S_H(c_H)$ is the precaution cost for HVs to exercise care level $c_H$, an increasing function with respect to $c_H$. 
	In addition, the upper limit of AVs' care level should be higher than that of HVs, mathematically, $ub_{c_A} \geq ub_{c_H}$. 
	In other words, AVs may have lower care level than HVs but the highest care level an AV can achieve (i.e., the best performance) should be higher than that of an HV, otherwise AVs will not bring traffic safety benefit. 
	\item 
	Human drivers share the same care level standard $\sigma_{H}$.  
	Also, all AVs share the same care level standard $\sigma_{A}$.  
	Accordingly, the share function in the $HH$ scenario can be simplified to $s_{H_i}^{(HH)} =\frac{(c_{H_i}^{(HH)})^{-1}}{(c_{H_1}^{(HH)})^{-1}+ (c_{H_2}^{(HH)})^{-1}},\ i=1,2$. 
	The share function in the $AA$ scenario can be reduced to $s_{A_1}^{(AA)}=s_{A_2}^{(AA)}=\frac{1}{2}$. 
	Because these two reduced share functions do not contain care level standard any more, it says that the lawmaker plays no role in risk apportion in neither the pure HV nor the pure AV markets.
	\item 
	Three vehicular encounter scenarios are independent of one another. In other words, one's care level change in one scenario does not influence her choice in other scenarios. 
	For example, 
	human drivers only adapt their care levels in the $AH$ scenario while encountering AVs  
	but maintain the same care levels in the $HH$ scenario while encountering other human drivers.
	\item 
	Human drivers have complete information about vehicle encounter scenarios. In other words, they know whether it is an AV or another HV they encounter.
\end{enumerate}

A mixed traffic system is comprised of hierarchical decision makers who make choices at different levels: 
\begin{enumerate}
	\item Law-maker: 
	the leader (i.e., the top level player) develops an optimal driver liability rule for human drivers and a product liability rule for AV manufacturers, 
	aiming to minimize the social cost (i.e., the total crash loss plus the total cost of care levels); 
	
	\item AV manufacturer: 
	the mid-level player 
	determines driving settings or AVs' care level 
	in order to minimize its expected payoff 
	over all possible traffic crashes involved with AVs (i.e., the sensor cost plus the expected crash loss). 
	In this paper, we assume there is only one AV manufacturer. Its generalization to multiple AV manufacturers will be left for future research. 
	
	The AV manufacturer makes a one-time investment regarding what types of sensors to purchase. 
	Once sensors are chosen and installed, AVs are put on public roads and may subject to potential risks of crashing into other vehicles. 
	
	
	
	
	\item Human driver: 
	the lower-level player (i.e., the follower) 
	selects the level of care 
	to maximize her own utility, 
	which is her cost of executing precaution plus the crash loss for one accident. 
	
	One driver can be involved in an accident with another HV or with an AV. 
	For the accident between two HVs, the driver liability rule is applied to apportion the crash loss. 
	For the accident between one HV and one AV, the driver liability rule is applied to the HV while the product liability rule is applied to the AV. 	
\end{enumerate}

\begin{figure}[H]
	\centering
	\subfloat[$0<p<1$ \label{fig:structure_hier}]{\includegraphics[width=4.5in]{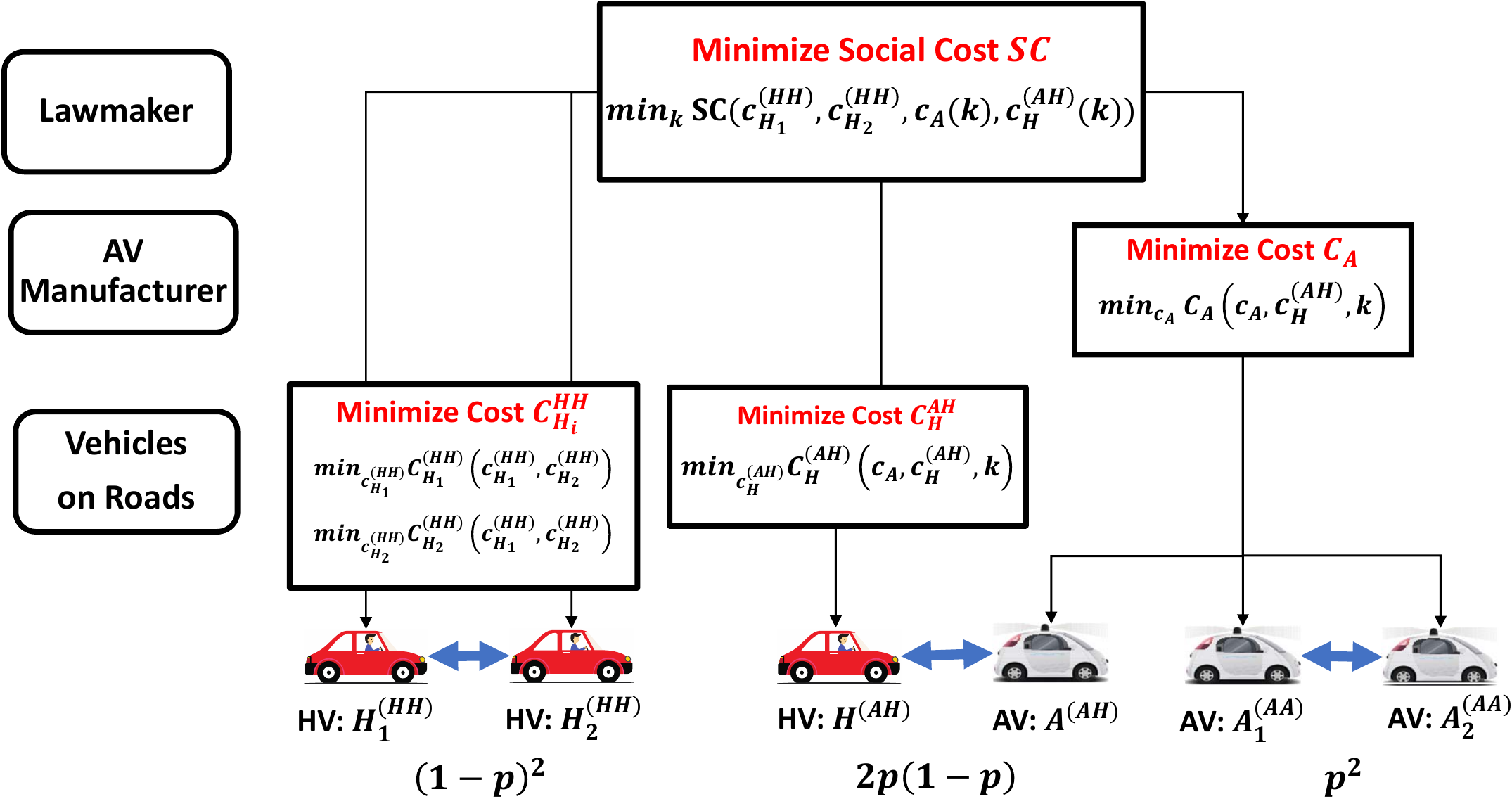}} 

	\subfloat[$p=0$ \label{fig:structure_purehv}]{\includegraphics[width=2in]{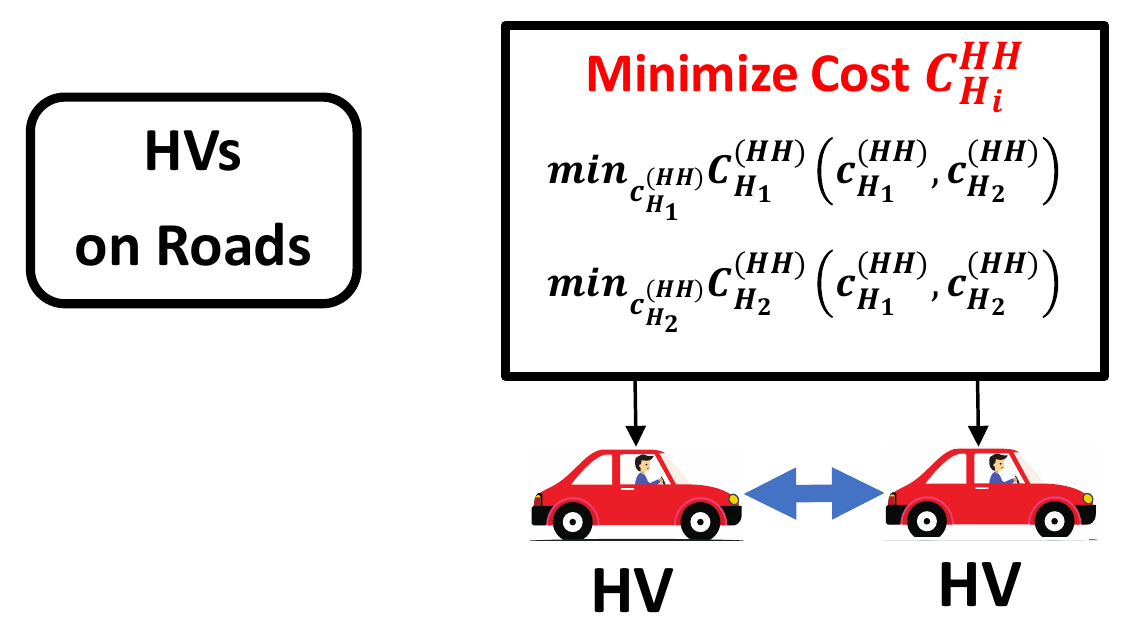}}
	\hspace{2.5cm}
	\subfloat[$p=1$ \label{fig:structure_pureav}]{\includegraphics[width=2in]{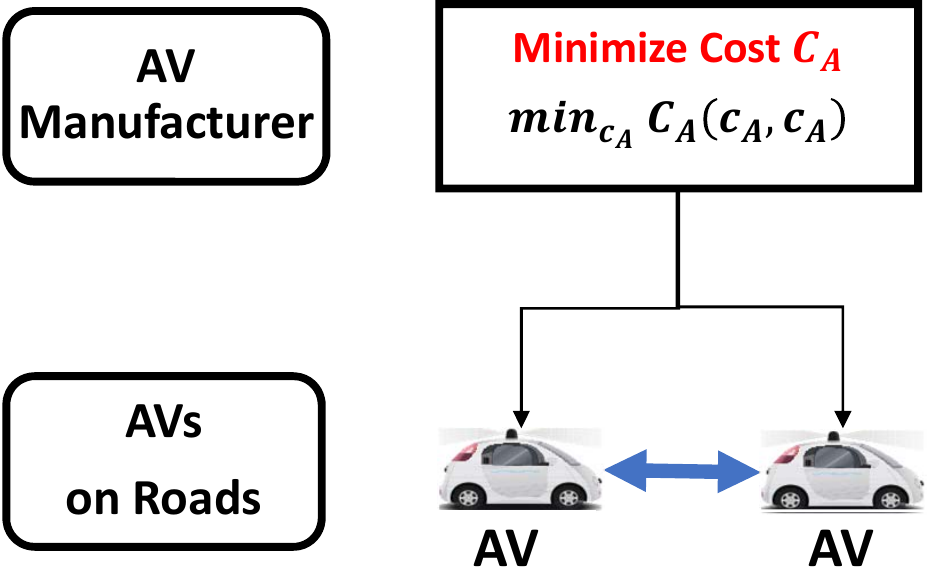}}
	\caption{Hierarchical game structure for three markets}
	\label{fig:structure}
\end{figure}

In summary, 
on the top level, the lawmaker is the leader to make a high-level decision on the driver liability rule and the product liability rule. 
Once the legal framework is formed, AV manufacturers decide on the strategic level as to what safety levels their AVs are by selecting sensors that determine one AV's level of care while driving on roads. 
On public roads, human drivers make an operational decision regarding the level of care while driving. 
To model decisions of various decision-makers on different levels, a hierarchical game-theoretic framework, leader-follower game in particular, needs to be developed. 
\figref{fig:structure} illustrates the structure of the game framework for three markets, respectively. 

In the next three subsections, we will introduce the game on each level, from the bottom level to the top level. 
In each game, we will define players' decision variable and disutility functions, 
the game formulation, 
and the equilibrium condition.  
The games will be parameterized by the AV penetration rate $p$, to reflect the impact of AV fleet size on each game. 

\subsection{Game Between Human Drivers}
\label{sec:humandriversgame}

On the lower level game, human drivers are strategic game players  
interacting with one another on public roads 
to minimize their own cost by choosing care levels.  
A two-player simultaneous Nash game is developed for the $HH$ scenario that could happen in both the pure HV market and the mixed AV-HV market.  
The set-up of this game is similar to the road safety game proposed by \cite{pedersen2001game}. 

\noindent
\textbf{Players:} 
Human drivers $H_i(i=1,2)$ play a symmetric pairwise Nash game. 

\noindent
\textbf{Decision Variables:} 
The driving decision the human driver $H_i$ chooses in a road game is her care level or the level of precaution in the $HH$ scenario, denoted as $c_{H_i}^{(HH)}$. Define ${\cal C}_{c_H}:(0,\bar{c_H}) \subset R^{+}$ as the feasible set for players' care-level and $\bar{c_H}$ is the upper bound. 

\noindent
\textbf{Cost Function:} While driving, human drivers aim to maintain traffic safety (i.e., to minimize total crash loss) 
and efficiency (i.e., to execute less care level if possible) \citep{pedersen2001game,chatterjee2013evolutionary}. 
The cost function of a human driver $H_i$ in the $HH$ scenario, denoted as $C_{H_i}^{(HH)}(c_{H_1}^{(HH)},c_{H_2}^{(HH)})$, 
is comprised of two parts: 
the cost of executing a care level and the crash loss share. Denote $S_H(c_{H_i}^{(HH)})$ as the cost of executing a care level for driver $H_i$:  
the higher the care level is executed, the higher the cost is, i.e.,  
$\frac{d S_H(c_{H_i}^{(HH)})}{d c_{H_i}^{(HH)}}>0$. 
Considering the crash loss share defined in Section \ref{sec:crashlossshare}, 
then the disutility function can be formulated as: 
\begin{linenomath*}
	\begin{equation}
	C_{H_i}^{(HH)}(c_{H_1}^{(HH)},c_{H_2}^{(HH)})=\underbrace{w_h \cdot S_H(c_{H_i}^{(HH)})}_{\text{cost of executing a care level}}+\ \ \ \ \ \ \ \ \underbrace{ (1-w_h) \cdot L(c_{H_1}^{(HH)},c_{H_2}^{(HH)}) \cdot s_{H_i}^{(HH)}(c_{H_1}^{(HH)},c_{H_2}^{(HH)})}_{\text{crash loss apportioned to $H_i$}},\ i=1,2. 
	\label{equ:tradeoffhh}
	\end{equation}
\end{linenomath*} 
where $w_h$ and $1-w_h$ are trade-off coefficients in human drivers' cost function. 
\begin{remark}
	\begin{enumerate}
	\item The coefficient ratio $\frac{w_h}{1-w_h}$ represents a marginal rate of substitution between precaution and crash loss, although it can also be treated as a marginal rate between risky driving and probability for a car crash. 
	The same explanation can be applied to other trade-off coefficients defined in the subsequent disutility functions. 
	\item Human drivers are generally heterogeneous, meaning that they could have different cost functions, resulting in different care levels in the same environment. In this paper, we assume human's heterogeneity is caused by the randomness in $w_h$. In other words, $w_h$ may be a random variable instead of a constant. How the probability distribution of $w_h$ influences HVs' care levels will be discussed in Section~\ref{sec:hetero_hv}. 
	\end{enumerate}
\end{remark}

\noindent
\textbf{Game:} 
Each human driver makes a trade-off between minimization of care level and minimization of crash loss. 
\begin{equation}
\begin{aligned}
    &[\mbox{\textbf{GameHH}}]\min \limits_{c_{H_i}^{(HH)} \in {\cal C}_{c_H}} && C_{H_i}^{(HH)}(c_{H_1}^{(HH)},c_{H_2}^{(HH)}),\ i=1,2
     \label{equ:simuta}
\end{aligned}
\end{equation}

The equilibrium of this game, denoted as $(c_{H_1}^{*(HH)},c_{H_2}^{*(HH)})$, represents the optimal care level for human drivers $H_1, H_2$. 

\begin{remark}
	According to Assumption 5, 
	the equilibrium care level solved from [GameHH] for the HH scenario remains the same in both the pure HV market and the mixed AV-HV market. 
	In other words, an HV selects the same care level whenever she meets another HV, regardless of the market.
\end{remark}



\noindent
\textbf{Equilibrium condition:}
At equilibrium, no human drivers can improve her utility by unilaterally switching her care level, i.e., 
$C_{H_i}^{(HH)}(c_{H_i}^{*(HH)}) \leqslant C_{H_i}^{(HH)}(c_{H_i}^{(HH)}), \forall c_{H_i}^{(HH)} \in {\cal C}_{c_H}$.

\subsection{Game Between the AV Manufacturer and HVs}
\label{sec:hieravmanu}

In the AV-HV mixed market, 
human drivers not only interact with one another in the $HH$ scenario, but also interact with AVs in the $AH$ scenario. 
Unlike human drivers, AVs are not strategic players because their care levels are predetermined by the AV Manufacturer. 
The AV Manufacturer trades off between investment cost of sensors and an average crash loss over all possible accidents involved with AVs.  
Accordingly, the AV Manufacturer and HVs play a leader-follower game 
in which the AV Manufacturer stipulates a care level for its AV fleet and human drivers select their care levels while encountering AVs on roads.

\begin{remark}
\begin{enumerate}
    \item We will specify the AV penetrate rate in two ways: one as an exogenous parameter and the other as an endogenous parameter. 
	In the latter case, the penetration rate $p$ is endogenized as an inverse function of crash loss \citep{henrik2005safety}, indicating that the higher the crash loss is, the lower market share the AV manufacturer would get. 	
	The rationale is, at the initial stage of the technology adoption, the most harmful loss for the AV manufacturer is the external loss from the reduced public opinion (after the crash is released to the society) rather than the direct loss of the crash. 
	The endogenous market will be discussed in Section~\ref{sec:sen5}.
    \item The AV penetration rate can also be explained as consumers' willingness to purchase AVs. Accordingly, $1-p$ represents the proportion of people who purchase HVs. Thus, HV manufacturers also play a crucial role in traffic safety. However, modeling consumers’ purchasing behavior will complicate the analysis and weaken the main message of this paper on AVs' safety impact, which will be left for future research. 
\end{enumerate}
\end{remark}

\noindent
\textbf{Players:} 
The AV Manufacturer is a leader and HVs are followers. 

\noindent
\textbf{Decision Variables:}
The AV manufacturer makes a one-time investment on what types of sensors to purchase,  
which is a trade-off between sensor investment and the total crash loss involved with AVs. 
Denote the care level of AVs as $c_A$ and the precaution cost associated with the care level $c_A$ as $S_A(c_A)$, named ``sensor cost" for short. 
Because AVs' care level is critically determined by sensors, the cost associated with care level will be also referred to as sensor (production) cost in the rest of the paper. 
Human drivers' decision variables are care levels $c_{H}^{(AH)}$ as defined before. In addition, feasible sets for $c_A,c_{H}^{(AH)}$ are ${\cal C}_{c_A}:(0,\bar{c_A}) \subset R^+$ and ${\cal C}_{c_H}:(0,\bar{c_H}) \subset R^+$, respectively. 

\begin{remark}
   Sensor cost $S_A(c_A)$ also depends on the AV penetration rate, i.e., $S_A(c_A,p)$, when the scaling effect of mass production is achieved. For notation simplicity, we will omit the parameter $p$ and use $S_A(c_A)$ to represent the sensor cost. The scaling effect of mass production is motivated by the fact that the production cost of AVs at the initial deployment stage is prohibitively high, for example, the estimated cost of one AV is $\$30,000 - 85,000$ \citep{shche2014avcost}. But it would not remain the same level always. Some estimated that AVs' cost may drop to $\$25,000 - 50,000$ each \citep{fag2015av} due to large-scale production.
\end{remark}

\noindent
\textbf{Cost Function:} The AV manufacturer's cost function, denoted as $C_{A}(c_A,c_{H}^{(AH)})$, is the average sensor cost plus the expected crash loss. To calculate the crash loss, two scenarios are relevant to the AV manufacture: $AH$ and $AA$ encounters. 
The average cost is calculated as: 
\begin{linenomath*}
\begin{equation}
\begin{aligned}
   \scriptsize{
   C_{A}(c_A,c_{H}^{(AH)}) = \underbrace{ w_a^{sen} \cdot p \cdot S_A(c_A)}_{\text{sensor cost}}+ \overbrace{ \underbrace{w_a^{loss} \cdot p^2 \cdot L(c_A,c_A) }_{\text{crash loss of $AA$ scenario}}+\underbrace{ (1-w_a^{sen}-w_a^{loss}) \cdot 2p(1-p) \cdot L(c_A,c_{H}^{(AH)}) \cdot s_{A}^{(AH)}(c_A,c_{H}^{(AH)})}_{\text{AV's loss share in the $AH$ scenario}}}^{\text{crash loss involved with AVs}}},
\label{equ:manutradeoff}
\end{aligned}
\end{equation}
\end{linenomath*}
where $w_a^{sen}$, $w_a^{loss}$, $1-w_a^{sen}-w_a^{loss}$ are trade-off coefficients, representing the weight the AV manufacturer places on sensor cost and crash loss. 

On public roads, non-strategic AV players and strategic HV players interact with one another. 
The care level for non-strategic AVs, i.e., $c_A$, is predetermined through minimization of the AV manufacturer's cost function. 
Strategic HV players choose care levels to minimize their own cost when they encounter AVs. Similar to Equation \ref{equ:tradeoffhh}, human drivers' disutility in the $AH$ scenario, denoted as $C_{H}^{(AH)}(c_{H}^{(AH)},c_{A})$, is defined as

\begin{linenomath*}
\begin{equation}
       C_{H}^{(AH)}(c_{H}^{(AH)},c_{A})=\underbrace{w_h \cdot S_H(c_{H}^{(AH)})}_{\text{cost of executing a care level}}+\ \ \ \ \ \ \underbrace{(1-w_h)\cdot L(c_{A},c_{H}^{(AH)}) \cdot s_{H}^{(AH)}(c_{A},c_{H}^{(AH)})}_{\text{crash loss apportioned to HV}}.
       \label{equ:tradeofhumanddriverinah}
\end{equation}
\end{linenomath*}

\begin{remark}
     The AV manufacture's care level is affected by human's care levels, i.e., $c_{H}^{(AH)}$, in the AV scenario. When human drivers are heterogeneous, they could select different care levels while encountering one AV. 
    In this case, the AV manufacturer's decision is affected by such heterogeneity, which will be discussed in Section~\ref{sec:hetero_hv}.
\end{remark}

\noindent
\textbf{Game:}
The AV manufacturer aims to select a set of optimal sensors to minimize its total cost, which depends on the care level of HV players in the $AH$ scenario.
Accordingly, the AV manufacturer and human drivers are the leader and the follower, respectively. Given $p \in (0,1)$, the game is formulated as: 
\begin{equation}
\begin{aligned}
    &[\mbox{\textbf{GameAH}}]\min \limits_{c_A \in {\cal C}_{c_A}} && C_{A}(c_A, c_{H}^{(AH)}) \\ 
    &\st &&c_{H}^{(AH)} \in
     \begin{aligned}[t]
        &\argmin_{c_{H}^{(AH)} \in {\cal C}_{c_H}} && C_{H}^{(AH)}(c_{H}^{(AH)}, c_A) 
     \end{aligned}
     \label{equ:stackel}
\end{aligned}
\end{equation}
The equilibrium of the game is denoted by $(c_A^{*}(p),c_H^{*(AH)}(p)),\ p \in (0,1)$. 

\textbf{Equilibrium condition:}
At equilibrium, the AV manufacturer and human drivers cannot lower their cost by unilaterally switching their respective decisions, i.e., 
$C_{A}(c_A^{*}(p),c_H^{*(AH)}(p)) \leqslant C_{A}(c_A,c_H^{*(HH)}(p)), \forall c_{A} \in {\cal C}_{c_A}$ and  
$C_{H}^{(AH)}(c_{H}^{*(AH)}(p),c_A) \leqslant C_{H}^{(AH)}(c_{H}^{(AH)},c_A), \forall c_{H}^{(AH)} \in {\cal C}_{c_H}$. 

In the pure AV market, there does not any HVs. 
According to Assumption 1, we assume there is only one AV manufacturer, so the pure AV market is essentially a monopoly market. 
Instead of solving the above game, the AV manufacturer solves an optimization problem for sensor investment. 
Given $p=1$, 
\begin{equation}
\begin{aligned}
   &[\mbox{\textbf{GameAV}}]\min \limits_{c_A \in {\cal C}_{c_A}}  && C_{A}(c_A)=w_a^{sen} \cdot S_A(c_A)+ w_a^{loss} \cdot L(c_A,c_A).
\label{equ:pureavstandard}
\end{aligned}
\end{equation}
The optimal solution is denoted by $c_A^{*}(p),\ p=1$. 




\subsection{Game Between the Lawmaker and Others}
\label{sec:hierlawmaker}

Individual's optimum may deviate from social welfare in most cases because of selfishness.
Each player involved in the hierarchical game has different and possibly opposing objectives. 
	This conflict may be accented when AVs are preprogrammed with algorithmic calculus and face dilemmas in decision-making process when social ethics are involved.
	The trade-off between social welfare and individual benefits is thus crucial in the AV-HV context.  

On the top level, the lawmaker develops a driver liability rule for human drivers 
and a product liability rule for the AV manufacturer, 
to minimize the social cost. 
Based on the liability policies, an optimal driving strategy has to be in-built within AVs' design guidelines. 
We would like to understand the impact of various liability policies on the decision-making of the AV manufacturer and road users
and identify a set of liability policies under which an optimal balance between efficiency and safety could be achieved for AVs.
For example, an overtly ``risk-averse" AV may potentially impact its own and other AVs driving efficiency, and create moral hazard effect on human drivers \citep{millard2016pedestrians}. 
On the other hand, an aggressively designed AV  (with lower weights on safety costs) has a higher risk of suffering penalty in the occurrence of an accident, thanks to product liability. 
Thus the future liability policies will heavily influence how AV manufacturers select cost coefficients and how human drivers react to AVs.

\noindent
\textbf{Players:} 
The lawmaker is a leader, while the AV Manufacturer and HVs are followers. 

\noindent
\textbf{Decision Variables:} In a liability rule, the lawmaker determines the care level standards $\sigma_H>0$ and $\sigma_A>0$ appearing in share functions.  
These two standards are defined in a comparative, not absolute scale, therefore we define 
the strategy of the lawmaker as the ratio of HV versus AV care level standards, denoted as $k \equiv \frac{\sigma_H}{\sigma_A}(k>0)$. 
The lawmaker's strategy $k$ can be interpreted as a punishment ratio for different road users in liability rules. 
$k>1$ implies that the standard care level of a human driver is higher than that of an AV. In other words, the lawmaker is stricter with HVs than AVs. 
$k<1$ implies that the lawmaker is stricter with AVs than HVs. 
$k=1$ implies that AVs and HVs have the same care level standard and the lawmaker treats them equally. Define ${\cal C}_k:(0,\bar{k})\subset R^+$ as the feasible set for the lawmaker's decision $k$ and $\bar{k}$ is the upper bound. 
According to Assumption 4, the lawmaker does not play a role in either the pure HV nor the pure AV market. Therefore the game developed on this level should only concern the mixed HV-AV market.

\noindent
\textbf{Cost Function:} 
The lawmaker aims to find an optimal care level standard ratio for HVs' and AVs' liability to minimize the social cost. 

Social cost is the total crash loss plus the total cost of care levels. Summarizing all three scenarios, the total crash loss $TL$ equals the sum of crash loss across three vehicular encounters, mathematically,  
\begin{equation}
\scriptsize{
\begin{aligned}
TL(c_{H_1}^{(HH)},c_{H_2}^{(HH)},c_A(k),c_H^{(AH)}(k))= \underbrace{p^2 \cdot L(c_{A}(k),c_{A}(k)) }_{\text{crash loss of $AA$ scenario}}+ \underbrace{2p(1-p) \cdot L(c_{A}(k),c_{H}^{(AH)}(k))}_{\text{crash loss of $AH$ scenario}} +\underbrace{(1-p)^2 \cdot L(c_{H_1}^{(HH)},c_{H_2}^{(HH)})}_{\text{crash loss of $HH$ scenario}},
\label{equ:totalcrashloss}
\end{aligned}}
\end{equation}

Note that $c_A$ and $c_H^{(AH)}$ are care levels for AVs and HVs, respectively in the AV-HV scenario. 
They both are affected by the lawmaker's decision $k$ via the cost functions of HVs and AVs in the $AH$ scenario defined in Equations \ref{equ:manutradeoff} and \ref{equ:tradeofhumanddriverinah}. 
Therefore, we will abuse the notations from now on by using $c_A(k)$ and $c_H^{(AH)}(k)$ and $C_A(c_A,c_H^{(AH)},k), C_H^{(AH)}(c_H^{(AH)},c_A,k)$.   

The total cost of care levels $TC$ is computed as the sum of all road users' precaution cost,  mathematically, 
\begin{equation}
\begin{aligned}
TC(c_{H_1}^{(HH)},c_{H_2}^{(HH)},c_A(k),c_H^{(AH)}(k))=\underbrace{2p^2S_A(c_{A}(k))}_{\text{precaution cost of $AA$ scenario}}+\underbrace{2p(1-p)(S_A(c_{A}(k))+S_{H}(c_{H}^{(AH)}(k)))}_{\text{precaution cost of $AH$  scenario}}\\ +\underbrace{(1-p)^2 (S_{H_1}(c_{H_1}^{(HH)})+S_{H_2}(c_{H_2}^{(HH)}))}_{\text{precaution cost of $HH$ scenario}},
\label{equ:totalcrashrate}
\end{aligned}
\end{equation}
To sum up, the social cost is computed as: 
\begin{equation}
\begin{aligned}
	\scriptsize{SC(c_{H_1}^{(HH)},c_{H_2}^{(HH)},c_{A}(k),c_{H}^{(AH)}(k))=\underbrace{w_l\cdot TC(c_{H_1}^{(HH)},c_{H_2}^{(HH)},c_{A}(k),c_{H}^{(AH)}(k))}_{\text{total cost of care levels}}+\underbrace{(1-w_l) \cdot TL(c_{H_1}^{(HH)},c_{H_2}^{(HH)},c_{A}(k),c_{H}^{(AH)}(k))}_{\text{total crash loss}},}
	\label{equ:lawmakertradeoff}
\end{aligned}
\end{equation}
where $w_l$ is a trade-off weight coefficient in the social cost function.

\textbf{Game:} The hierarchical game contains two levels of Stackelberg games. In the upper level, the lawmaker is the leader and all others are followers; while in the lower level, 
the AV manufacturer is the leader and human drivers are followers. 
Summarizing the three levels of games defined in Equations~\ref{equ:simuta} and \ref{equ:stackel} gives us a hierarchical game: 
\begin{equation}
\begin{aligned}
    &[\mbox{\textbf{GameSum}}] 
    \min \limits_{k \in {\cal C}_k} && SC(c_{H_1}^{(HH)},c_{H_2}^{(HH)},c_{A}(k),c_{H}^{(AH)}(k)),\ 0<p<1\\
    &\st\ [\mbox{\textbf{GameHH}}] && {c_{H_i}^{(HH)}} \in \argmin_{{c_{H_i}^{(HH)}}\in {\cal C}_{c_H}} \ \ C_{H_i}^{(HH)}(c_{H_1}^{(HH)},c_{H_2}^{(HH)}),i=1,2\\
    &\ \ \ \ \ \ [\mbox{\textbf{GameAH}}] && c_A \in
     \begin{aligned}[t]
        &\argmin_{c_A \in {\cal C}_{c_A}} && C_{A}(c_A, c_H^{(AH)}, k) \\
        &\st &&c_H^{(AH)} \in
         \begin{aligned}[t]
            &\argmin_{c_H^{(AH)} \in {\cal C}_{c_H}} && C_{H}^{(AH)}(c_H^{(AH)},c_A, k) \\
         \end{aligned}
     \end{aligned}
      \label{equ:hierproblem}
\end{aligned}
\end{equation}

\noindent The equilibrium of the hierarchical game is denoted by $c^{*} = (c_{H_1}^{*(HH)},c_{H_2}^{*(HH)},c_{A}^{*}(k^{*},p),c_{H}^{*(AH)}(k^{*},p),k^{*}(p)), p\in(0,1)$. 

\noindent
\textbf{Equilibrium condition:}
At equilibrium, no player can improve her total cost by unilaterally switching strategies, i.e., 
$SC(c_{H_1}^{*(HH)},c_{H_2}^{*(HH)},c_{A}^{*},c_{H}^{*(AH)}) \leqslant SC(c_{H_1}^{(HH)},c_{H_2}^{(HH)},c_{A},c_{H}^{(AH)})$, 
\\ $C_{A}(c_{A}^{*},c_H^{*(AH)},k)\leqslant C_{A}(c_{A},c_H^{*(AH)},k)$, $C_{H}^{(AH)}(c_H^{*(AH)},c_A, k)\leqslant C_{H}^{(AH)}(c_H^{(AH)},c_A, k)$,\\ $C_{H_i}^{(HH)}(c_{H_1}^{*(HH)},c_{H_2}^{*(HH)})\leqslant C_{H_i}^{(HH)}(c_{H_1}^{(HH)},c_{H_2}^{(HH)}), i=1,2$.

%
\subsection{Performance Measures} 
\label{sec:puremarket}

In this subsection, we will define three performance measures to evaluate the impact of AV technologies on traffic. 

\begin{definition} 
	\begin{enumerate}
		\item (Moral Hazard.) We say that a moral hazard happens to a road user $i$ if the following condition holds: 
		\begin{equation}
		    c_{i}^{*}(x)> c_{i}^{*}(x^{'})
		\end{equation}
		where $c_{i}^{*}$ is road user $i$'s equilibrium care level. In this paper, $x$ represents road environment or other road users' care level and $x^{'}$ represents improved road environment or other road users' care level. In other words, a moral hazard happens if a road user chooses a lower care level when others' care level or road environment is improved.
		
		
		
		
		\item (Road Safety.) We say that a pure AV market improves road safety if the following condition holds: 
		\begin{equation}
			TR^{(mixed)}(c_{H_1}^{*(HH)},c_{H_2}^{*(HH)},c_A^{*},c_H^{*(AH)}) > TR^{(pureAV)}(c_{A}^{*}), 
		\end{equation}
		where $TR^{(mixed)}, TR^{(pureAV)}$ are total crash rate of a mixed AV-HV and a pure AV markets, respectively.  
		
		\item (Social Welfare Improving Property.) We say that a pure AV market improves total social welfare if the following condition holds: 
		\begin{equation}
		\small
			SC^{(mixed)}(c_{H_1}^{*(HH)},c_{H_2}^{*(HH)},c_A^{*},c_H^{*(AH)}) > SC^{(pureAV)}(c_{A}^{*}), 
		\end{equation}
		where $SC^{(mixed)}, SC^{(pureAV)}$ are equilibrium social cost of a mixed AV-HV and a pure AV markets, respectively.

	\end{enumerate}	
\label{def:moralhazard}
\end{definition}

\section{Analytical Properties and Algorithm}
\label{sec:property}

In this section, we show existence and uniqueness of the equilibrium and develop an algorithm to solve the proposed game. 

\subsection{Solution Existence and Uniqueness}
\label{sec:uniqueness andd existence}

The hierarchical game [\textbf{GameSum}] defined in Equation \ref{equ:hierproblem} is essentially a tri-level game, with a simultaneous game in the lower level. 
The leader, the sub-leader, and the follower in this tri-level game correspond to the lawmaker, the AV manufacturer, and human drivers, respectively. 
The simultaneous game is the symmetric game between human drivers. 
To show solution existence and uniqueness, we will first introduce lemmas related to a bi-level game in the context of road safety and a simultaneous game, respectively. 
Then we apply these conditions to our tri-level game with the lower-level simultaneous game.  

\begin{lemma} 
	\label{lemma:leitpedersen}
	\citep{leit1978sse,pedersen2003moral} A general Stackelberg game is defined as follows:
	\begin{equation}
	\begin{aligned}
	&\max \limits_{x_l} && g^{l}(x_l, x_f) \\
	&\st &&x_f \in
	\begin{aligned}[t]
	&\argmax_{x_f} && g^{f}(x_l, x_f) 
	\end{aligned}\\
	& &&x_f\in {\cal X}_{f},\ x_l\in {\cal X}_{l}
	\end{aligned}
	\end{equation}
	where $x_l, x_f$ are decisions of a leader and a follower, $g^{l},g^{f}$ are their payoff functions and ${\cal X}_{l},{\cal X}_{f}$ are feasible sets for players' decisions. The sufficient conditions for the existence and uniqueness of equilibrium $(x_f^{*},x_l^{*})$ are: 
	\begin{enumerate}
		\item[A] The solution to $\frac{\partial g^{f}(x_{f}, x_l)}{\partial x_f}=0$ is a singleton, denoted by $x_{f} \equiv m_{l}(x_{l})$. 
		\item[B] The solution to $\frac{d g^{l}(x_{l}, m_{l}(x_{l}))}{d x_{l}}=0$ is a singleton, denoted by $x_{l}^{*}$. 
	\end{enumerate}
	
	Furthermore, \cite{pedersen2003moral} applied the general Stackelberg game to a road safety leader-follower game and provided more specific sufficient conditions for existence and uniqueness of the equilibrium. Mathematically,
	\begin{enumerate}
		\item $\frac{\partial^2 g^{f}(x_{l}, x_f)}{\partial x_f^2}>0$ where $g^{f}$ is the follower's cost function. 
		\item $\frac{d^2 g^{l}(x_l,m_l(x_l))}{d(x_l)^2}>0$ where $x_{f}^{*}\equiv m_l(x_l)$ is the solution to $\frac{\partial g^{f}(x_{l}, x_f)}{\partial x_f}=0$ where $g^{l}$ is the leader's cost function.
	\end{enumerate}
\end{lemma}

\begin{lemma} 
	\label{lemma:rosen}
	\citep{rosen1965exist} A simultaneous game between two players who want to minimize their own cost functions has a unique solution if the symmetric matrix $J+J^{T}$ is positive definite, where $J$ is the Jacobian matrix for the first derivative of cost functions.  
\end{lemma}

Based on Lemma \ref{lemma:leitpedersen} and Lemma \ref{lemma:rosen}, we provide sufficient conditions of existence and uniqueness of the equilibrium in the hierarchical game.
For notational simplicity, we will omit the parameter $p$ because it does not affect the analytical properties below. 

\begin{corollary}
	\label{prop:sse}
	Recall that the hierarchical game [\textbf{GameSum}] defined in Equation \ref{equ:hierproblem} is a tri-level game and a simultaneous game. The equilibrium is $(c_{H_1}^{*(HH)},c_{H_2}^{*(HH)},c_{H}^{*(AH)},c_{A}^{*},k^{*})$ where $c_H^{*(AH)} \equiv m_{sl}(c_A^{*})$ and $c_A^* \equiv m_l(k^*)$. In the tri-level game, the lawmaker's equilibrium is denoted by $k^*$, the AV manufacturer's equilibrium by $c_{A}^{*}$ and human drivers' equilibrium by $c_{H}^{*(AH)}$, respectively. Cost functions of players in the tri-level game are $SC(c_{A}(k),c_{H}^{(AH)}(k))$, $C_A(c_{A},c_{H}^{(AH)},k)$ and $C_H^{(AH)}(c_{A},c_{H}^{(AH)},k)$, respectively. In the simultaneous game, human drivers' equilibrium is denoted as $(c_{H_1}^{*(HH)},c_{H_2}^{*(HH)})$. Cost functions of players in the simultaneous game are $C_{H_i}^{(HH)}(c_{H_1}^{(HH)},c_{H_2}^{(HH)}),i=1,2$. Sufficient conditions of existence and uniqueness of $(c_{H_1}^{*(HH)},c_{H_2}^{*(HH)},c_{H}^{*(AH)},c_{A}^{*},k^{*})$ are:
	\begin{enumerate}
		\item $\frac{\partial^2 C_H^{(AH)}(c_A,c_H^{(AH)},k)}{\partial (c_H^{(AH)})^2}>0$. 
		\item $\frac{\partial^2 C_{A}(c_A,m_{sl}(c_A),k)}{\partial c_A^2}>0$ where $c_H^{(AH)} \equiv m_{sl}(c_A)$  is the solution to $\frac{\partial C_H^{(AH)}(c_A,c_H^{(AH)},k)}{\partial c_H^{(AH)}}=0$.
		\item $\frac{d^2 SC(m_l(k),m_{sl}(m_l(k)))}{dk^2}>0$ where $c_A \equiv m_{l}(k)$ is the solution to $\frac{\partial C_{A}(c_A,m_{sl}(c_A),k)}{\partial c_A}=0$. 
		\item $J+J^{T}$ is positive definite, where $J=\begin{bmatrix}
		\frac{ \partial^2 C_{H_1}^{(HH)}(c_{H_1}^{(HH)},c_{H_2}^{(HH)})}{\partial c_{H_1}^{2(HH)}}       & \frac{ \partial^2 C_{H_1}^{(HH)}(c_{H_1}^{(HH)},c_{H_2}^{(HH)})}{\partial c_{H_1}^{(HH)}\partial c_{H_2}^{(HH)}}  \\
		\frac{ \partial^2 C_{H_2}^{(HH)}(c_{H_1}^{(HH)},c_{H_2}^{(HH)})}{\partial c_{H_1}^{(HH)}\partial c_{H_2}^{(HH)}}       & \frac{ \partial^2 C_{H_2}^{(HH)}(c_{H_1}^{(HH)},c_{H_2}^{(HH)})}{\partial c_{H_2}^{2(HH)}}
		\end{bmatrix}$.
	\end{enumerate}
\end{corollary}

\subsection{Algorithm Design}
\label{sec:algorithmdesign}

A computational algorithm is developed leveraging a special structure in the proposed hierarchical game [\textbf{GameSum}], that is, 
the lawmaker's strategy, denoted by $k$, implicitly impacts the social cost function via the care levels of AVs and HVs. 
Thus the upper level game and the game between human drivers and the AV manufacturer can be decoupled once the value of $k$ is fixed. Based on the SSE property (defined in Section \ref{sec:uniqueness andd existence}) of this hierarchical game, lawmaker's game in the upper level is a convex optimization with respect to $k$.  
Given an AV penetration rate $p$, we first initialize the lawmaker's decision variable $k=k_0 >0$ and apply Frank–Wolfe algorithm to solve $k$ by updating $k$ as $k_{n+1}=k_{n}-\gamma \nabla SC(c_{A}^*(k_n),c_{H}^{*(AH)}(k_n)),\ \gamma=\frac{2}{n+2}$, where $n$ is the iteration index. We then solve the equilibrium care levels for the AV manufacturer and human drivers within each iteration using the conditions (1,2,4) in Corollary \ref{prop:sse}. The algorithm convergence is guaranteed for a convex optimization and is also demonstrated in Section~\ref{sec:sen3}.

\begin{algorithm}[H]
\SetAlgoLined
	\DontPrintSemicolon
	\KwInput{Payoff Functions: $C_{H_1}^{(HH)},C_{H_2}^{(HH)},C_{H}^{(AH)},C_{A},SC$}
	\KwOutput{Players' Strategies: $k_n,\ c_{A}^*(k_n),\ c_{H}^{*(AH)}(k_n),\ c_{H_1}^{*(HH)}(k_n),\ c_{H_2}^{*(HH)}(k_n)$}
	Initialization: $n=0,\ max\_iter = 10000$\\
	\While{n < max\_iter }
	{
		\For{$c_A \in {\cal C}_{c_A}$}
		{ $c_H^{(AH)}=\argmax_{c_H^{(AH)}}\ C_H^{(AH)}(c_A,c_H^{(AH)},k)$\;
			\If{$\frac{\partial C_{A}(c_{H}^{(AH)}(c_A),c_A,k)}{\partial c_A}=0$}
			{ $c_A^*(k_n)=c_A,\ c_H^{*(AH)}(k_n)=c_H^{(AH)}$;  \tcp*{Stackelberg Game between AV manufacturer and HVs}
			}
		}
		\For{$c_{H_1}^{(HH)} \in {\cal C}_{c_H}$}
		{ $c_{H_2}^{(HH)}=\argmax_{c_{H_2}^{(HH)}}\ C_{H_2}^{(HH)}(c_{H_1}^{(HH)},c_{H_2}^{(HH)})$\;
			\If{$\frac{\partial C_{H_1}^{(HH)}(c_{H_1}^{(HH)},c_{H_2}^{(HH)})}{\partial c_{H_1}^{(HH)}}=0$}
			{ $c_{H_1}^{*(HH)}(k_n)=c_{H_1}^{(HH)},\ c_{H_2}^{*(HH)}(k_n)=c_{H_2}^{(HH)}$;   \tcp*{Simultaneous Game between HVs}
			}
		}
		$k_{n+1}=k_{n}-\gamma \nabla SC(c_{A}^*(k_n),c_{H}^{*(AH)}(k_n)),\ n=n+1$\\ 
	}
	\caption{Hierarchical Game}
	\label{alg:hier}
\end{algorithm}

\tikzstyle{decision} = [diamond, draw, fill=white!20, aspect=5, node distance=1cm, minimum height=1em]
\tikzstyle{decision2} = [diamond, draw, fill=white!20, aspect=4, node distance=1cm, minimum height=1em]
\tikzstyle{block} = [rectangle, draw, fill=white!20, 
    text width=5em, text centered, rounded corners, minimum height=4em]
\tikzstyle{block0} = [rectangle, draw, fill=white!20, 
    text width=45em, text centered, rounded corners, minimum height=1em]
\tikzstyle{block1} = [rectangle, draw, fill=white!20, 
    text width=30em, text centered, rounded corners, minimum height=2em]
\tikzstyle{block2} = [rectangle, draw, fill=white!20, 
    text width=15em, text centered, rounded corners, minimum height=1em]
\tikzstyle{block3} = [rectangle, draw, fill=white!20, 
    text width=8em, text centered, rounded corners, minimum height=1em]
\tikzstyle{block4} = [rectangle, draw, fill=white!20, 
    text width=15em, text centered, rounded corners, minimum height=2em]
\tikzstyle{line} = [draw, -latex']
\tikzstyle{cloud} = [draw, ellipse,fill=red!20, node distance=3cm,
    minimum height=2em]

\section{Numerical Examples}
\label{sec:numerical} 
To investigate the behaviors of human drivers, the AV manufacturer, and the lawmaker in the AV-HV mixed traffic, numerical examples are provided as the AV penetration rate increases. The goals of these examples are to understand:
(1) under what conditions human drivers develop a moral hazard,
(2) under what conditions the AV manufacturer's cost in the pure AV market is lower than that of the AV-HV mixed market, 
and (3) under what conditions an optimal liability rule reduces total social cost. 
The outcomes are evaluated on three performance measures, namely, 
the care levels of AVs determined by the AV manufacturer and that of HVs, 
road safety in light of total crash rate, 
and social welfare. 
Accordingly, we would like to test the following hypotheses as the penetration rate of AVs grows: 
\begin{enumerate}
	\item Human drivers take advantage of the AV manufacturer's choice on AVs' care level.
	\item The AV manufacturer chooses a lower care level for AVs in a pure AV market than in a mixed one.
	\item A strategic lawmaker performs better than a non-strategic lawmaker in reducing social cost. 
\end{enumerate}

We will first fix the liability rule (i.e., assuming that the lawmaker is not a strategic player and $k=1$) in Sections \ref{sec:sen1} and \ref{sec:sen2}.  
A baseline model is first tested to investigate how road users and the AV manufacturer may change their care levels and how road safety and social welfare may change, as the number of AVs increases. 
Then sensitivity analysis is performed over a set of cost and crash related parameters.   
In Section \ref{sec:sen3}, we relax our assumption and allow the lawmaker to select an optimal set of liability rules. 
We will then compare the social cost when the lawmaker is a non-strategic player and a strategic player, 
to demonstrate the critical role the lawmaker plays in the AV-HV mixed traffic. All these analyses assume the AV market share is exogenous. In Section 5.4, we endogenize the market share of AVs as a variable of their crash loss. We will show how the law maker's decision would vary with the reduced public opinion towards AVs. 


The parameters, cost functions, crash probability functions, and crash severity functions used in the numerical examples are listed in Table \ref{table:parameter}. Note that the parameters in the cost functions are carefully selected to fulfill the existence and uniqueness criteria of equilibria in Corollary \ref{prop:sse}. 
\begin{table}[H]
\centering
\scriptsize{
\begin{tabular}{c|c|c|c|c}
\hline Name &Notation &Function Form &Parameter &Parameter Value\\
\hline Precaution cost & \makecell{$S_A(c_A)$\\$S_H(c_H)$} & \makecell{$((1-\alpha c_A)^{-1}-1) \cdot p^{-\frac{1}{2}}$\\$-(\beta c_H-1)^{-1}-1$} & $[\alpha;\beta]$ &\makecell{$[0.4,0.5]$\\$[0.3;0.5]$}\\
\hline Crash probability & \makecell{$P(c_{A},c_{A})$\\$P(c_{H_1}^{(HH)},c_{H_2}^{(HH)})$\\$P(c_{A},c_{H}^{(AH)})$}& \makecell{$(a c_{A}+ a c_{A}+1)^{-1}$ \\$(h c_{H_1}^{(HH)}+ h c_{H_2}^{(HH)}+1)^{-1}$\\$(a c_{A}+ h c_{H}^{(AH)}+1)^{-1}$}& $[a;h]$ &\makecell{$[10;10]$\\$[5;10],[15;10]$\\$[10;5],[10,15]$}\\ 
\hline Crash severity & \makecell{$T(c_{A},c_{A})$\\$T(c_{H_1}^{(HH)},c_{H_2}^{(HH)})$\\$T(c_{A},c_{H}^{(AH)})$}& \makecell{$M-s c_{A}- s c_{A}$ \\$M-t c_{H_1}^{(HH))}- t c_{H_2}^{(HH)}$\\$M-s c_{A}-t c_{H}^{(AH)}$}& $[s;t]$ &\makecell{$[5;5]$}\\ 
\hline Lawmaker's trade-off coefficient& $w_l$ & - &  - & 0.16  \\ 
\hline AV manufacturer's trade-off coefficient& $[w_a^{sen},w_a^{loss}]$ & - &  - & \makecell{$[0.125;0.25]$\\$[0.2;0.4],[0.67;0.17]$\\$[0.14;0.14],[0.1;0.4]$}  \\ 
\hline Human drivers' trade-off coefficient& $w_h$ & - &  - & \makecell{$0.5$\\ $\mathcal{N}(0.5,0.1) \cap (0,1)$}  \\ 
\hline parameter of endogenous $p$ function & $\eta$ & - &  - & \makecell{$1.67,2,2.86$}  \\ 
\hline
\end{tabular}}
\caption{\small{Parameter settings}}
\label{table:parameter}
\end{table} 


\subsection{Game Between Humans and the AV Manufacturer}
\label{sec:sen1} 

In the game between humans and the AV manufacturer, a base model is developed where $\alpha=0.4,\ \beta=0.5,\ a=h=10,\ M=20,\ s=t=5, w_l=0.16,\ w_a=[0.125,0.25,0.625],\ w_h=0.5$. We will vary $p$ and test the impact of a growing AV penetration rate on the system performance. 

\begin{figure}[H]
	\centering
	\subfloat[Cost functions of care level \label{fig:costfunction}]{\includegraphics[width=2.7in]{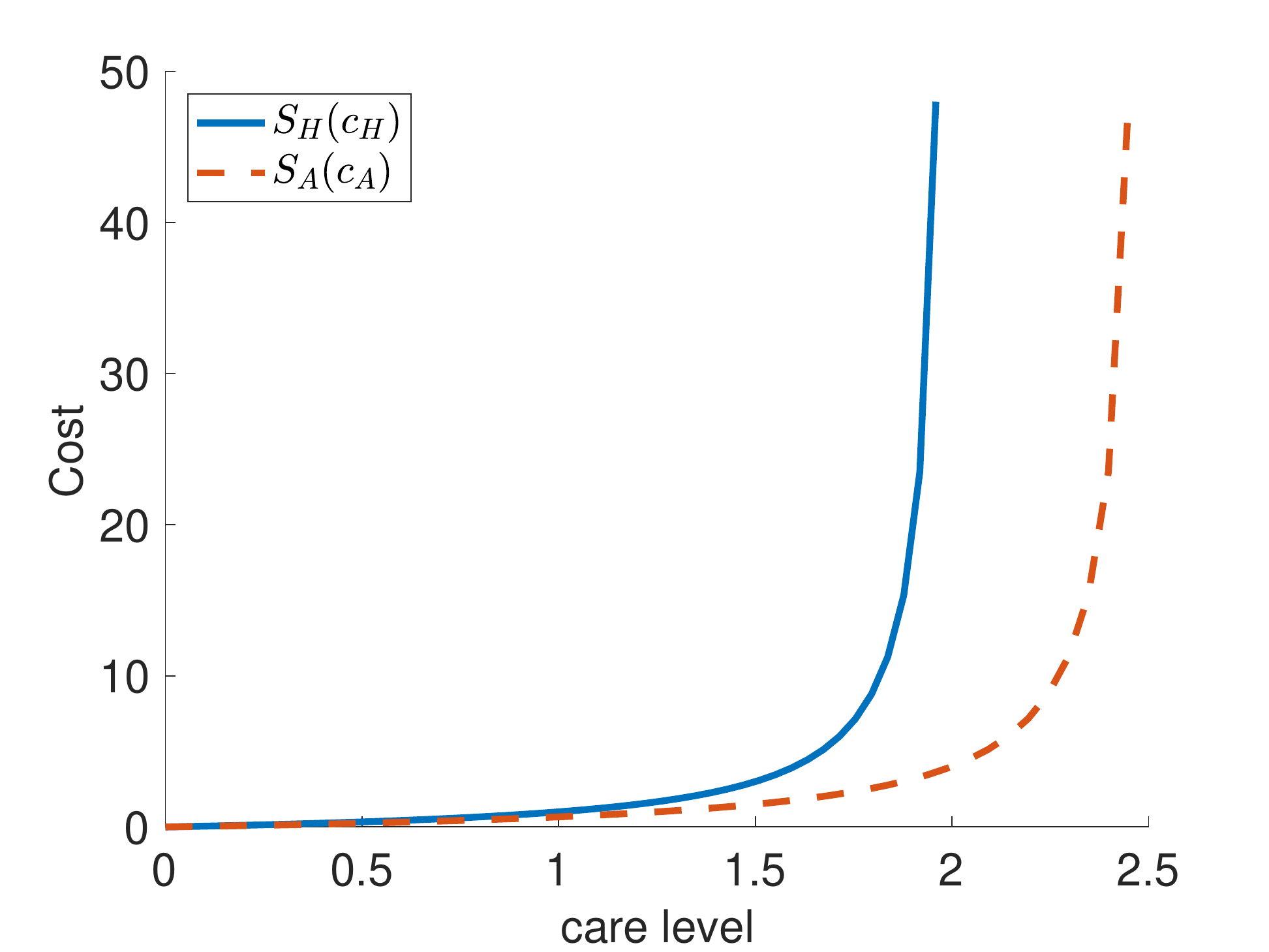}} 
	\subfloat[Road users' care level \label{fig:basicmodelcarelevel}]{\includegraphics[width=2.7in]{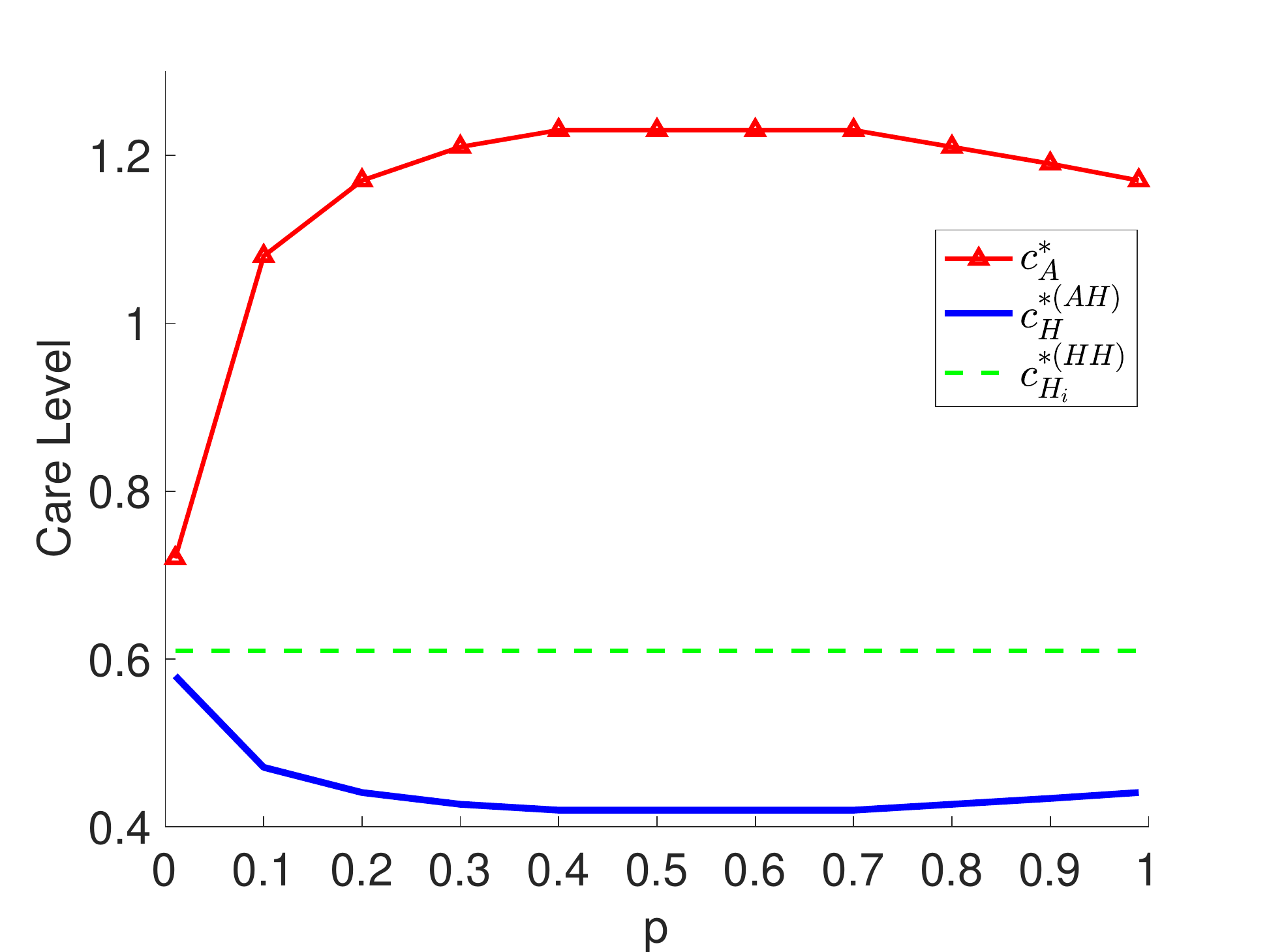}}
	\caption{Care level analysis}
	\label{fig:basicmodel}
\end{figure}

We first plot HVs' and AVs' cost functions with respect to care levels in \figref{fig:costfunction} to validate our Assumption 3 that AVs can achieve the same care level with a lower cost than humans.  
The blue solid line is HVs' precaution cost while the red dashed line is AVs' sensor cost. 
We can see two trends: (1) humans require higher cost than AVs to achieve the same care level on the x-axis, and (2) AVs' care level (determined by sensor specification) can be much larger than that of humans  \citep{kalra2016driving,fag2015av}, demonstrating the fact that AVs have much more perception and reaction power than humans \citep{chatterjee2016understanding}.  

Given the care level cost functions, we can then compute the equilibrium care levels for HVs and AVs. 
\figref{fig:basicmodelcarelevel} illustrates how human drivers and the AV manufacturer adjust equilibrium care levels (y-axis) as AVs' penetration rate $p \in (0,1)$ (x-axis) increases. 
Human's care levels in the $AH$ scenario and the $HH$ scenario are indicated by a blue line and a dashed green line, respectively.  
AVs' care level is represented by a red line with triangle markers.  

We first focus on the trend of care levels for HVs and AVs in the $AH$ scenario.
As the penetration rate $p$ increases, 
AVs' care level first increases and then slightly decreases after $p>0.7$. 
HVs' care level first decreases and then slightly increases again after $p>0.7$.
When $p<0.7$, as the number of AVs increases, the AV manufacturer becomes more attentive while HVs become less attentive. Because the growing number of AVs on the road increases the crash loss of the AV manufacturer, to reduce the total crash loss, the AV manufacturer has to improve safety specification of AVs to increase the care level. 
When AVs become more attentive, human drivers begin to lower their care levels by taking advantage of increasing AVs' care level. 
This trend is usually observed in a leader-follower game \citep{pedersen2001game}. When the leader - AVs - execute higher precaution, the follower - HVs - lower precaution accordingly. 
Note that HVs' care levels in the $HH$ scenario remain constant, regardless of AVs' penetration rate. 
We can see that HVs execute higher care levels to other HVs than to AVs. 
According to Definition~\ref{def:moralhazard}, moral hazard happens because human drivers excute less care when encountering AVs than when meeting other HVs. In addition, AVs' care level is always higher than that of HVs, which is an analogue of the existing situation when AVs act conservatively on public roads.

Nevertheless, the AV manufacturer cannot always increase its safety specification, because its cost increases exponentially as sensor specifications increase.  
To balance total cost, the AV manufacturer has to select relatively cheaper sensors when $p>0.7$. 
This is because when AVs dominate the market and the portion of human drivers becomes smaller, the overall traffic environment is improved and thus the sensor cost dominates AVs' crash loss. Reducing sensor cost slightly would not increase crash loss that much. 


In summary, AVs' and HVs' care levels are correlated in the leader-follower game and AVs are more conservative than HVs.
When the AV manufacturer increases AVs' care level, human drivers tend to be less attentive by decreasing their care levels as the number of AVs increases.

\begin{figure}[H]
	\centering
	\subfloat[Crash probability \label{fig:crashprob}]{\includegraphics[width=3in]{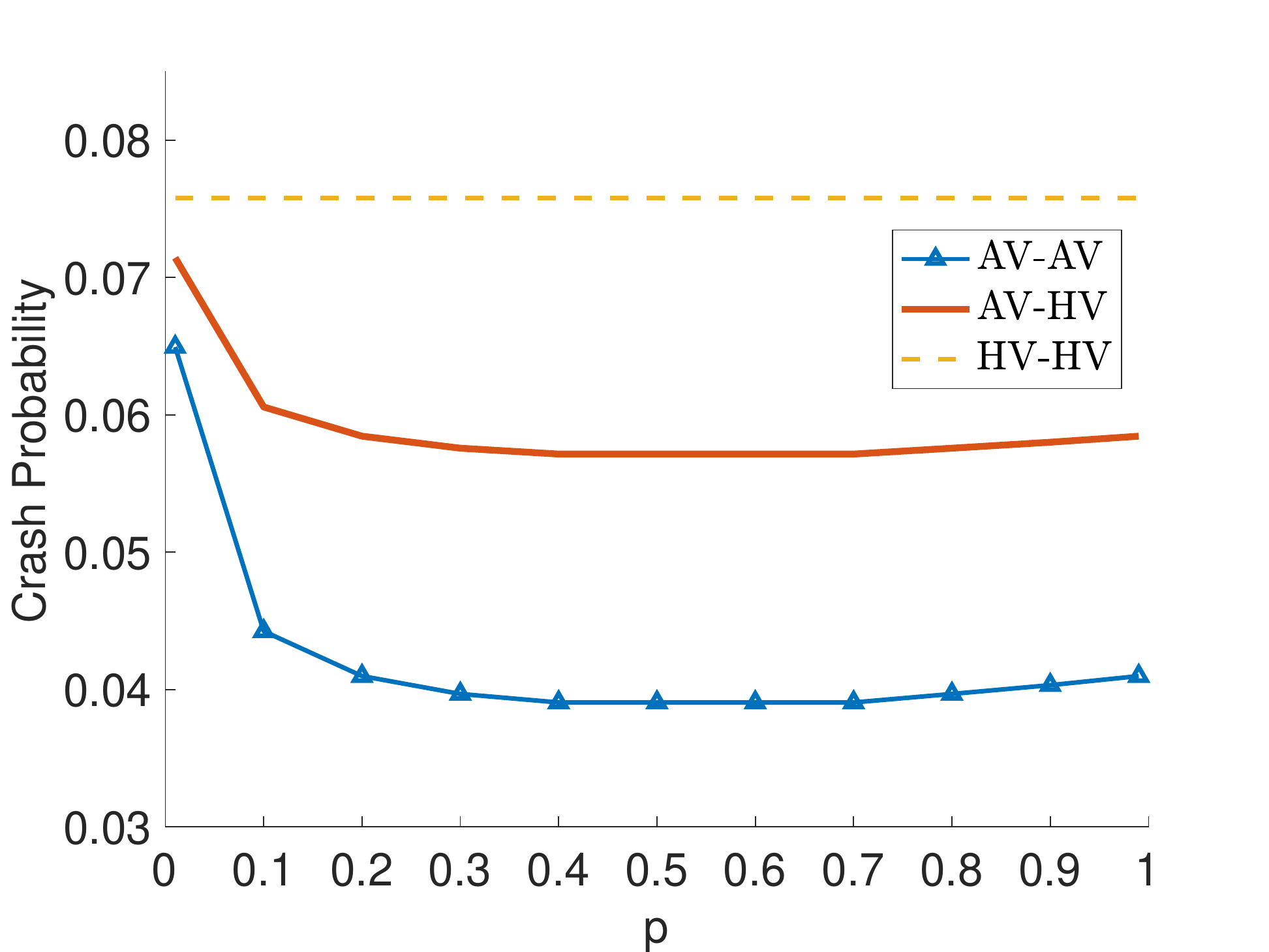}}
	\subfloat[Crash rate \label{fig:crashrate}]{\includegraphics[width=3in]{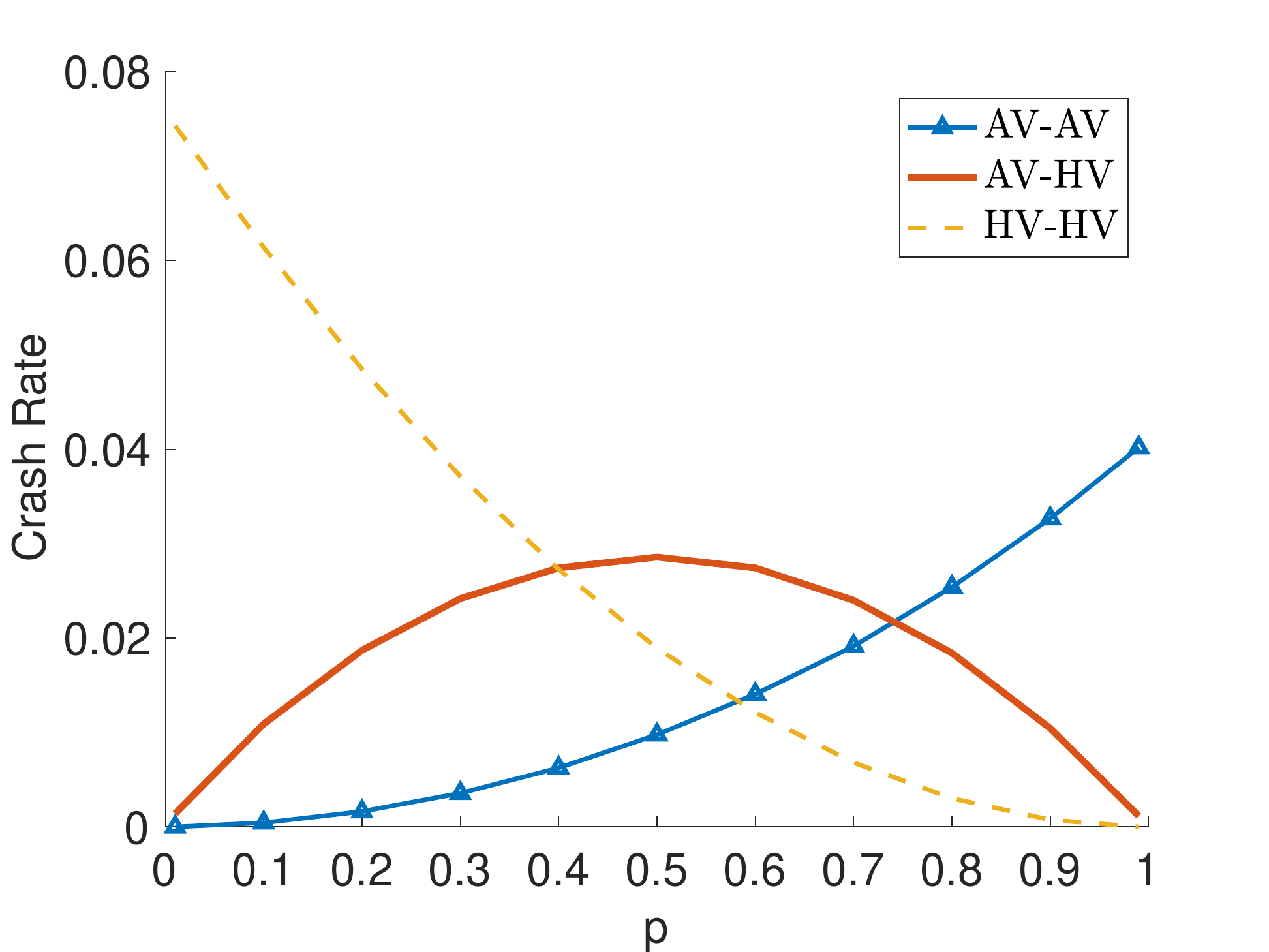}}
	
	\subfloat[Performance measures \label{fig:socialwelfare}]{\includegraphics[width=3in]{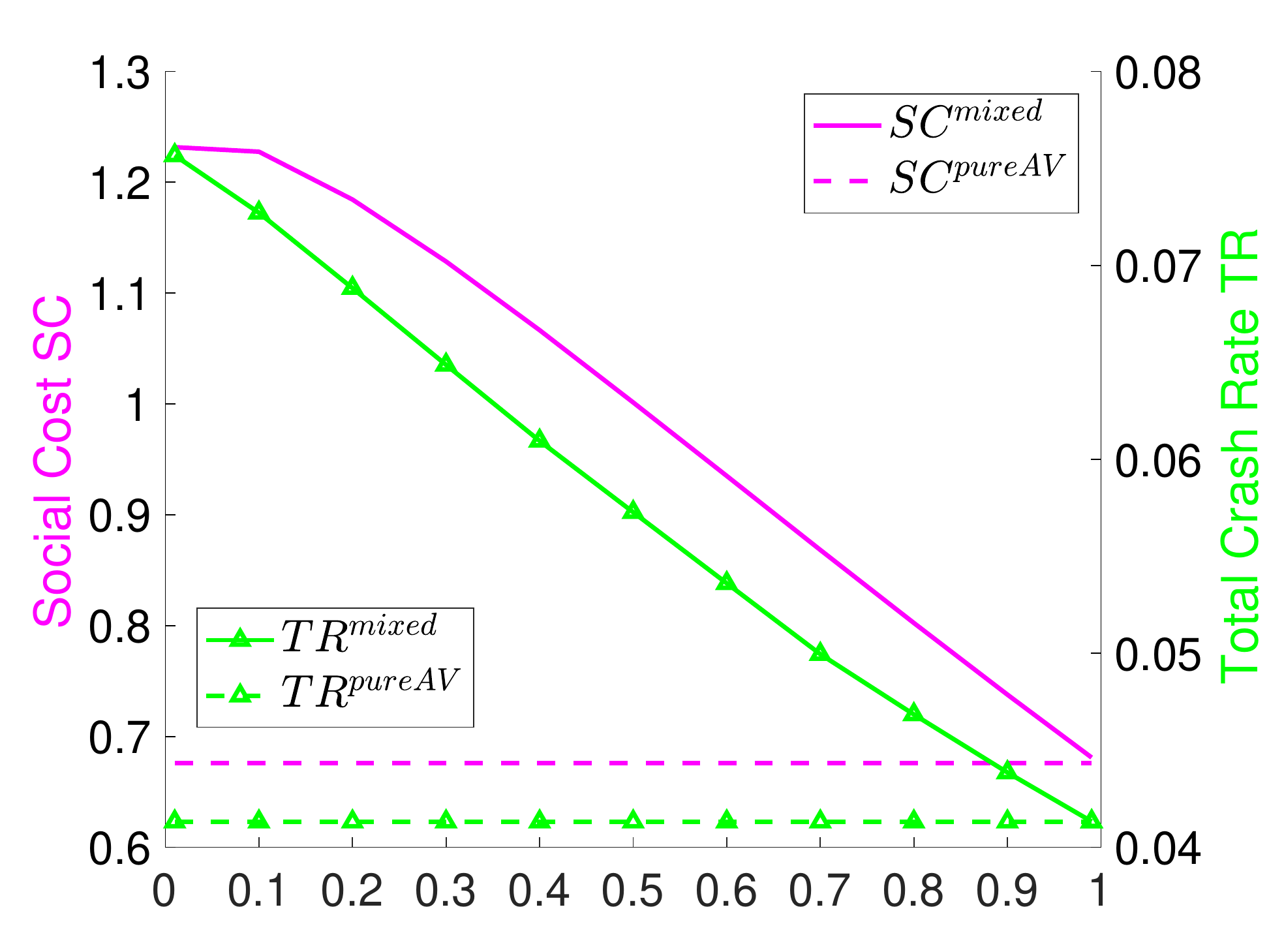}}
	\caption{Performance analysis in the base model}
	\label{fig:crash}
\end{figure}

In \figref{fig:crashprob} and \figref{fig:crashrate}, blue lines with triangle markers represent the performance measures for the AA scenario, red lines represent those for the AH scenario, and the yellow dotted lines represent those for the HH scenario. 
As the AV penetration rate increases, crash probability for both the $AA$ and $AH$ scenarios first decreases and then slightly increases after $p>0.7$. The crash probability for the $HH$ scenario remains constant, because one HV's care level when facing other HVs remains constant regardless of the AV penetration rate.
In \figref{fig:crashrate}, 
the crash rate for the $AA$ scenario increases, 
that for the $HH$ scenario decreases, 
and that for the $AH$ scenario increases first and then decreases.  
The first two trends are reasonable because as AVs increase and HVs decrease, the average crash rate between two AVs increases while that between two HVs decreases. 
The interesting one is the $AH$ scenario. The maximum crash rate is attained around $p=0.5$, i.e., half AVs and half HVs,  
when traffic exhibit the highest heterogeneity and there are the most encounters between AVs and HVs. The difference in care levels lead to higher crash rate between AVs and HVs.   
In either an HV-dominated (i.e., $p < 0.5$) or an AV-dominated market (i.e., $p > 0.5$), crash rate is lower because vehicular encounters happen mostly between the same types of vehicles. 
\newpage
\begin{remark}
To match the crash rate for the HH and AH scenarios to real data, we calibrate the weighting parameters in the cost functions of HVs and AVs, respectively. Crash rate in our model is defined as the number of crashes over that of vehicle encounters. Using naturalistic driving data collected from 109 drivers for a period of 12 to 13 months \citep{klauer2006nearcrash}, 69 crashes happened out of 8,295 incidents. Here incidents can be treated as a surrogate of vehicle encounters. Therefore, the crash rate for the HH scenario in the pure HV market can be approximated as 69/8295=0.08. The crash rate for the AH scenario data can also be calibrated using the AV data released by California DMV \citep{CA_DMV}, the only open data source about Level-4 AVs. California is analogous to a HV-dominated market with AVs’ market share of approximately zero, given there are 15 million total registered vehicles in California \citep{CA_DMV_register}, among them 769 are AVs \citep{CA_DMV_register_av} accounting for a penetration rate of 0.005\%. We will only use Waymo data and exclude other AV fleets, because Waymo provides a more complete information about its fleet size and disengagement events. Here we use disengagement as a surrogate of vehicle encounters. A disengagement event happens when this AV is interacting with other road users but requires human’s intervention. By far Waymo has had 25 collisions \citep{CA_DMV} and 9,359 disengagement events \citep{CA_DMV_register_av}. Accordingly, the AV crash rate is calculated as 25/9,359=0.0027.
\end{remark}


In \figref{fig:socialwelfare}, 
both total crash rate, represented by a green line with triangle makers, 
and social cost, represented by a maroon line, 
have a decreasing trend, implying that the growing number of AVs improves road safety and social welfare.  
We also compare social cost and total crash rate with those in the pure AV market. 
In the pure AV market, social cost is represented by a dashed maroon line, 
and total crash rate is represented by a dashed green line with triangle markers. 
Note that in the pure AV market, these two values are only attained when $p=1$. We draw a constant line over $p$ for the purpose of identifying the critical penetration rate that worsens the system performance in the presence of human drivers. Social cost and total crash rate in the pure AV market are always smaller than those in the mixed market. It means that the presence of AVs improves the system performance. With the AV fleet size growing, the traffic system becomes safer and socially better off.

\subsection{Sensitivity Analysis}
\label{sec:sen2}

In this subsection, we would like to discuss how parameters in cost and crash related coefficients influence the system outcome, compared to our base model. 
There are three classes of parameters: 
\begin{enumerate}
	\item Coefficients $\alpha,\beta$ in AVs' and HVs' precaution cost functions, respectively: 
	it reflects how much AVs' (or HVs') precaution cost changes if AVs' (or HVs') care level is increased by one unit. 
	The physical meaning of $\alpha$ is the marginal production cost of AV sensors. 
	$\beta$ indicates HVs' marginal cost of executing a care level, which depends on both drivers perception and reaction time, judgment, maneuvering capability, and vehicles' characteristics.   
	These two coefficients directly influence care levels and thus the system performance. 
	\item Coefficients $a,h$ in crash probability functions for three scenarios and coefficients $s,t$ in crash severity functions for three scenarios: 
	it reflects the external environment factors that contribute to the crash probability of AVs (or HVs), given that AVs' (or HVs') care level remains the same. 
	The parameter $a$ (or $h$) can be interpreted as the marginal change in the crash probability for AVs (or HVs), if the road environment is improved for AVs (or HVs). 
	For example, government builds more roadside infrastructure that can communicate with AVs so that AVs have better awareness of the environment. 
	For human drivers, the driving environment may become safer because of traffic safety measures such as red light camera enforcement \citep{retting2008reducing}. 
	A more important environment change to human drivers is the increasing number of AVs. If these AVs create a safer road environment for HVs, the crash probability contributed by HVs can be reduced. We find that parameters in crash severity functions basically play the same role as those in crash probability functions, so its sensitivity analysis is skipped here. 
	Another important external road environmental change for AVs is to build an AV-dedicated lane or zone in which the movement of AVs and HVs is separated, removing the AV-HV encounter scenario. We will also discuss how this countermeasure influences care levels and the system performance. 
	\item Coefficients $w_a^{sen},w_a^{loss}$ in the disutility function of the AV manufacturer, respectively: 
	It reflects how much weight the AV manufacturer allocates to the precaution cost and the crash loss. 
	These weights highly depend on the external liability rule. 
	When the liability rule is fixed, the varying weights could change AVs' care level and thus the system performance.
	Here we mainly discuss the impact of $w_a^{sen},w_a^{loss}$ on traffic performance, because it is crucial to regulate how AV manufactures weigh in traffic efficiency and safety.
	\item Coefficient $w_h$ in the disutility function of human drivers. It reflects how much weight human drivers allocate to the precaution cost, which depends on individual preference. Here we assume human drivers are heterogeneous and they can have different $w_h$.
\end{enumerate}

\subsubsection{Precaution Cost Coefficient for AVs}
\label{sec:sensorcost}

\begin{figure}[H]
	\centering
	\subfloat[Sensor cost \label{fig:sensor_avsensorcost}]{\includegraphics[width=3in]{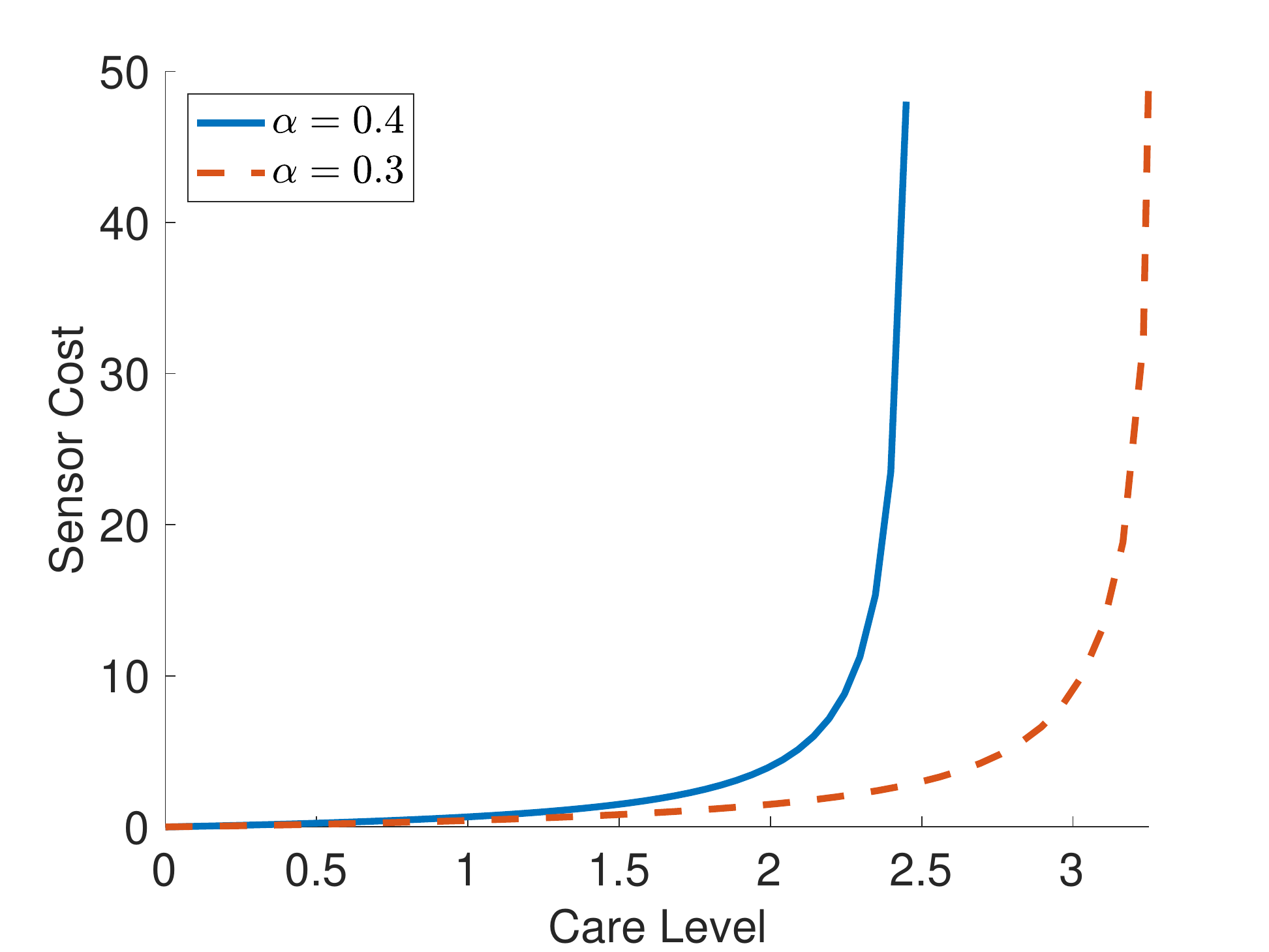}}
	
	\subfloat[Care level \label{fig:sensor_care}]{\includegraphics[width=3in]{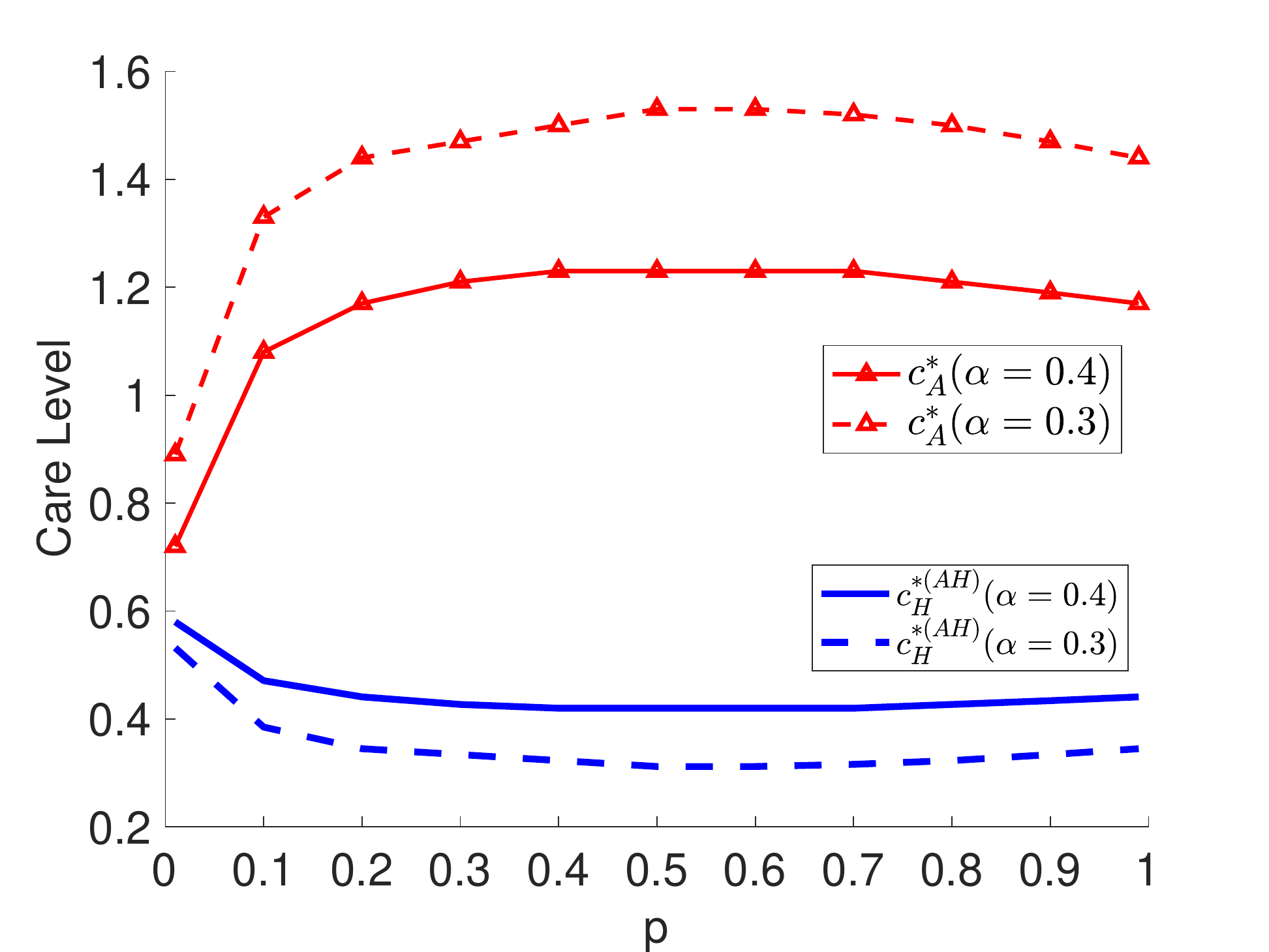}}
	\subfloat[Performance measures \label{fig:sensor_measure}]{\includegraphics[width=3in]{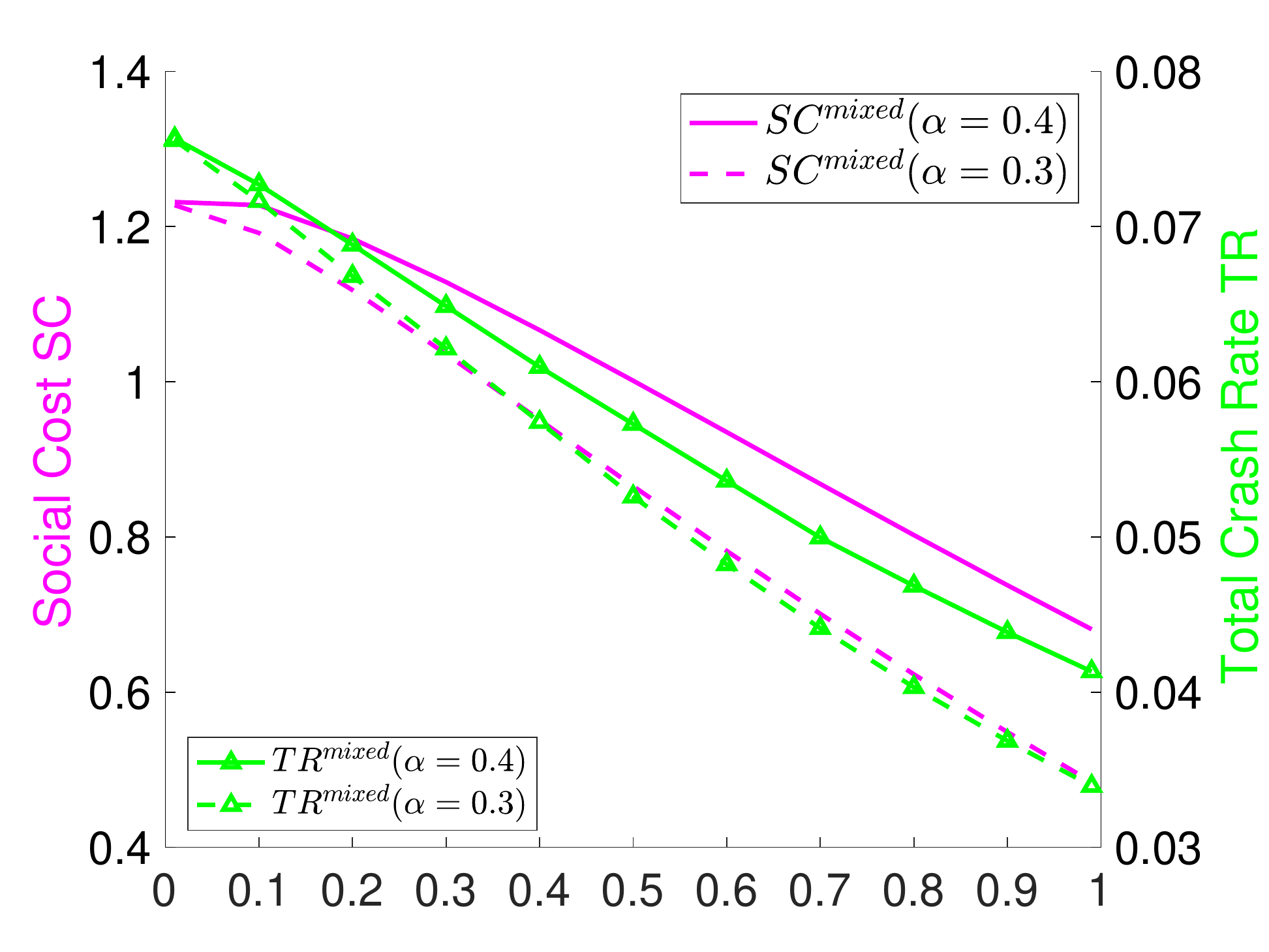}}
	\caption{Sensitivity analysis for the marginal cost coefficient in sensor cost}
	\label{fig:sensorcost}
\end{figure}

In the base model, AVs' sensor cost is defined as $S_A(c_A) = ((1-\alpha c_A)^{-1}-1)\cdot p^{-\frac{1}{2}}$ (Table \ref{table:parameter}). 
When $\alpha$ is reduced, the sensor cost becomes cheaper for the same sensor quality. This may happen because of the government subsidy.
We first compare the case when $\alpha= 0.3$ to the case when $\alpha= 0.4$ (baseline) in \figref{fig:sensorcost}. 

In \figref{fig:sensor_avsensorcost}, AVs' precaution cost functions at $\alpha=0.4$, $0.3$ are plotted in a blue and a dashed red line, respectively, as the care level increases. 
To execute  the same care level, the sensor cost with a higher $\alpha$ is larger.
\figref{fig:sensor_care} shows AVs' and HVs' care levels in red lines and blue lines, respectively. 
The solid red line and the solid blue line with triangle markers represent the case when $\alpha=0.4$ while their corresponding dashed lines represent the case when $\alpha=0.3$. 
The trend of AVs' and HVs' care levels remains the same as $p$ increases. 
AVs become more attentive while HVs' care level is reduced for $\alpha=0.3$, compared to $\alpha=0.4$. 
This is because the AV manufacturer can afford sensors with higher safety specifications at the same cost, thus increasing AVs' care level. 
HVs, as followers, take advantage of an increase in AVs' care level and exhibit moral hazard. 
Nevertheless, both social cost and total crash rate, represented by solid maroon lines and green lines with markers, respectively, in \figref{fig:sensor_measure}, 
decrease as $p$ increases. 
More importantly, the performance with $\alpha=0.3$ is better off than those with $\alpha=0.4$. 
In other words, the traffic system is better off with a lower marginal cost of AV production. 
In summary, if government could subsidy the AV manufacturer \citep{luo2019accelerating} to reduce production cost, it would greatly encourage the AV manufacturer to increase AVs' care level and improve the overall traffic system performance.

\subsubsection{External Road Environment}
\label{sec:externalfactor}
\begin{figure}[H]
	\centering
	\subfloat[Care level when varying $a$ \label{fig:extrenal_care}]{\includegraphics[width=3in]{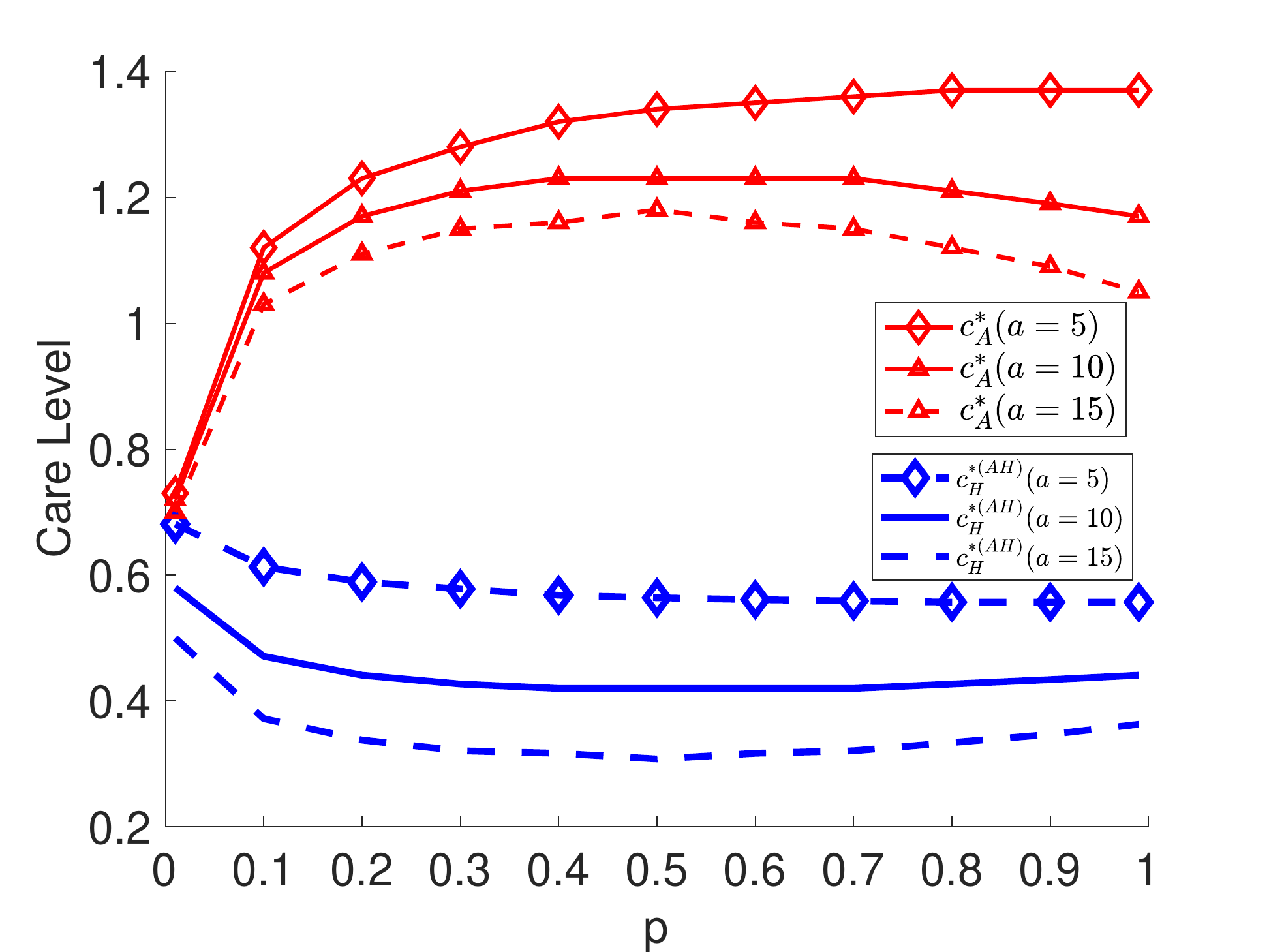}}
	\subfloat[Performance measures when varying $a$ \label{fig:external_cost}]{\includegraphics[width=3in]{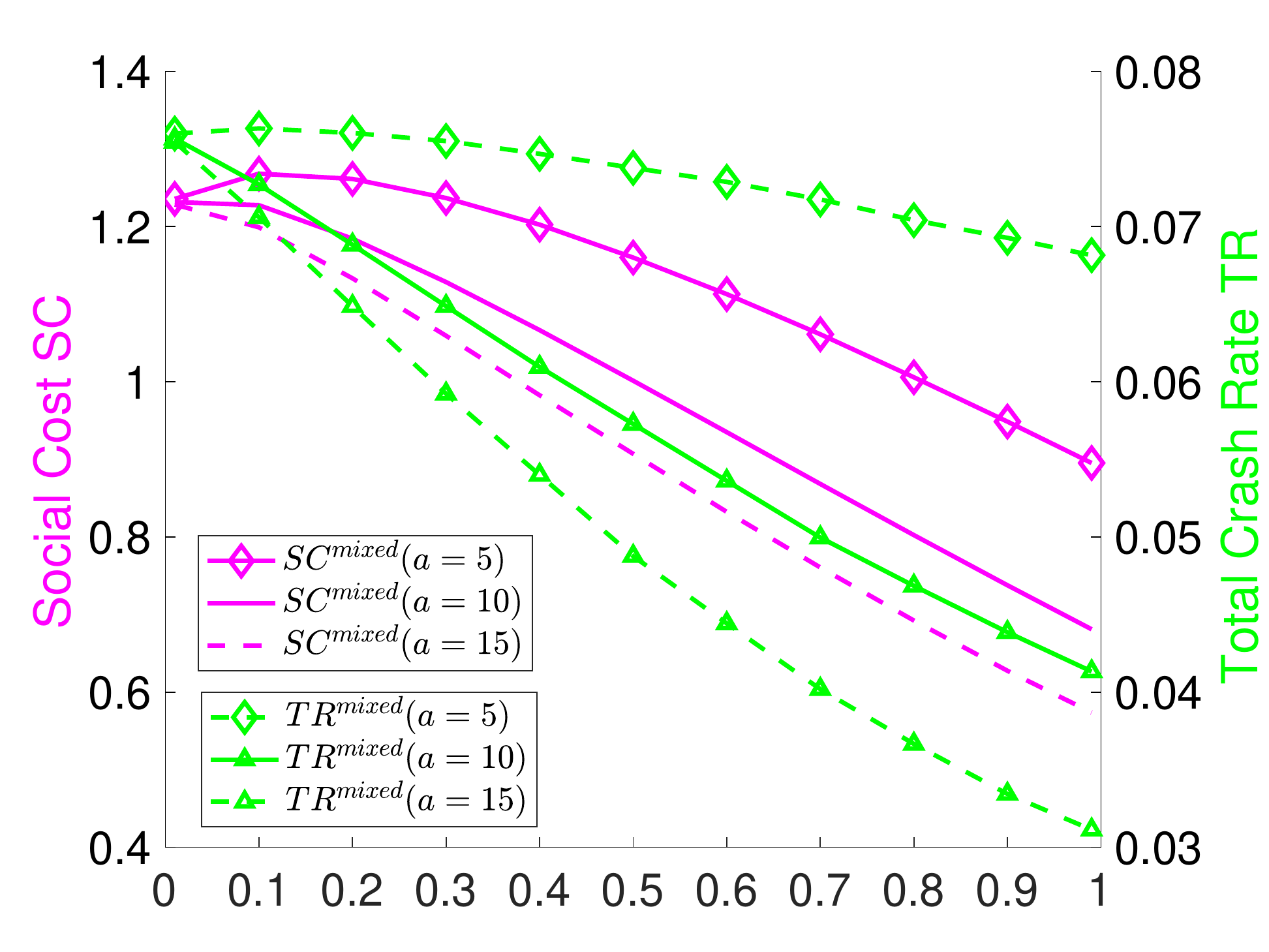}}
	
	\subfloat[Care level when varying $h$ \label{fig:extrenal_care_1}]{\includegraphics[width=3in]{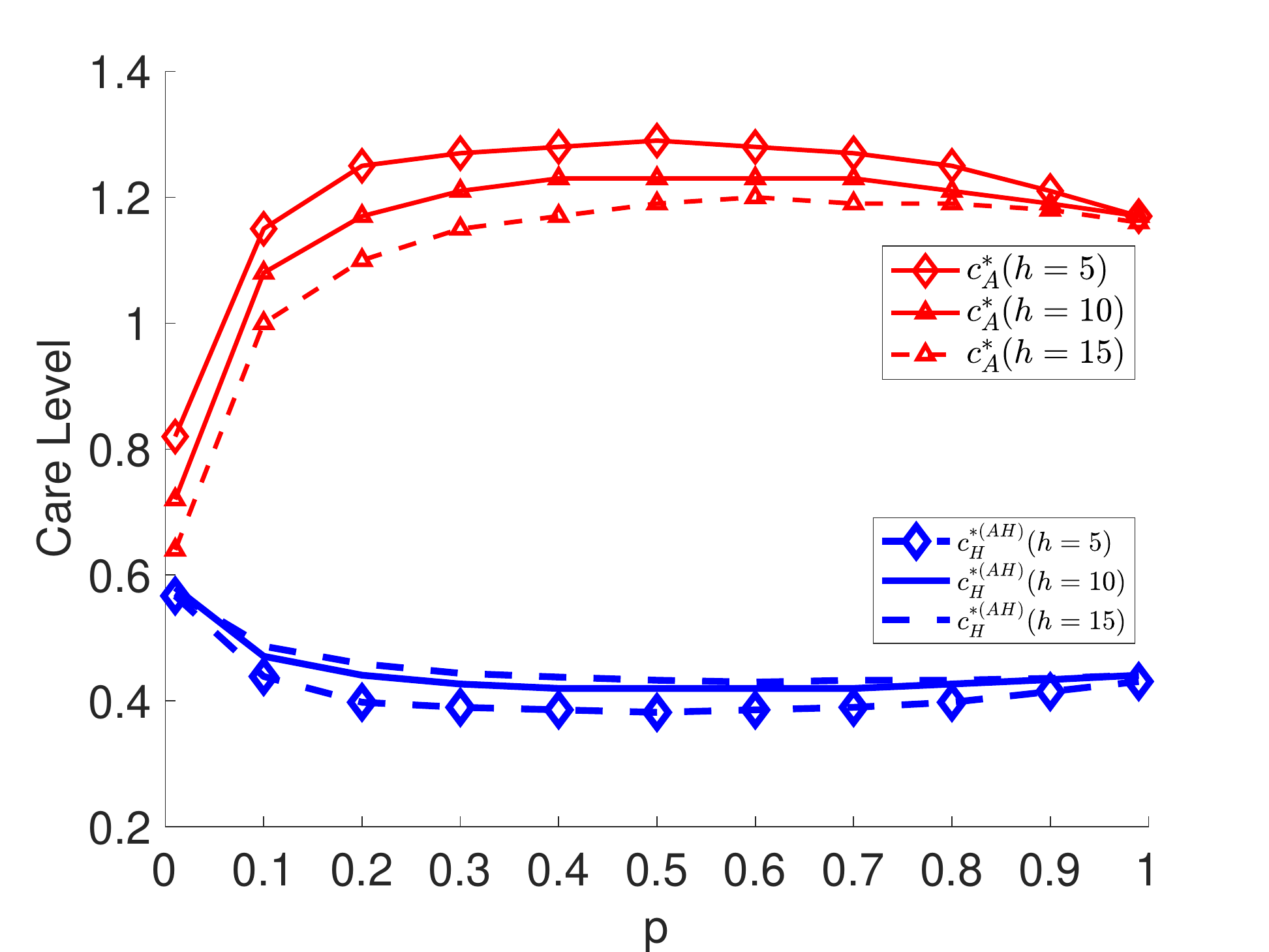}}
	\subfloat[Performance measures when varying $h$ \label{fig:external_cost_1}]{\includegraphics[width=3in]{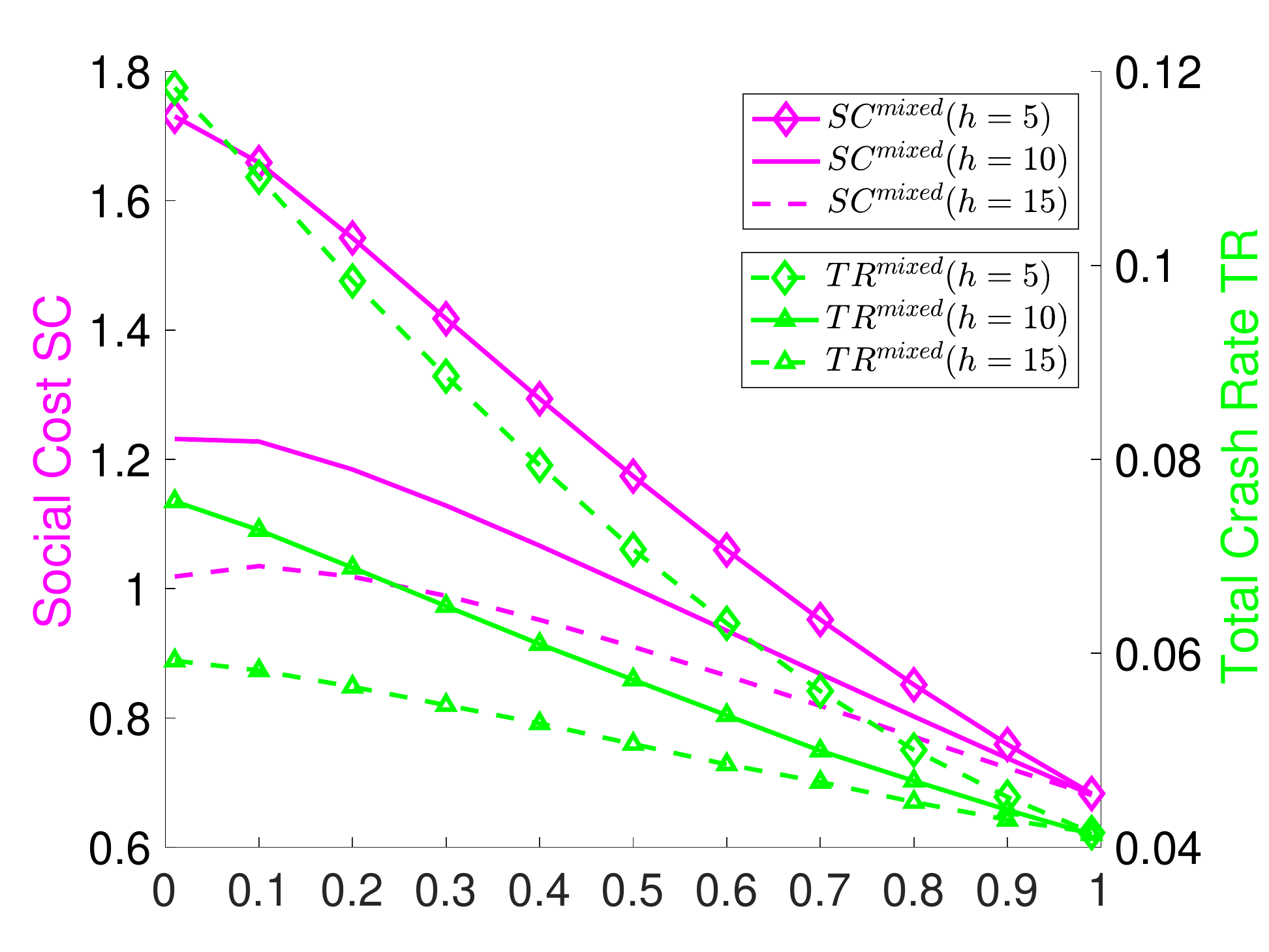}}
	\caption{Sensitivity analysis for external road environment parameters}
	\label{fig:sensiecternal}
\end{figure}
Here we investigate the impact of external environment parameters $a,h$ on the system performance in \figref{fig:sensiecternal}. 
The upper two figures represent when $a$ is varied while the lower two figures represent when $h$ is varied. 
Two solid lines represent the base model ($a, h=10$). In the remaining four lines,
the lines with diamond markers represent the scenario when road environment is worse off ($a, h=5$) for AVs (or HVs) and the dashed lines represent the scenario when road environment is improved ($a, h=15$) for AVs (or HVs).

\figref{fig:extrenal_care} and \figref{fig:extrenal_care_1} plot care levels when varying $a$ and $h$, respectively. Within each case, red lines represent AVs' care levels while blue lines represent HVs'. \figref{fig:external_cost} and \figref{fig:external_cost_1} plot performance measures when varying $a$ and $h$, respectively. Within each case, maroon lines represent social cost while green lines represent total crash rate.

The case with varying the external environment for AVs, i.e., the parameter $a$, is more straightforward to interpret. 
When the road environment is improved for AVs ($a$ is increased from 10 to 15), in other words, the crash probability for AVs is lower at the same care level, the AV manufacturer lowers AVs' care level because of lower crash loss. Accordingly, HVs' crash loss in the $AH$ scenario is also lower, thanks to the lower crash probability incurred by AVs.  On the other hand, as followers, HVs should increase their care levels when AVs become less attentive. Driven by two opposite forces, in this case, HVs' care level also decreases.  This indicates that the improved road conditions allow the AV manufacturer and human drivers to drive with lower care levels. Both social welfare and road safety are enhanced. When the road environment is worse off for AVs ($a$ is decreased from 10 to 5), the care levels and performance measures show the opposite trend from those in the improved case.

The case with varying the external environment for HVs, i.e., the parameter $h$, is more complicated.  
When the road environment is improved for HVs ($h$ is increased from 10 to 15), HVs' care levels increase. 
Although the crash probability for HVs decreases due to the improve road environment for HVs, it also reduces the crash probability for AVs in the $AH$ encounter scenario. 
As the leader, the AV manufacturer would reduce its precaution cost by taking advantage of such improvement for HVs. HVs, as followers, have to increase their care levels if the benefit of environment improvement cannot counteract the reduction of AVs' precaution cost. 
Both social welfare and road safety are enhanced when the road environment is improved for HVs. 
When the road environment is worse off for HVs ($h$ is decreased from 10 to 5), the care level and performance measures show the opposite trend from those in the improved case.

Parameter $a$ reflects an abstract representation of the external road environment for AVs. Now we will discuss a specific scenario when the road environment is improved for AVs, which is the construction of “dedicated AV lanes or zones”. Assuming that AVs and HVs use completely separate lanes or zones while driving in a road network, accordingly, the right-of-way between AVs and HVs are separated as well, leading to the removal of the encounter between AVs and HVs. The equilibrium care levels are altered, because the AV manufacturer only needs to take the $AA$ scenario into consideration.
Therefore, the game is reformulated as:
   \begin{equation}
   \begin{aligned}
   &[\mbox{\textbf{GameAV}}]\min \limits_{c_A \in {\cal C}_{c_A}}  && C_{A}(c_A)=w_a^{sen} \cdot p \cdot S_A(c_A)+ w_a^{loss} \cdot \frac{p^2}{p^2+(1-p)^2}\cdot L(c_A,c_A)\\
   &[\mbox{\textbf{GameHH}}]\min \limits_{c_{H_i}^{(HH)} \in {\cal C}_{c_H}} && C_{H_i}^{(HH)}(c_{H_1}^{(HH)},c_{H_2}^{(HH)}),\ i=1,2
   \nonumber
   \end{aligned}
   \end{equation}
\begin{figure}[H]
	\centering
	\subfloat[Care level \label{fig:exclusive_carelevel}]{\includegraphics[width=3in]{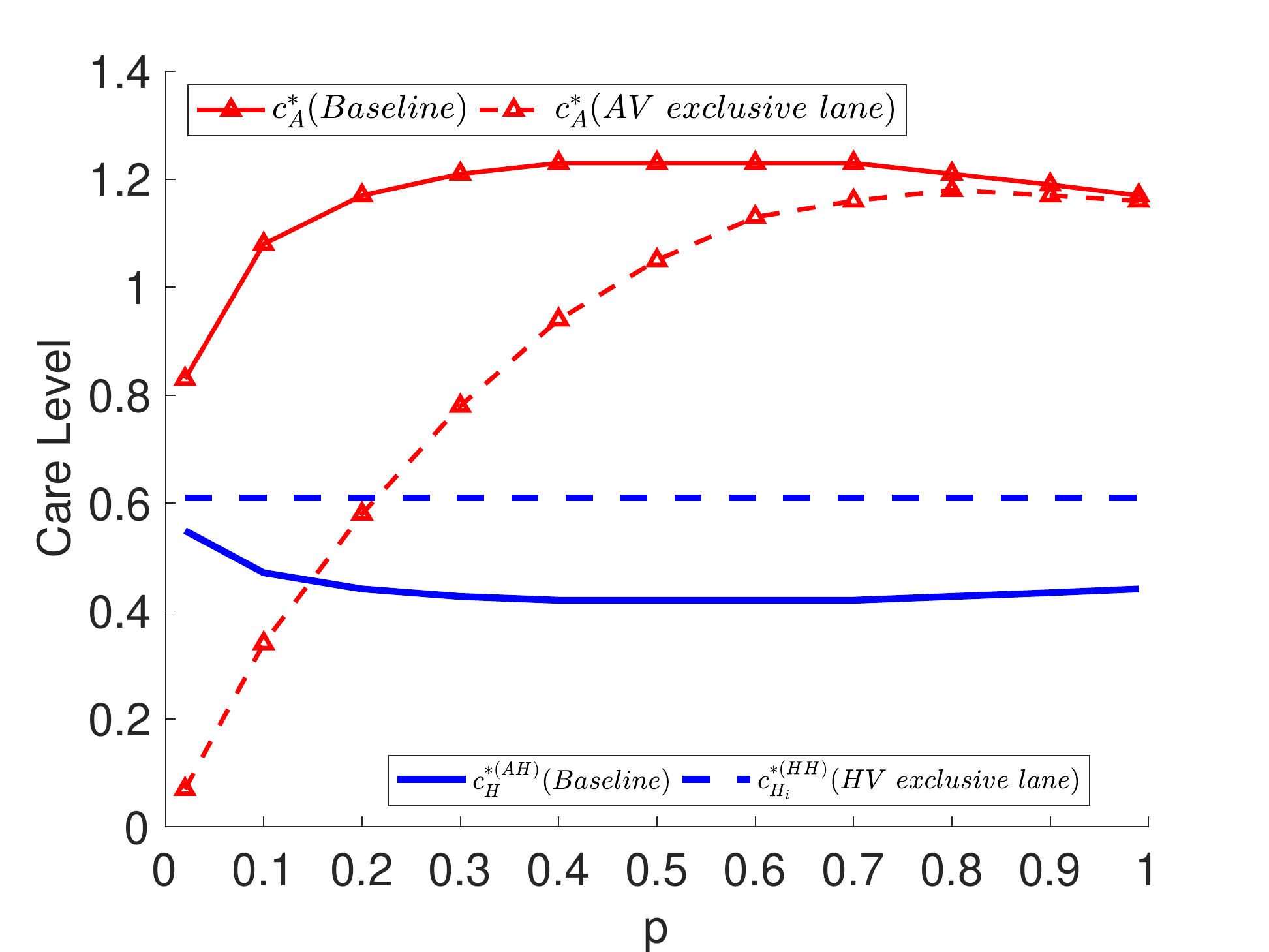}} 
	\subfloat[Performance measures \label{fig:exclusive performance}]{\includegraphics[width=3in]{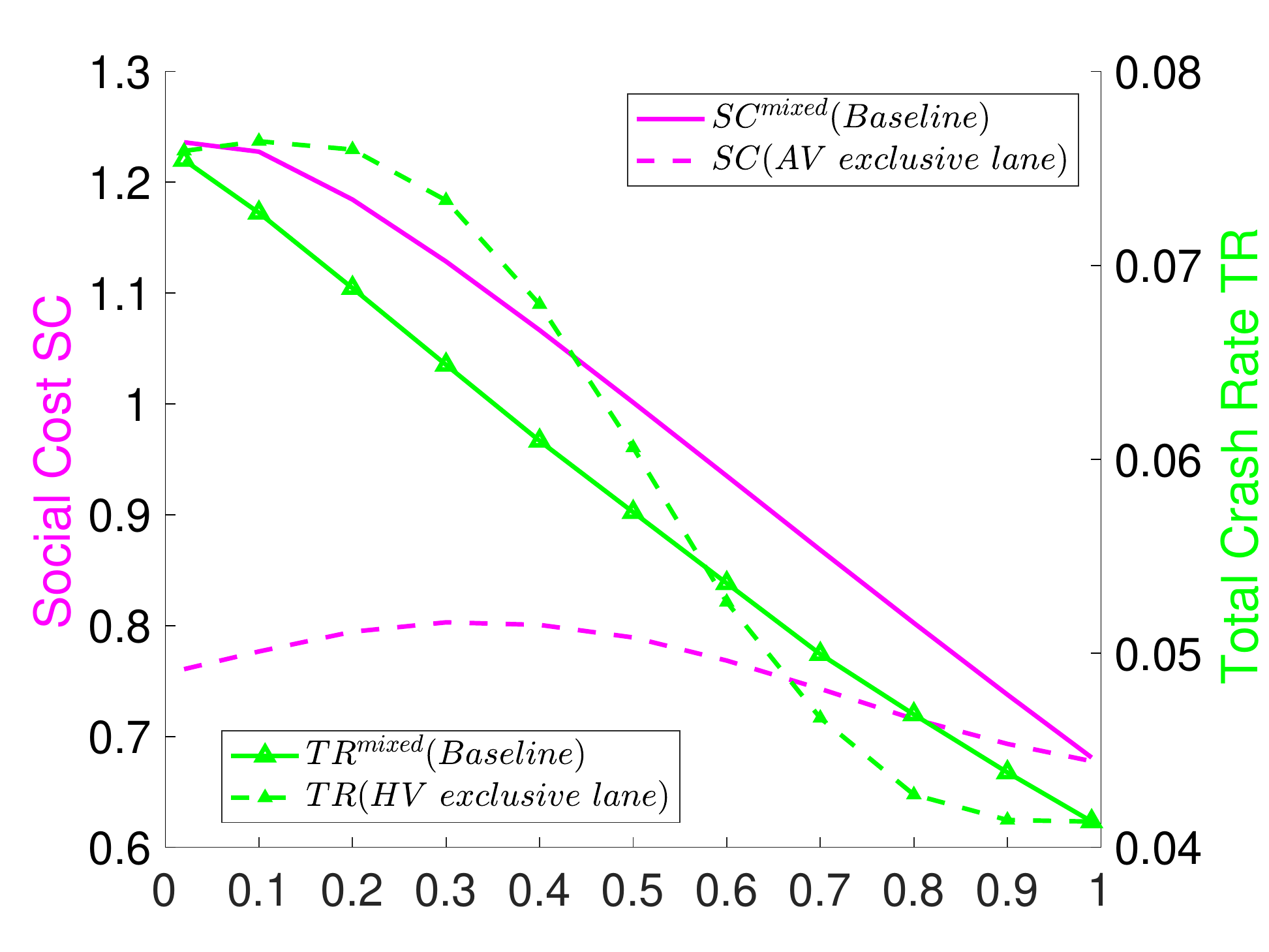}}
	\caption{AV exclusive lane}
	\label{fig:exclusive}
\end{figure}

In \figref{fig:exclusive_carelevel}, two solid lines represent care levels in the base model. The dashed red line with triangle markers represents AVs' care level in AV exclusive lanes and the dashed blue one represents HVs' care level in HV exclusive lanes. 
It is not surprising that AVs' care level in AV exclusive lanes is lower than that in mixed lanes, because the AV manufacturer does not need to worry about the crash loss of the $AH$ scenario.

In \figref{fig:exclusive performance}, two solid lines represent performance measures in the base model and the dashed lines represent those in AV/HV exclusive lanes. 
Social cost with AV/HV exclusive lanes is always smaller than that in mixed lanes because there is no crash between AVs and HVs. In addition, social cost with AV/HV exclusive lanes first slightly increases when $p<0.4$ and then decreases. This is because AVs' care level is quite low when the market is not dominated by AVs. Similarly, the total crash rate with AV/HV exclusive lanes is higher than that with mixed lanes when $p<0.6$, also thanks to a low care level selected by the AV manufacture with AV/HV exclusive lanes. 
Accordingly, AV dedicated lanes or zones do not necessarily improve the system performance when the AV penetration rate is low.

\subsubsection{Cost Weighting Coefficient in the AV Manufacturer's Cost Function}
\label{sec:tradeoffcoefficent}

In this subsection, the discussion is motivated by the existing situation the AV industry faces, that is,  
many AV manufacturers are unclear to what extent they would be punished if their AVs are involved in any accidents due to the uncertainty in the liability system involved with AVs. 
To minimize the expected loss, most AV manufacturers set their AVs' driving algorithms highly conservatively. 
It is interesting to find that Californians have uploaded numerous Youtube videos to share their opinions of how conservatively Waymos drive \citep{waymoYoutube}. 
We would like to explore how different cost designs would influence traffic in general and how this impact will evolve as the AV penetration rates grows.  


\begin{figure}[H]
	\centering
	\subfloat[Care level \label{fig:weight_care}]{\includegraphics[width=3in]{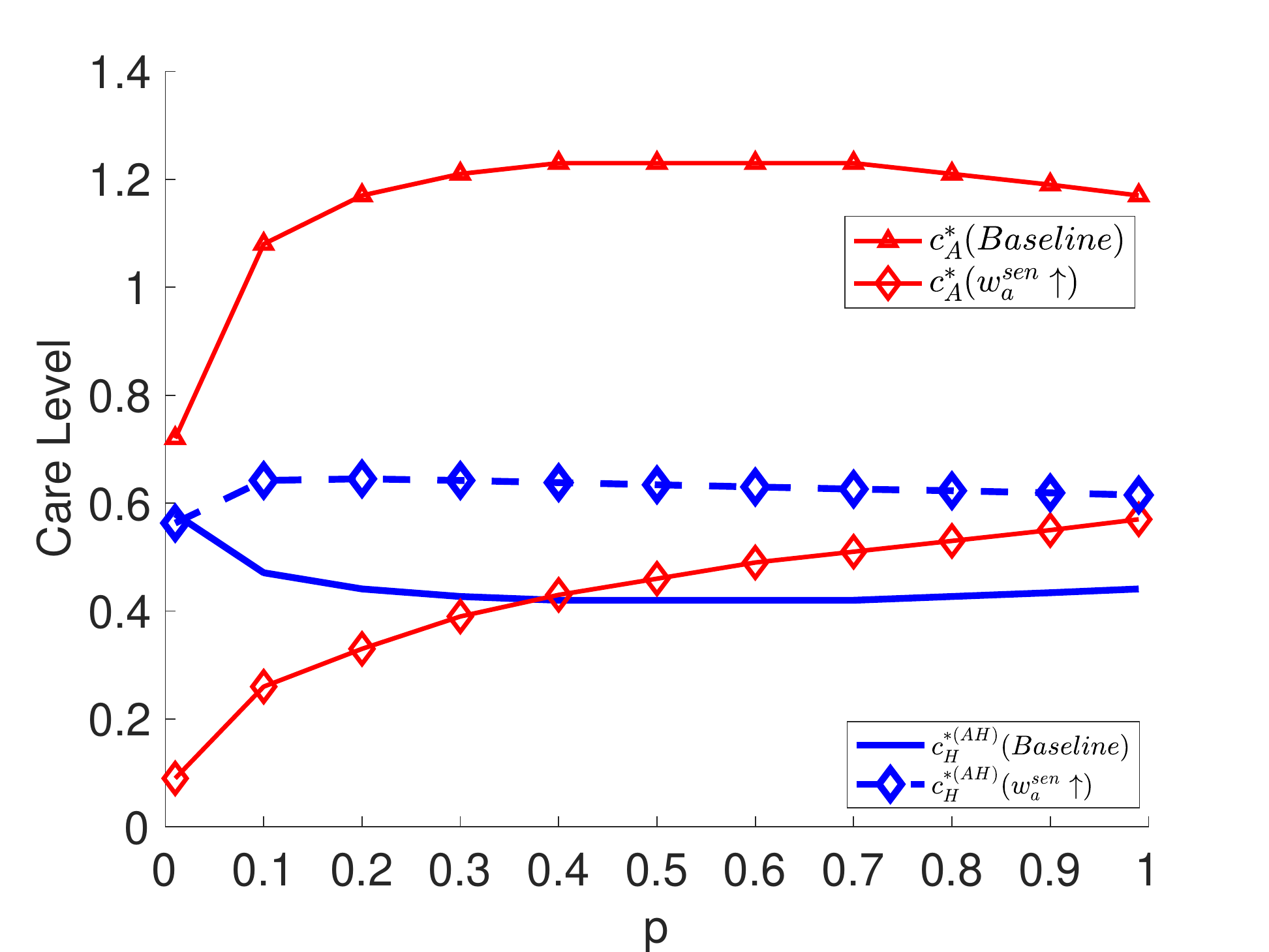}}
	\subfloat[Performance measures \label{fig:weight_cost}]{\includegraphics[width=3in]{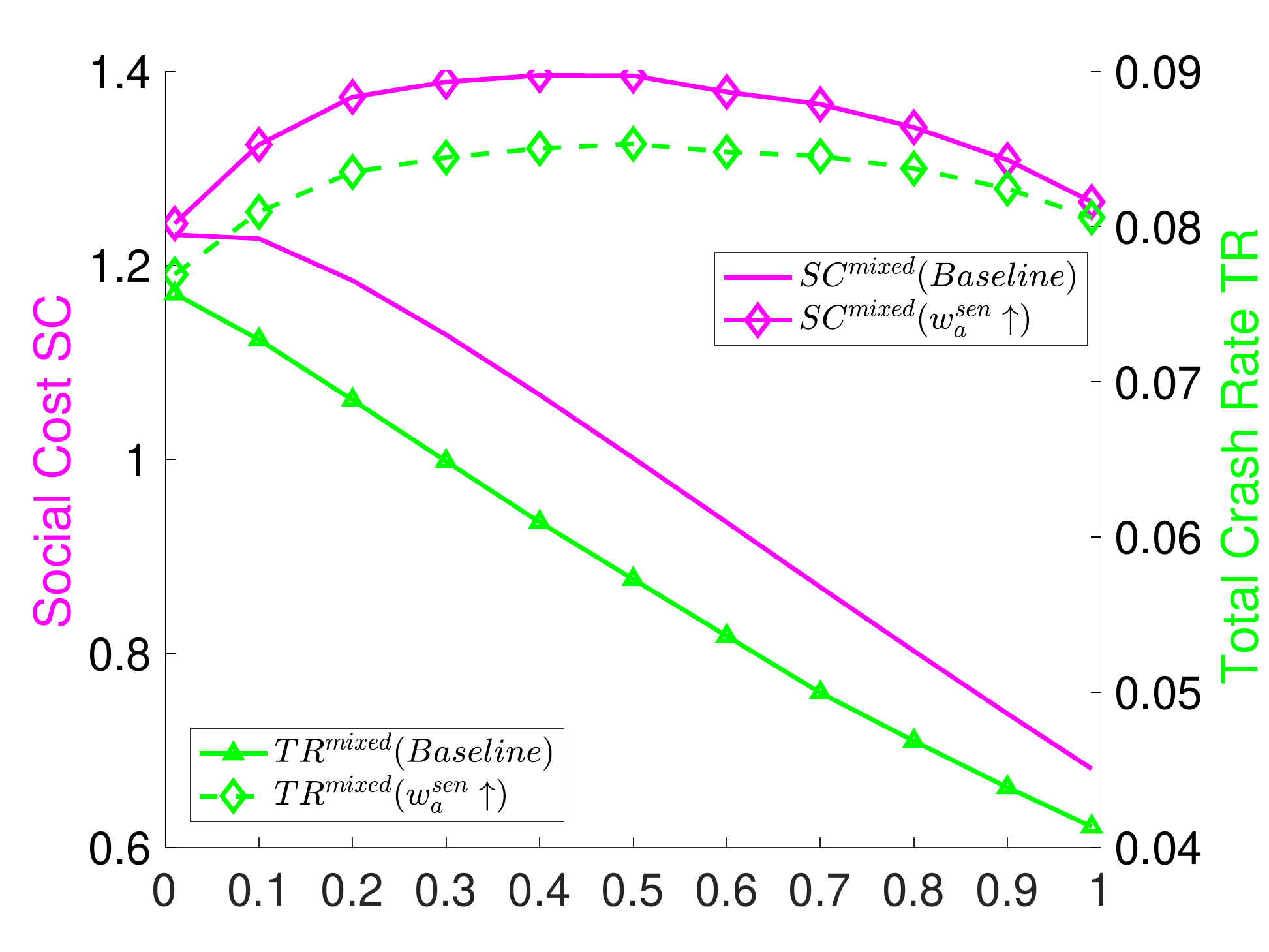}}
	\caption{Sensitivity analysis for the weighting coefficient in the AV manufacturer’s cost}
	\label{fig:weight}
\end{figure}

\figref{fig:weight} shows how the trade-off coefficient $w_a^{sen}$ in the AV manufacturer's cost affects AVs' equilibrium care level and road safety. 
The solid lines represent the base model ($w_a^{sen}=0.125, w_a^{loss}=0.25$) and the lines with diamond markers represent the scenario when $w_a^{sen}=0.67$ and $w_a^{loss}=0.16$. In \figref{fig:weight_care}, red lines represent AVs' care levels while blue lines represent HVs'. In \figref{fig:weight_cost}, maroon lines represent social cost while green lines represent total crash rate.

In \figref{fig:weight_care}, we observe unexpected behavior of AVs when $w_a^{sen}$ is increased from 0.125 (baseline) to 0.67. AVs' care level is much lower than HVs', because the AV manufacturer needs to reduce production cost as much as they can and care less about crash loss. 
This makes the AV manufacturer less attentive and accordingly, human drivers have to drive much more careful when they encounter AVs than in the base model. 
Accordingly, social welfare and crash rate increase first and then decrease, as $p$ increases. 
There is a penetration regime when $p\in(0, 0.4)$, social welfare is compromised and road safety is worsened off, because of the increasing vehicle encounters between AVs and HVs. 
This case may not be realistic, but its take-home message is that, if the AV manufacturer is not regulated in terms of AV technology specifications or is not properly subsidized, the AV manufacturer can be purely profit-oriented and harm the overall traffic system. 





\begin{figure}[H]
    \centering
	\subfloat[Care level when varying $w_a^{loss}$ \label{fig:weight_care_1}]{\includegraphics[width=3in]{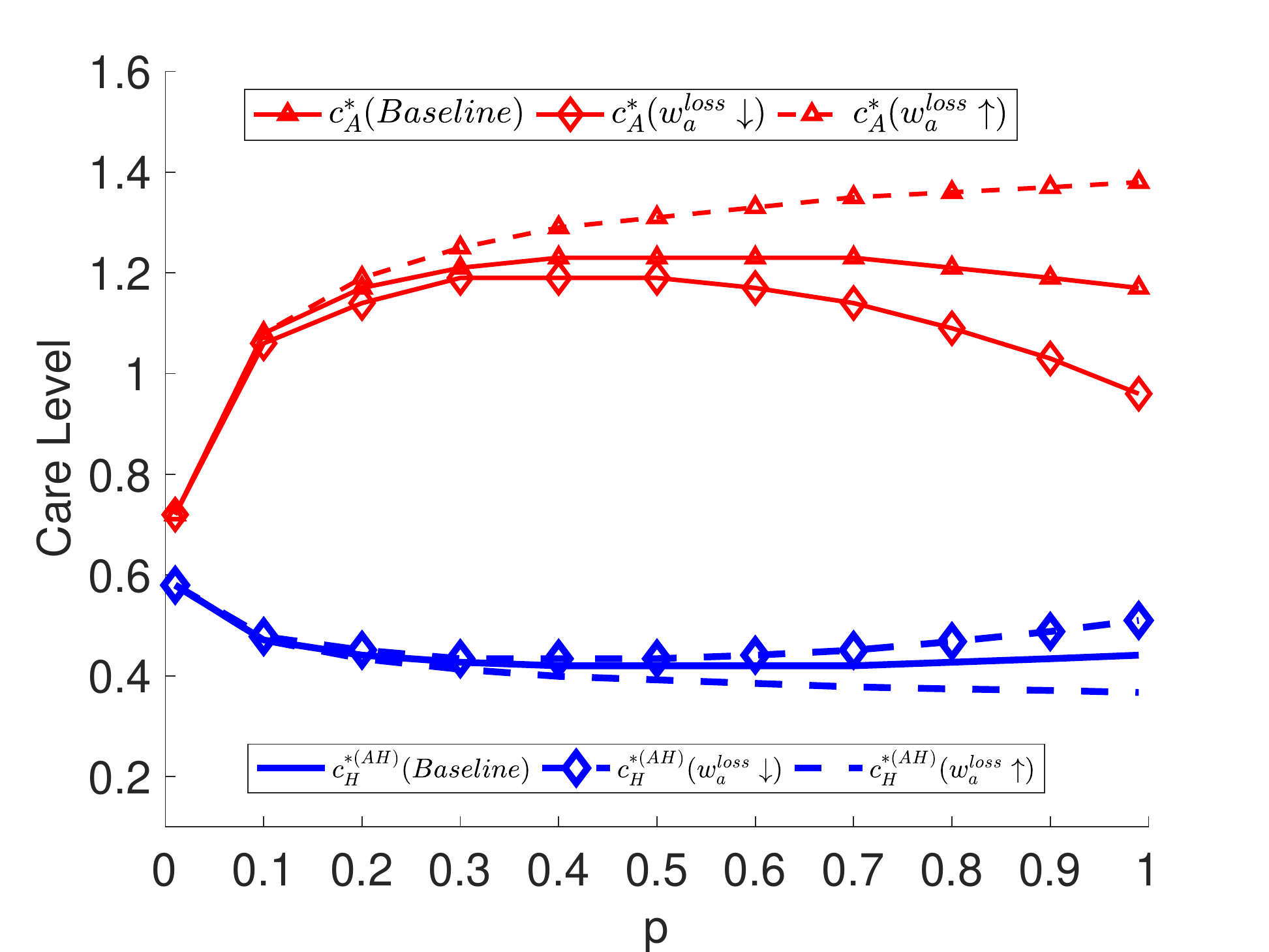}}
	\subfloat[Performance measures when varying $w_a^{loss}$ \label{fig:weight_cost_1}]{\includegraphics[width=3in]{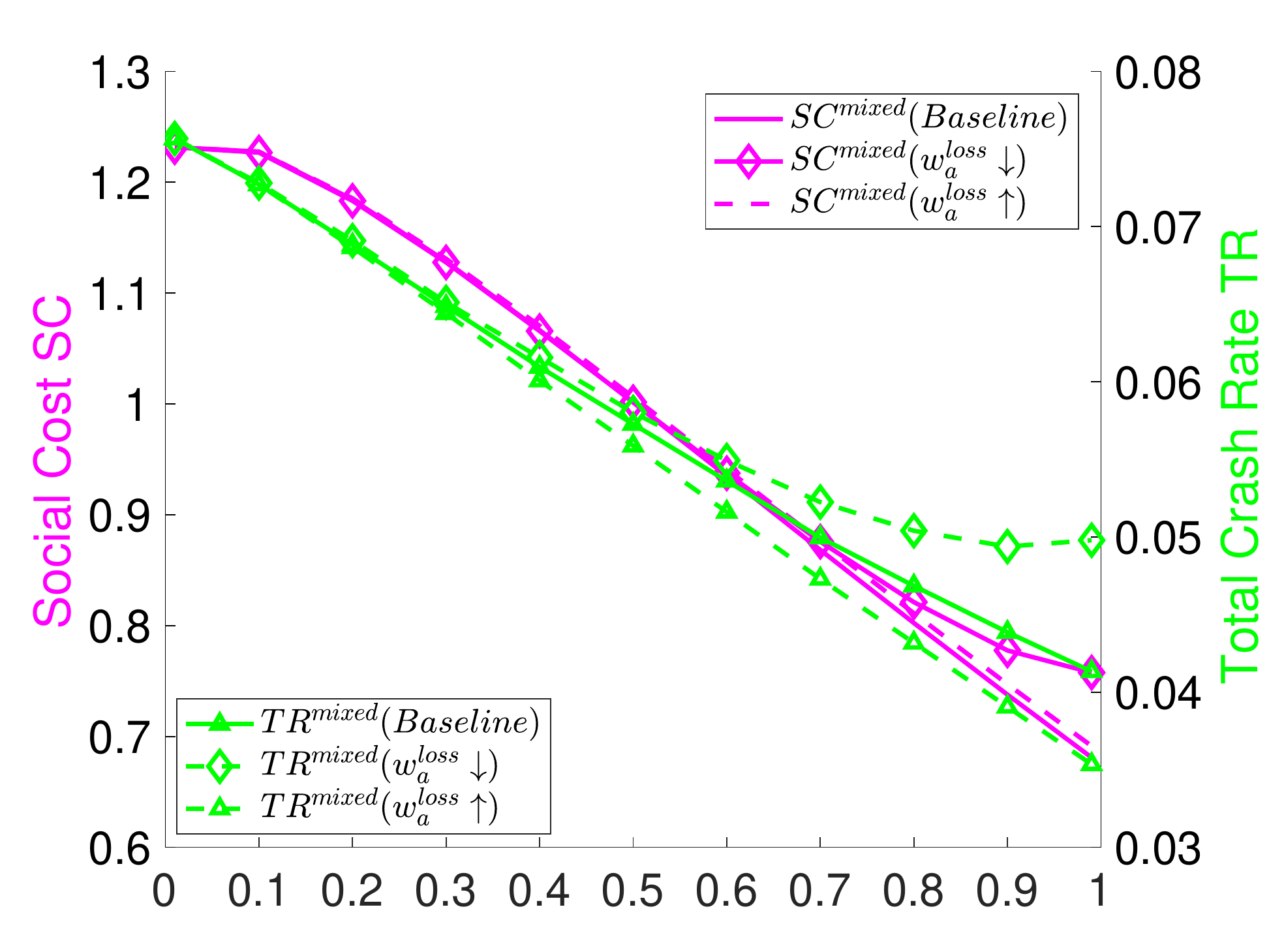}}
	\caption{Sensitivity analysis for the weighting coefficient in the AV manufacturer’s cost}
	\label{fig:avmanu_weight_aa}
\end{figure}

\figref{fig:avmanu_weight_aa} shows how the trade-off coefficients $w_a^{loss}$ in the AV manufacturer's cost affects AVs' equilibrium care level and road safety. Two solid lines represent the base model ($w_a^{loss} = 0.25$). In the remaining four lines, the lines with diamond markers represent the scenario when $w_a^{loss} = 0.15$ and the dashed ones represent the scenario when $w_a^{loss} =0.4$. Red lines in \figref{fig:weight_care_1} represent AVs' care levels while blue lines represent HVs'. Maroon lines in \figref{fig:weight_cost_1} represent social cost while green lines represent total crash rate.

When $w_a$ increases from 0.25 (baseline) to 0.4, the AV manufacturer allocates more weight (represented by the dashed red line with triangle markers) to crash loss over production cost and AVs drive more attentively. 
Accordingly, human drivers exhibit moral hazard (represented by the dashed blue line) because they perceive their environment comprised of AVs becomes safer. 
The social welfare and road safety are improved compared to the base model and as AVs grow.

When $w_a$ decreases from 0.25 (baseline) to 0.15, the AV manufacturer allocates less weight (represented by the dashed red line with diamond markers) to the crash loss over the production cost and AVs drive less attentively. Accordingly, human drivers increase their care level as the follower because they perceive their environment comprised of AVs becomes worse. The social cost and total crash rate become higher than those in the base model after $p>0.6$. Based on above observations, we can say that AVs' attentive attitude helps improves road safety and social welfare at the beginning with a smaller number of AVs. 
However, too conservative driving may jeopardize the overall traffic efficiency in the later stage. 
That is consistent with how California drivers feel while encountering AVs on roads \citep{waymoYoutube}. 
A gradual relaxation on $w_a$ will help improve traffic efficiency in the mixed AV-HV system.

\subsubsection{Heterogeneous Human Drivers}
\label{sec:hetero_hv}
\begin{figure}[H]
	\centering
	\subfloat[Heterogeneous human drivers' care level ($p=0.5$) \label{fig:heter_base_hist}]{\includegraphics[width=3in]{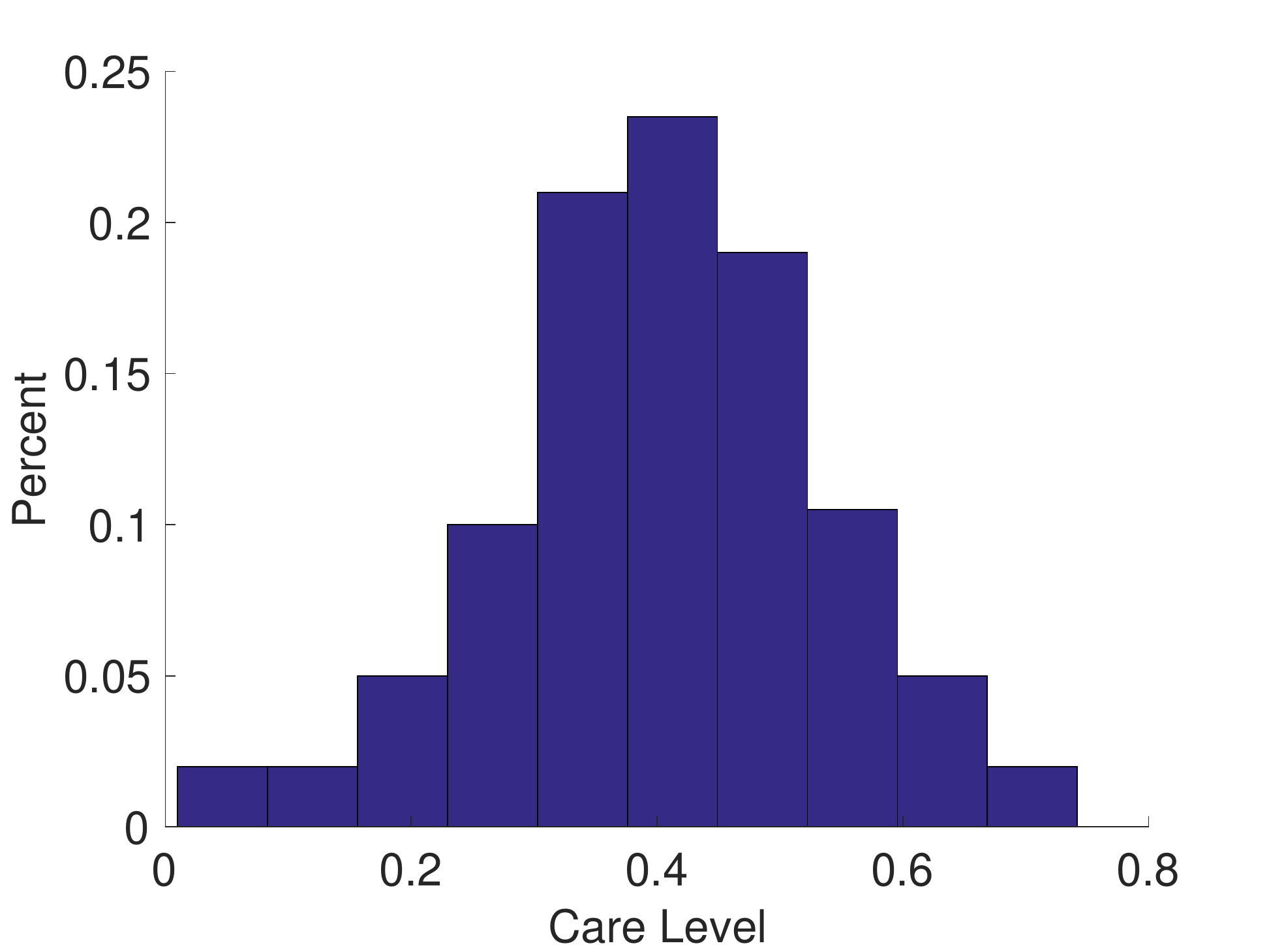}}
    \subfloat[Care level \label{fig:heter_base_carelevel}]{\includegraphics[width=3in]{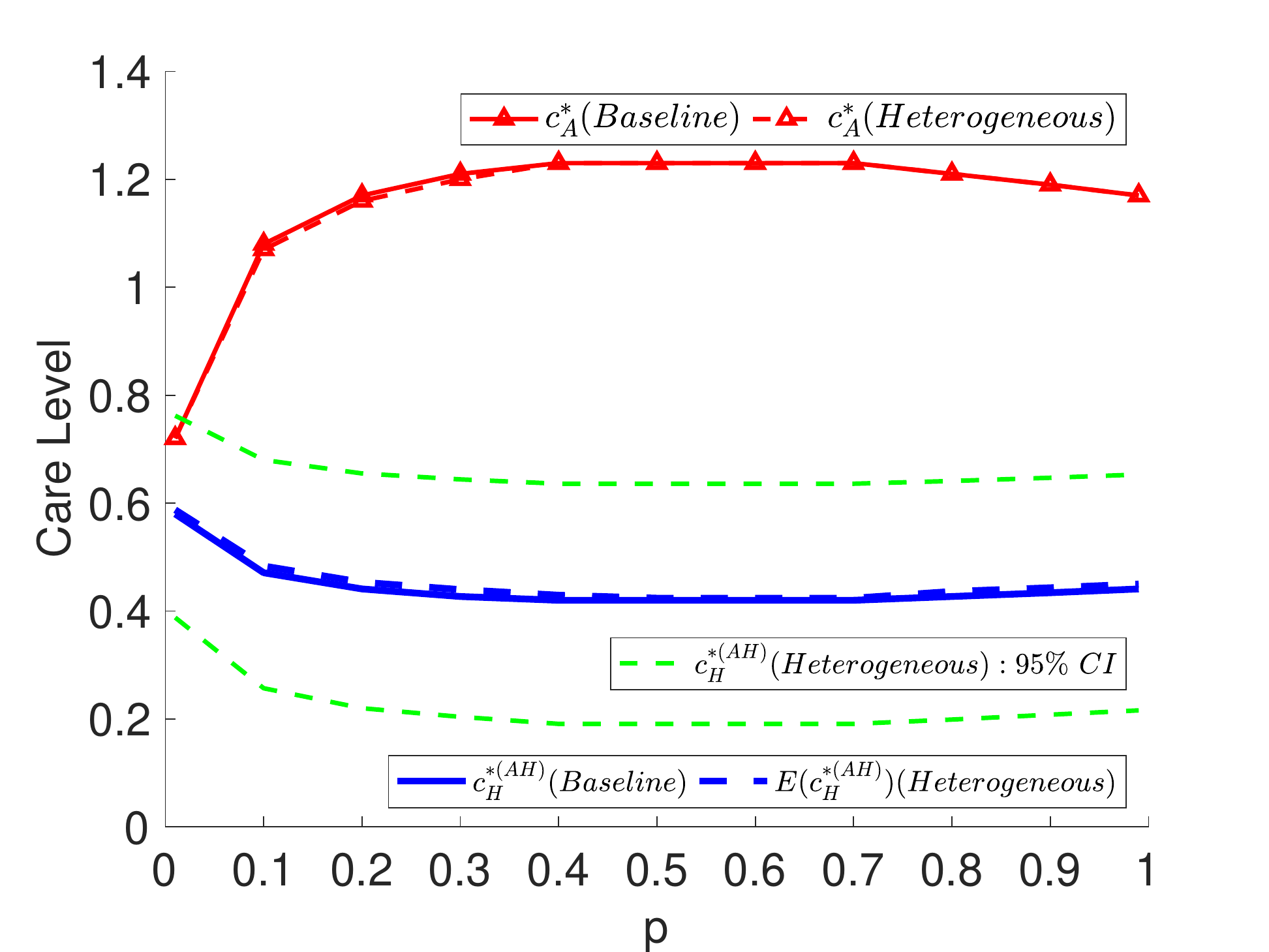}}
	\caption{Heterogeneous human drivers}
	\label{fig:heter1}
\end{figure}

Here we investigate how heterogeneous human drivers affect the AV manufacturer. We make a comparison of  the base model ($w_h=0.5$) and the model with heterogeneous human drivers where $w_h$ is sampled from a truncated normal distribution $\mathcal{N} (0.5,0.1)$ over a closed interval $[0,1]$. It is practical for the AV manufacturer to consider the expected crash loss arising from interacting with a driver population in the $AH$ scenario, instead of frequently changing AVs' care levels every time to deal with different human drivers. Therefore, the AV manufacturer's cost is modified as:
\begin{linenomath*}
    \scriptsize{
    \begin{equation}
    C_{A}(c_A,c_{H}^{(AH)})= \underbrace{ w_a^{sen} \cdot p \cdot S_A(c_A)}_{\text{sensor cost}}+ \overbrace{ \underbrace{w_a^{loss} \cdot p^2 \cdot L(c_A,c_A) }_{\text{crash loss of $AA$ scenario}}+\underbrace{ (1-w_a^{sen}-w_a^{loss})\cdot 2p(1-p)  E_{w_h}[L(c_A,c_{H}^{(AH)}(w_h)) \cdot s_{A}^{(AH)}]}_{\text{AV's loss share in the $AH$ scenario}}}^{\text{crash loss involved with AVs}}.
    \nonumber
\end{equation}}
\end{linenomath*}
The new game with heterogeneous human drivers is solved using Monte Carlo simulation. In each run, we sample $w_h$ randomly from the truncated normal distribution and solve the human drivers' care level. We then average human drivers' care levels from 1000 runs to calculate the expected crash loss share, i.e.,  $E_{w_h}[L(c_A,c_{H}^{(AH)}(w_h)) \cdot s_{A}^{(AH)}]$, and solve the equilibrium care level for the AV manufacturer. In \figref{fig:heter_base_hist}, we plot the histogram of HVs' care levels at a penetration rate of $p=0.5$. In \figref{fig:heter_base_carelevel}, two solid lines represent care levels in the base model. The dashed red line represents AVs' care level when facing heterogeneous HVs. The dashed blue line represents the mean 
and $95\%$ standard deviation 
of humans' care levels. Note that AVs' care levels facing a population of homogeneous and heterogeneous drivers are almost the same, indicating that heterogeneity of human drivers would not change our conclusions significantly.

\subsection{Lawmaker's Decision}
\label{sec:sen3} 
In the previous examples, the liability rule, i.e., the care level standard ratio $k$, is fixed. 
In this part, we assume the lawmaker is a strategic game player who optimizes $k$ as the penetration rate of AVs increases. 

\begin{figure}[H]
	\centering
	\subfloat[Social cost \label{fig:socialcost_k}]{\includegraphics[width=3in]{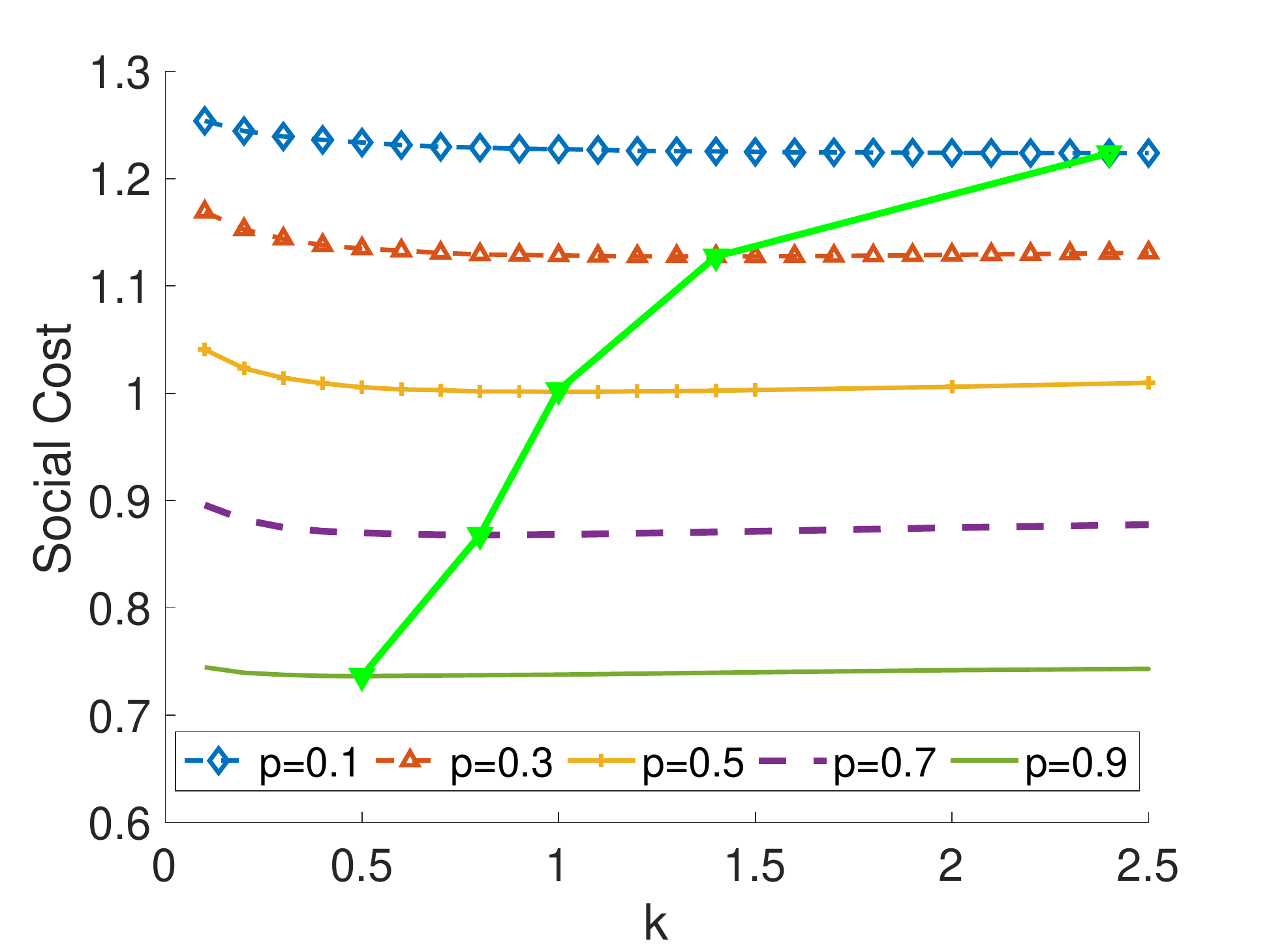}}~
    \subfloat[Optimal care level standard ratio \label{fig:k_p}]{\includegraphics[width=3in]{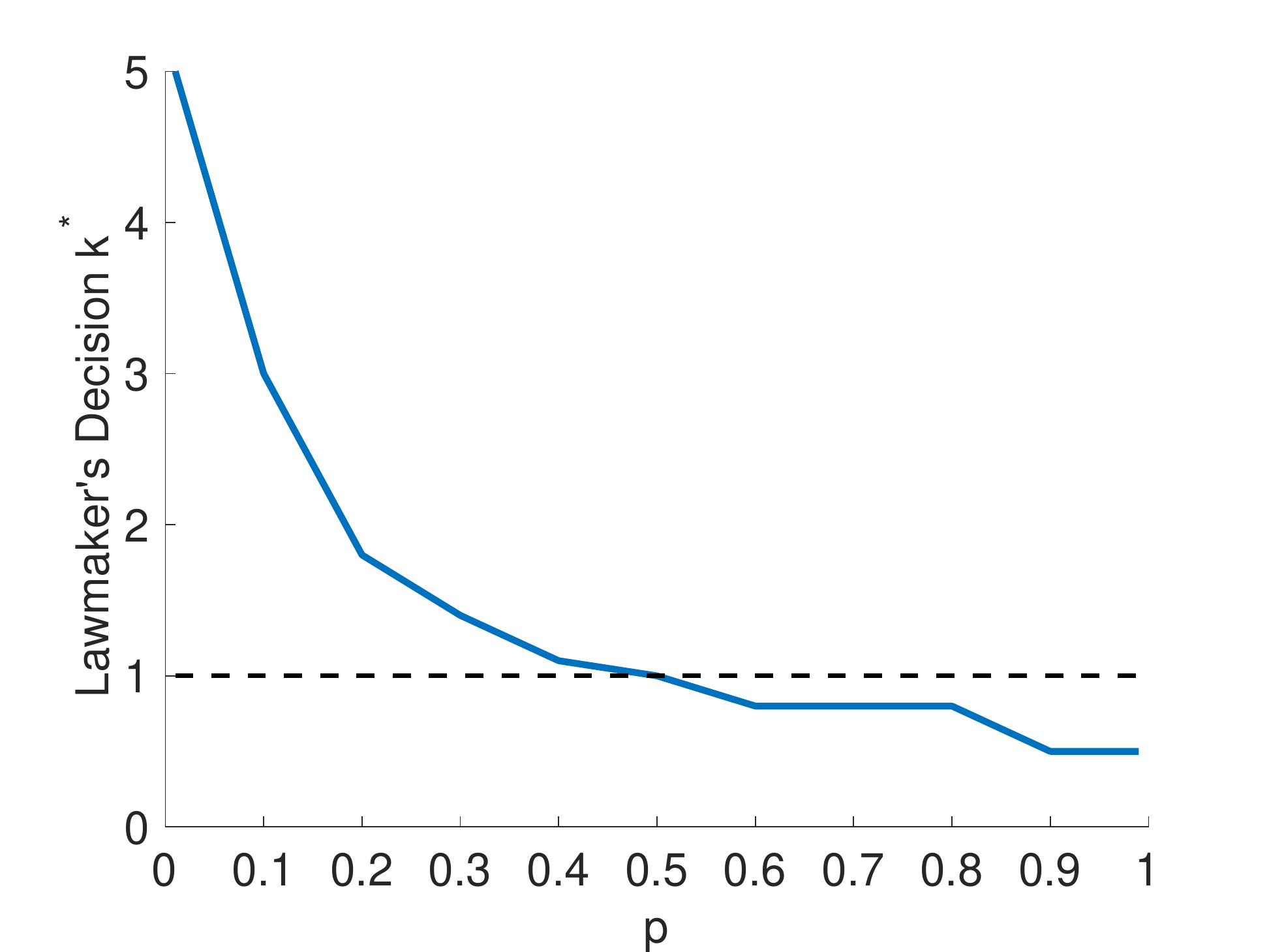}}
    
    \subfloat[Algorithm convergence $p=0.3$ \label{fig:converp_0_3}]{\includegraphics[width=3in]{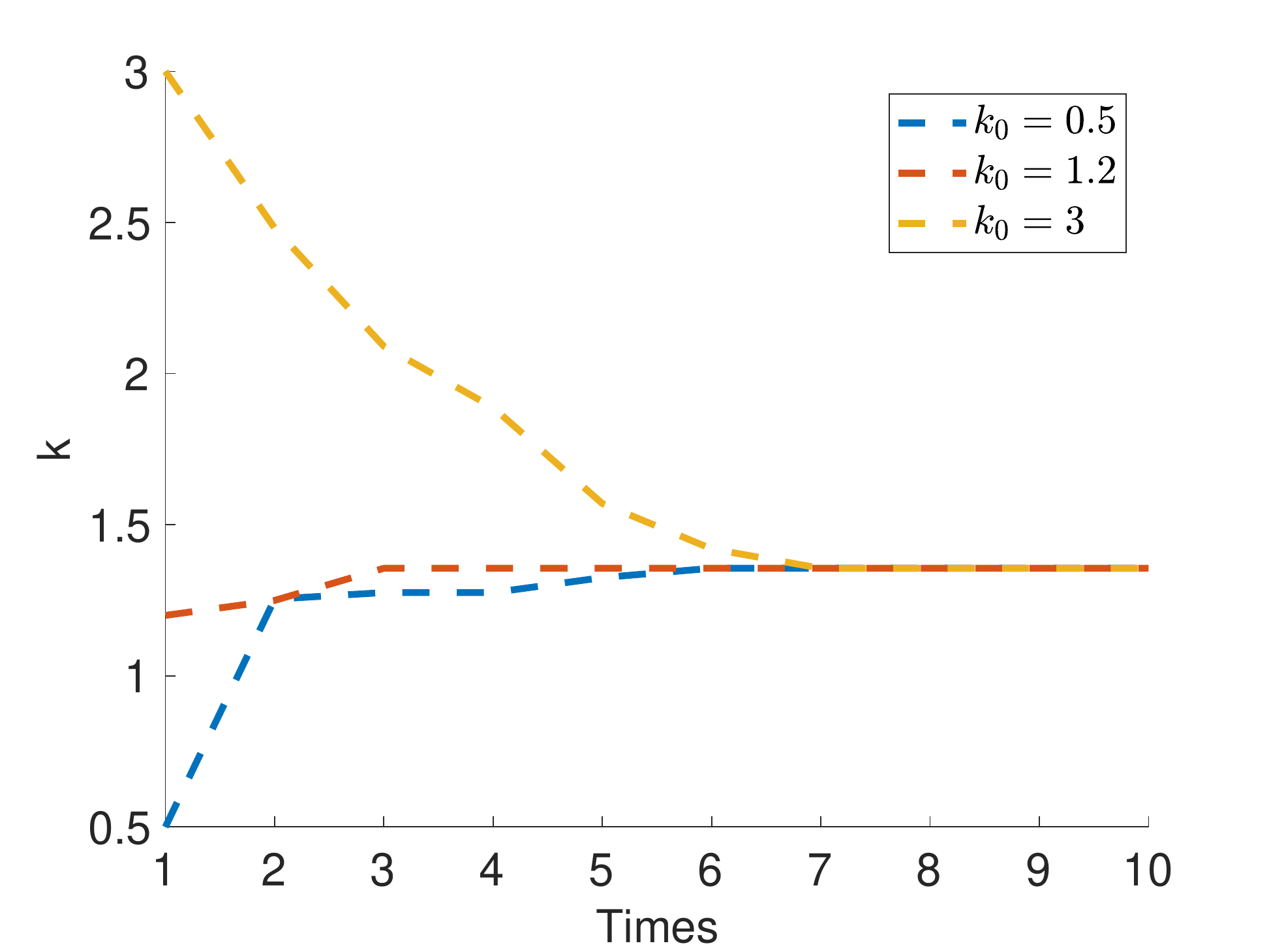}} 
	\subfloat[Algorithm convergence $p=0.7$ \label{fig:converp_0_7}]{\includegraphics[width=3in]{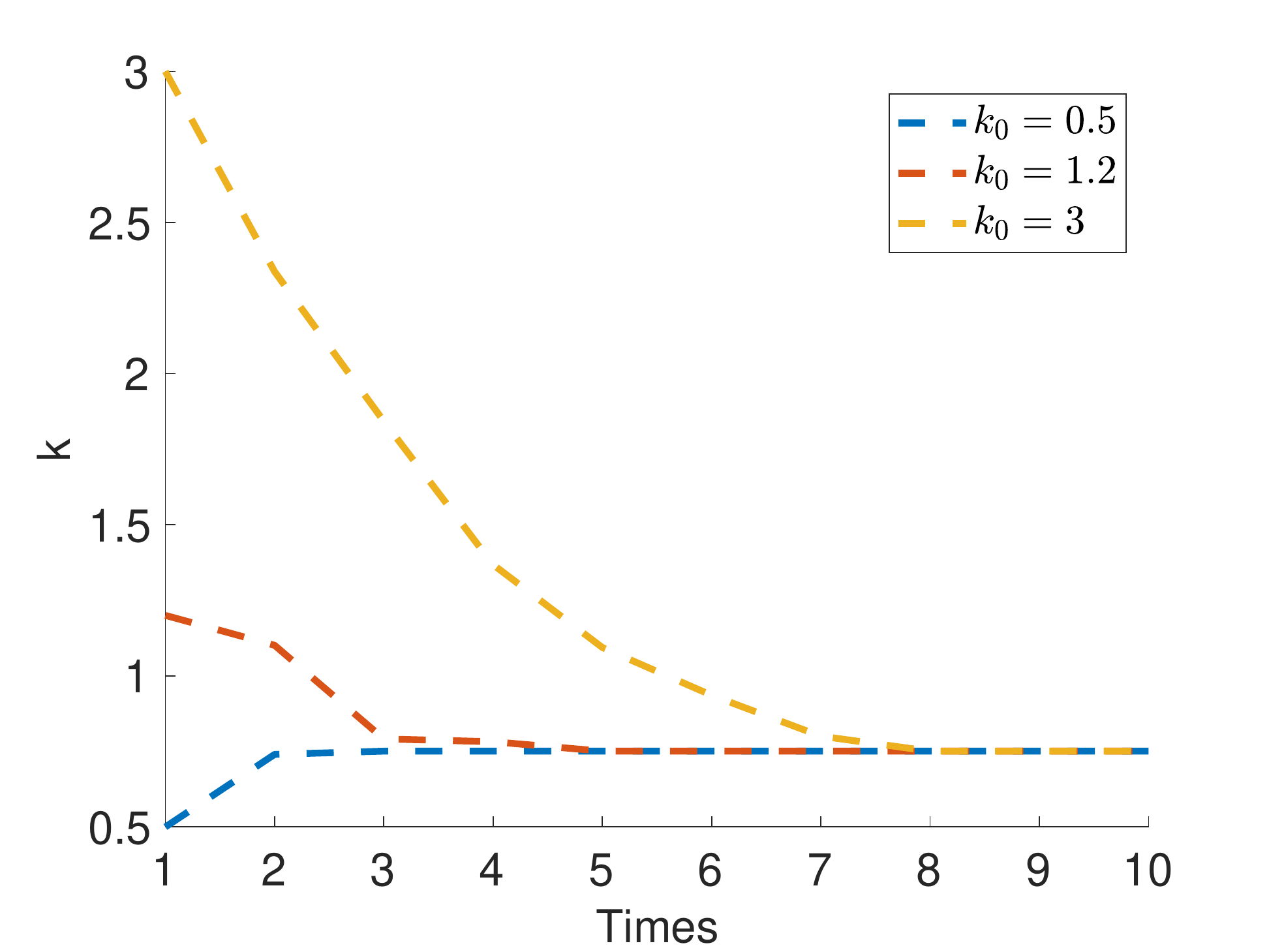}}
	\caption{Lawmaker's decision when varying $p$}
	\label{fig:stralawsocialcost}
\end{figure}

As we enumerate all the $k$'s, the social cost can be plotted as a function of $k$. \figref{fig:socialcost_k} demonstrates how the social cost changes as $k$ varies. 
Each curve represents a social cost function with respect to $k$, given $p$. 
The optimal care level standard ratio $k^*$ is obtained at the minimal of a social cost curve. We also solved optimal $k^*$'s using our proposed algorithm and plot it against $p$ in \figref{fig:k_p}. 
The proposed algorithm to solve the lawmaker's decision $k^{*}$ is implemented in Matlab R2017a. \figref{fig:converp_0_3} and \figref{fig:converp_0_7} show the algorithm convergence given $p=0.3$ and $p=0.7$, respectively. 
We can see that an optimal $k$ is attained in less than 10 iterations. The solid line represents the solved optimums $k^*$ and the dashed line represents $k=1$ in the base model.
We can see the solved optimums in \figref{fig:k_p} are consistent with those identified from \figref{fig:socialcost_k}. 
As the AV penetration rate increases, the lawmaker decreases its care level standard ratio first at a faster rate and then at a slower rate. 
When $p> 0.5$, the optimal care level standard ratio is below one, indicating that the lawmaker sets a much lower standard for human drivers as the AV population grows. 

\begin{figure}[H]
	\centering
    \subfloat[Care level \label{fig:lawmaker_carelevel_1}]{\includegraphics[width=3in]{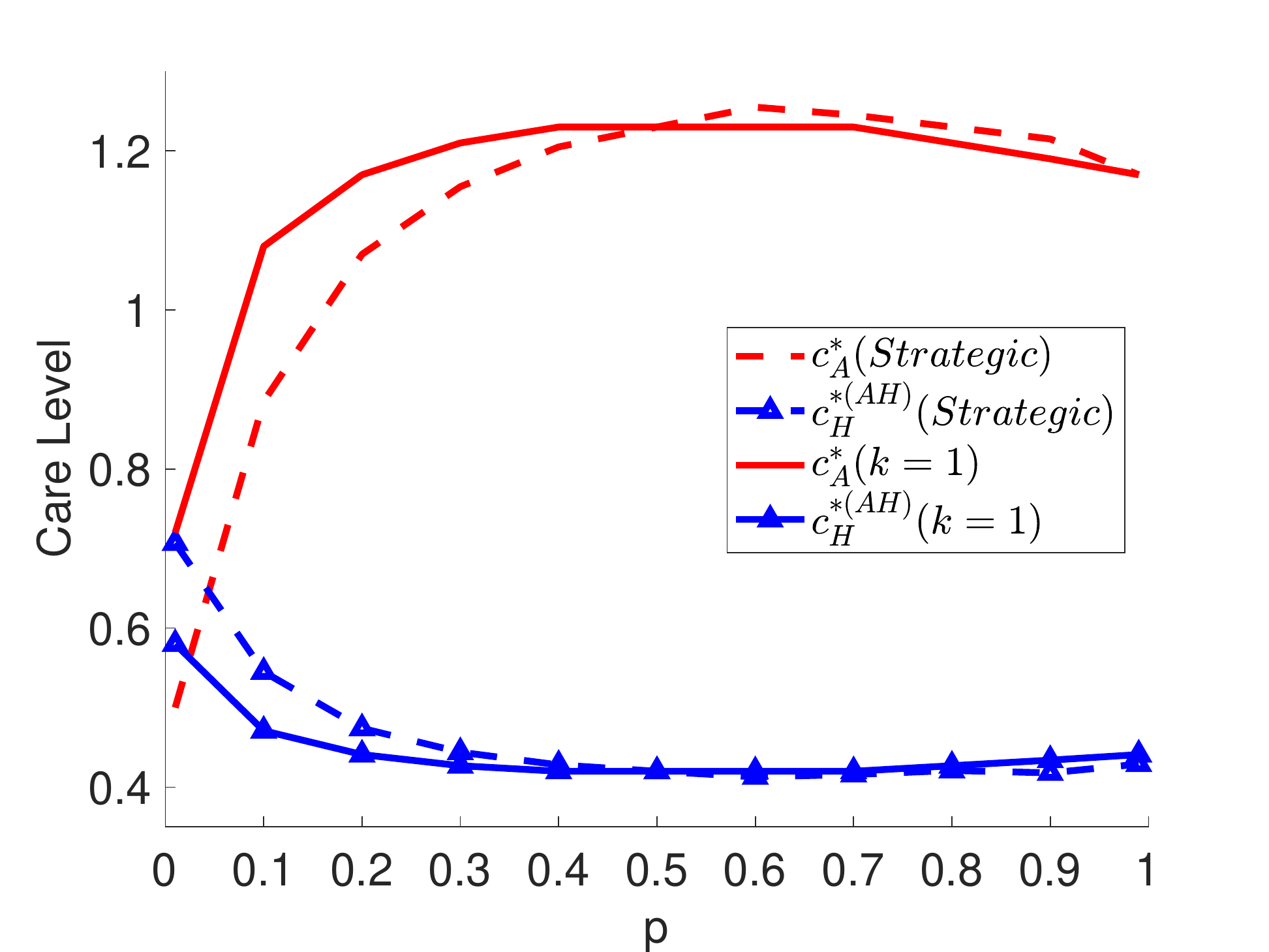}}
	\subfloat[Performance measures \label{fig:lawmaker_performance_1}]{\includegraphics[width=3in]{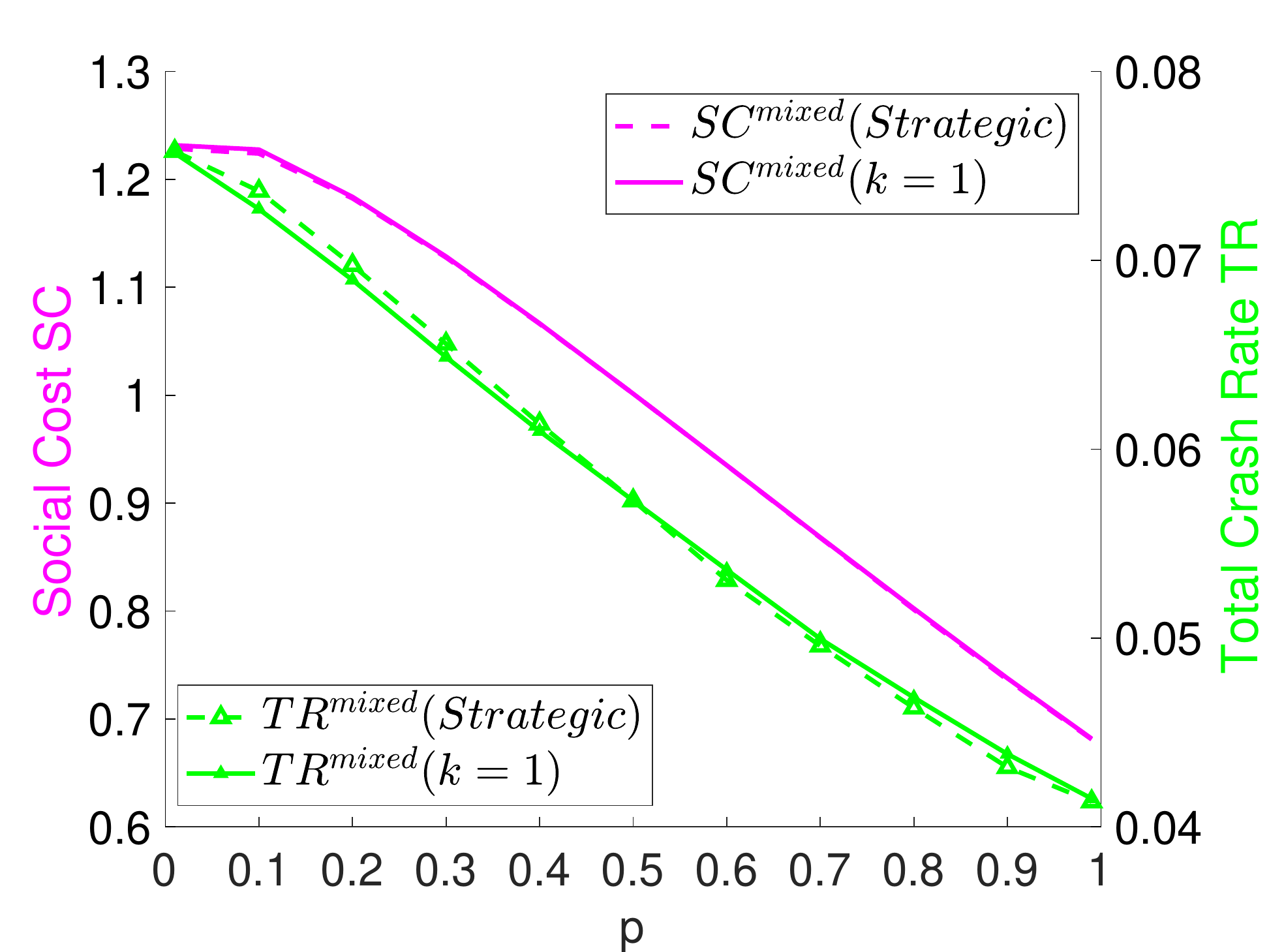}}
	\caption{Strategic lawmaker versus non-strategic lawmaker ($k=1$)}
	\label{fig:stralawperformancewithbasemodel}
\end{figure}

We first compare the base model when the lawmaker is not a strategic decision-maker in the base model ($k=1$)
and when the optimal $k^*$ is implemented. 
In \figref{fig:lawmaker_carelevel_1}, 
the solid lines represent care levels when the lawmaker is strategic and the dashed lines represent care levels in the base model. 
Red lines represent AVs' care levels while blue lines represent HVs'.

When $p>0.5$, AVs' care level under the strategic law-maker is higher than that under the non-strategic lawmaker, 
and HVs' care levels under the strategic law-maker is lower than that under the non-strategic lawmaker. 
The strategic lawmaker punishes AVs more than the non-strategic lawmaker when the AV penetration rate is relatively high, so the AV manufacturer has to increase its care level. 
Accordingly, HVs take advantage of AVs' increasing care level and exhibit moral hazard. 
When $p<0.5$, the trend is reversed, because the market is dominated by HVs, the strategic lawmaker punishes HVs more than AVs, leading to less attentive AVs. 
In \figref{fig:lawmaker_performance_1}, 
the solid lines represent performance measures when the lawmaker is strategic and the dashed lines represent performance measures in the base model. Maroon lines represent social cost while green lines represent total crash rate. 
Social welfare is slightly lower under the strategic law-maker, but total crash rate is slightly higher under the strategic law-maker when $p<0.5$. 
Because the goal of the strategic law-maker is to minimize social cost, not crash rate, it may compromise road safety a bit so that advanced transportation technologies can be adopted.   

\begin{figure}[H]
	\centering
	\subfloat[Care level \label{fig:lawmaker_carelevel2}]{\includegraphics[width=3in]{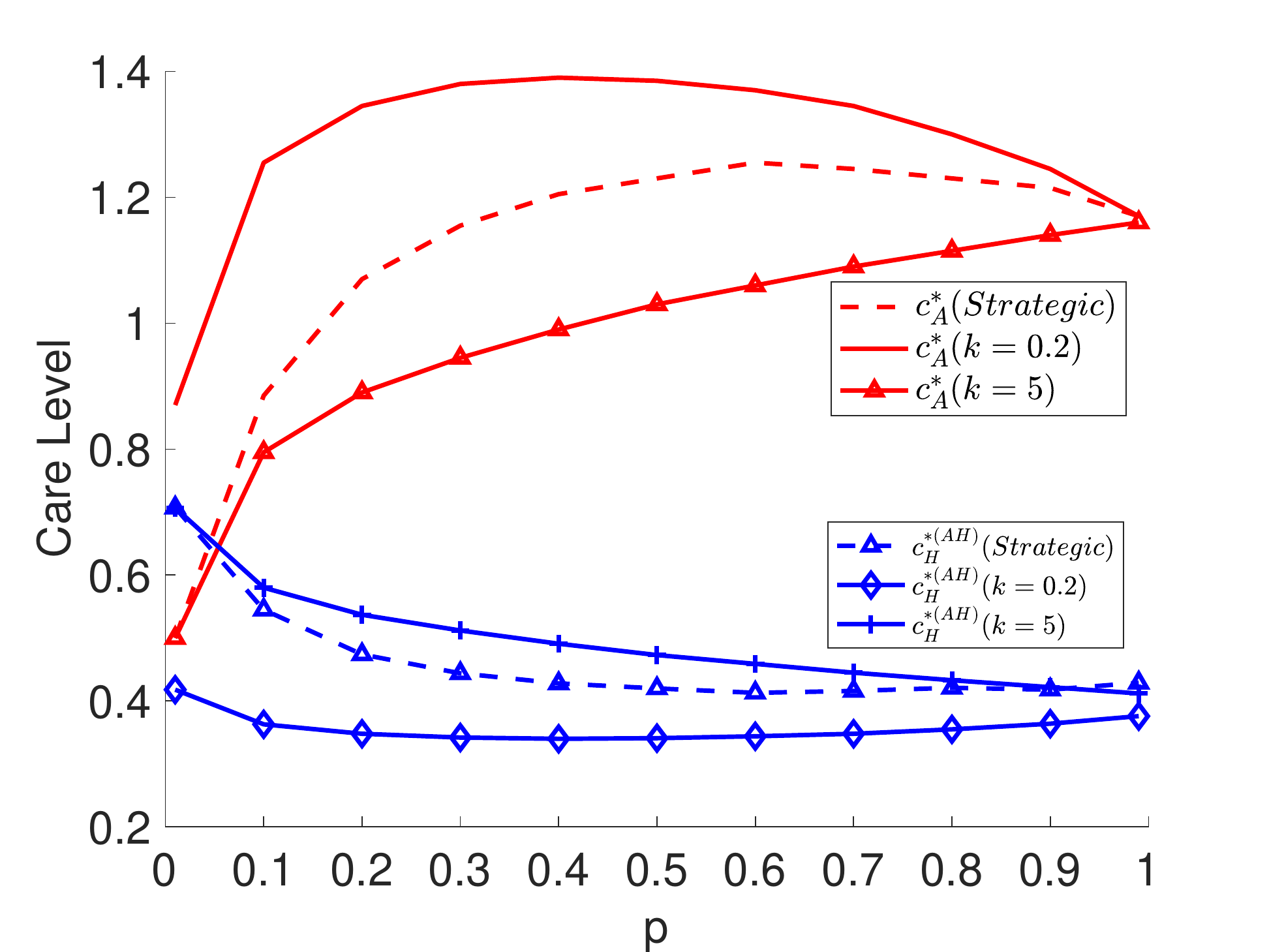}}
	\subfloat[Performance measures \label{fig:lawmaker_performance2}]{\includegraphics[width=3in]{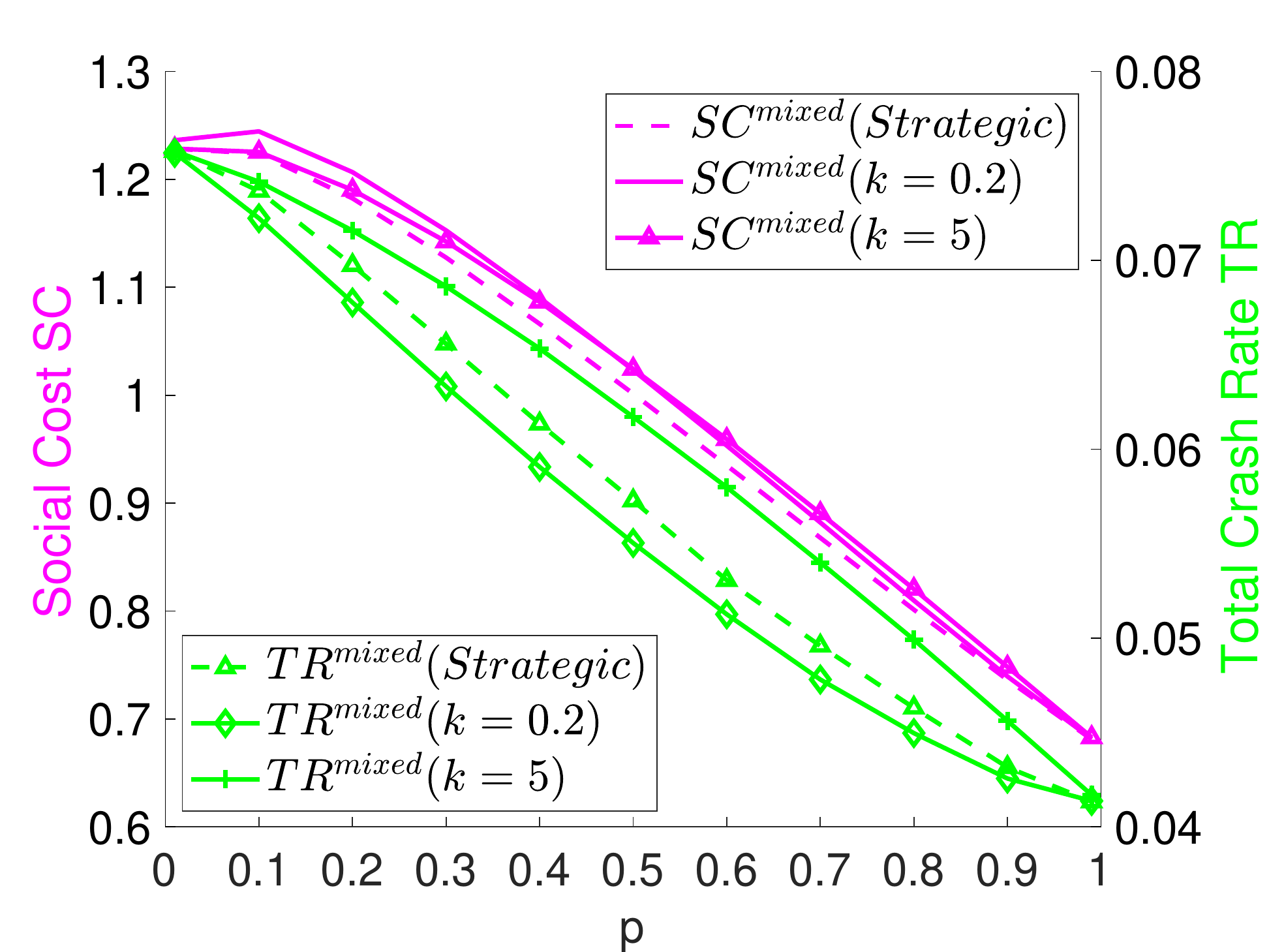}}
	\caption{Non-optimal care level standard ratios in comparison with the optimal one}
	\label{fig:stralawperformance2}
\end{figure}

We then compare how varying $k$ below or above optimal values influences care levels and the system performance. 
When the lawmaker is non-strategic, we assume the liability policy is given, which are $k=0.2$ and $5$, respectively. 
In each sub-figure of \figref{fig:stralawperformance2}, two lines with diamond markers (one is solid and the other is dashed) represent the scenario when $k=0.2$. In the remaining four lines, the solid ones represent the scenario with the optimal liability rule and the dashed ones represent the scenario when $k=5$.
\figref{fig:lawmaker_carelevel2} plot care levels in each case. Red lines represent AVs' care levels while blue lines represent HVs'. \figref{fig:lawmaker_performance2} plot performance measures in each case. Maroon lines represent social cost while green lines represent total crash rate.


When $k=0.2 (<1)$, AVs have high care levels while human drivers always choose low care levels. This is because the non-strategic lawmaker punishes AVs more than humans so that humans take advantage of the AV manufacturer when AVs' care level goes up. 
When $k=5 (>1)$, human drivers have a higher care level standard than AVs, leading to higher care levels for human drivers. 
As $p$ increases, human's care level goes down while that of AVs goes up. Humans also take advantage of AVs' increasing care levels.
With the stragic lawmaker, the optimal care level for humans is also inversely correlated to that for AVs.

Under the stragic lawmaker, human drivers tend to have higher care levels than under the non-strategic lawmaker. This indicates that the non-strategic lawmaker punishes human drivers to a lesser degree. 
AVs' care level under the strategic lawmaker is between those under the non-strategic lawmaker. 
In other words, the non-strategic lawmaker punishes AVs more than necessary if $k<1$, forcing the AV manufacturer to select a quite high care level in order to reduce crash losses. 
This stringent policy can ensure traffic safety in the mixed traffic but may halt the development of technologies at the initial stage. 

\subsection{Endogenous Penetration Rate}
\label{sec:sen5} 
Our previous analysis implicitly assumes that the AV penetration rate, $p$, is an exogenous parameter. 
At the initial stage of AV deployment, 
however, 
the most harmful loss for the AV manufacturer could be likely the external loss from the reduced public opinion (after the crash is released to the society) rather than the direct crash loss. 
To accommodate such understanding,  
we endogenize the AV penetration rate as a function of AVs’ crash loss arising from AV related vehicle encounters, which are AV-AV and AV-HV scenarios. 
Our goal is to explore how the equilibrium penetration rate, care levels, and lawmaker’s liability rule would vary.

Compared to the original game model with an exogenous $p$, the only modification here is to include an endogenous constraint: $p=f(L_{AV})$,
where $L_{AV}$ is the AV-related crash loss, computed as the sum of two losses for AA and AH scenarios, which is, $L_{AV}=p^2\cdot L(c_A,c_A)+2p(1-p) \cdot L(c_A,c_{H}^{(AH)})$. 
The function $f(L_{AV})$ is monotonically decreasing with respect to the AV related crash loss, representing how public concerns of AVs' safety aspect might lower their market share. Note that $L_{AV}$ is an implicit function of $p$, i.e., $L_{AV}\equiv L_{AV}(p)$. Thus the optimal $p$ is a fixed point of this constraint. 

According to \cite{henrik2005safety}, we employ a decreasing linear function for the right-hand side function. 
Then the endogenous constraint becomes:  
\begin{linenomath*}
\begin{equation}
    p= -\eta L_{AV}+1,
    \nonumber
\end{equation}
\end{linenomath*}
where $\eta$ represents the reduction rate of the AV market share given a one unite increase in the AV related crash loss. 
When $L_{AV}=0$, meaning there is no crash loss caused by AVs, the AV penetration rate is one. 

\begin{figure}[H]
	\centering
    \subfloat[Endogenous constraint \label{fig:endogen_baseline}]{\includegraphics[width=2.7in]{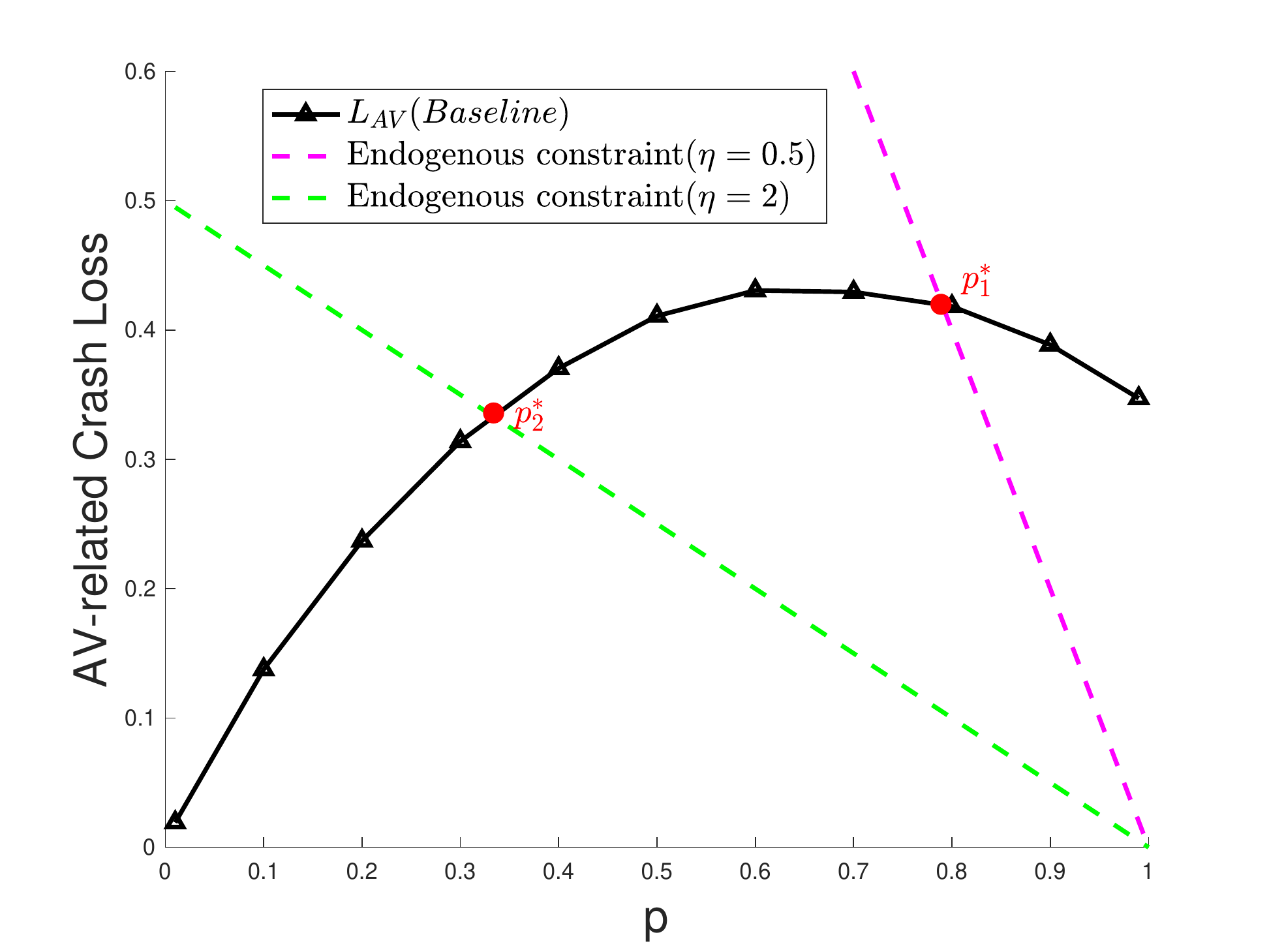}}
    
	\subfloat[Care level \label{fig:endogen_base_carelevel}]{\includegraphics[width=2.7in]{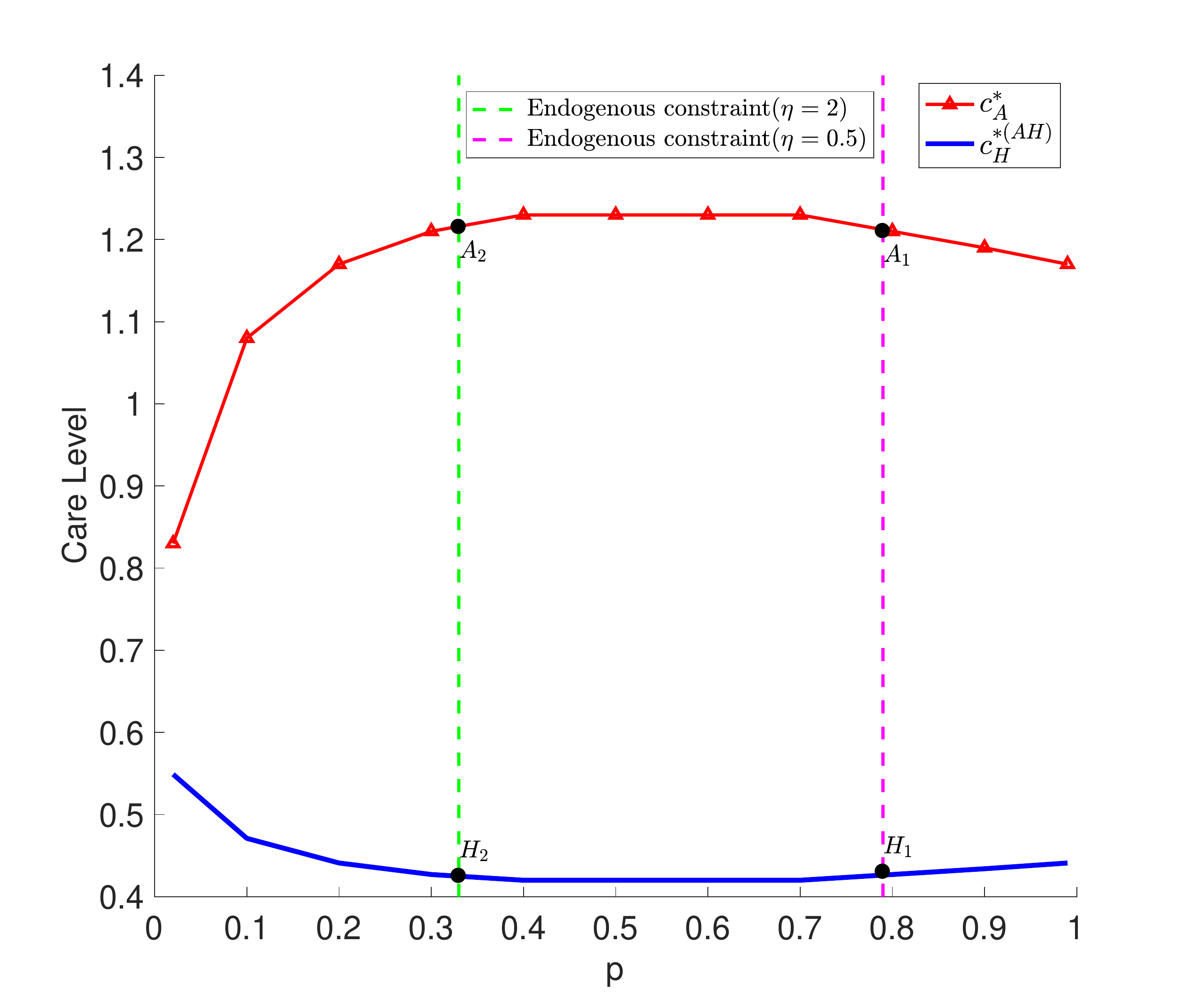}}
	\subfloat[Performance measures \label{fig:endogen_base_performance}]{\includegraphics[width=2.7in]{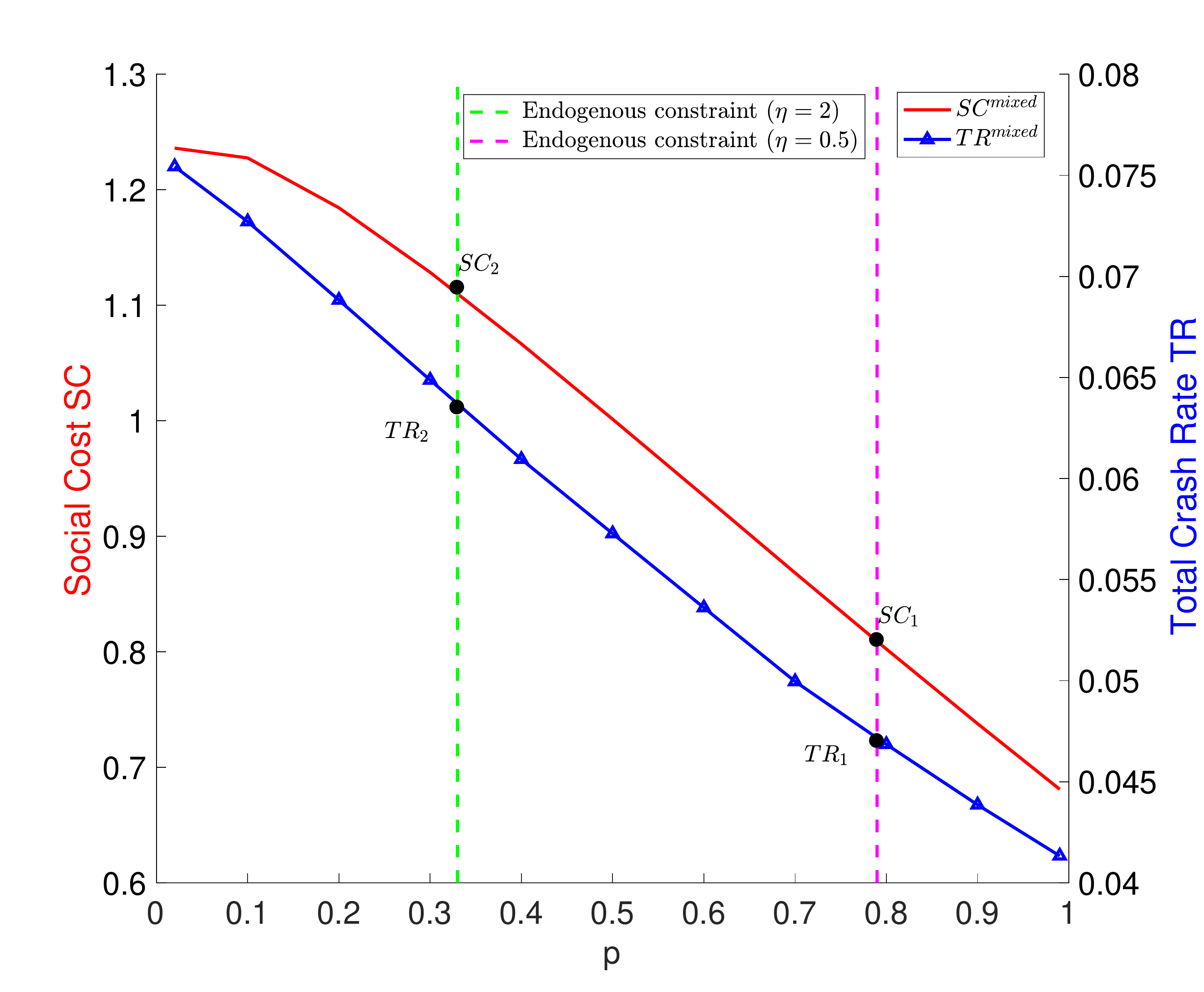}}
	\caption{Endogenous penetration rate}
	\label{fig:endogen_base}
\end{figure}

To understand the impact of endogenous $p$ on the equilibrium care level and the lawmaker’s decision, we need to discuss how the curve, which is the AV related crash loss, intersects with the new constraint.
The intersection point determines the equilibrium penetration rate and consequently, the equilibrium care levels. 
We first investigate how different $\eta$'s influence the game equilibrium in \figref{fig:endogen_base}.  

In \figref{fig:endogen_baseline}, the x-axis is $p$ and the y-axis is the AV related crash loss. 
The parabolic-shaped curve represents $L_{AV}$ without the endogenous constraint. 
Two different endogenous constraints with different $\eta$'s are plotted in dashed lines. 
The green dashed line ($\eta=2$) represents a market where people are more worried about AV crashes than the maroon one ($\eta=0.5$). 
The intersection points of these two endogenous constraints with the loss curve (solid line) indicate the equilibrium market shares, which are  $p^{*}_1=0.79$ to $p^{*}_2=0.33$, respectively. It is obvious that the more concerned people are about AV related crashes, the lower the equilibrium market share is.

Now we would like to inspect the equilibrium care levels of AVs and HVs and the system performance measure, at the equilibrium market share. 
\figref{fig:endogen_base_carelevel}-\figref{fig:endogen_base_performance} are essentially \figref{fig:basicmodelcarelevel} and \figref{fig:socialwelfare} with two additional endogenous constraints, indicated by vertical dashed lines. 
In \figref{fig:endogen_base_carelevel}, 
the intersection points, i.e., $A_1$ and $H_1$, $A_2$ and $H_2$, are the equilibrium care levels with endogenous constraints $\eta=0.5, 2$, respectively. 
It is interesting to find that neither the equilibrium care level of AVs nor HVs change much when the equilibrium penetration rate decreases from $p^{*}_1=0.79$ to $p^{*}_2=0.33$. 
In \figref{fig:endogen_base_performance}, the intersection points $SC_1$ ($SC_2$) and $TR_1$ ($TR_2$) are the equilibrium social cost and total crash rate at $p^{*}_1=0.79$ to $p^{*}_2=0.33$, respectively. Consistent with our finding from the baseline, the increasing AV market share improves social welfare, i.e., $SC_1<SC_2$, $TR_1<TR_2$. 
This example tells us that if media can ease the public concern towards the autonomous driving technology, the penetration rate of AVs may increase greatly, so is the social welfare.

\begin{figure}[H]
	\centering
	\subfloat[Endogenous constraint \label{fig:endogen_lawmaker}]{\includegraphics[width=3in]{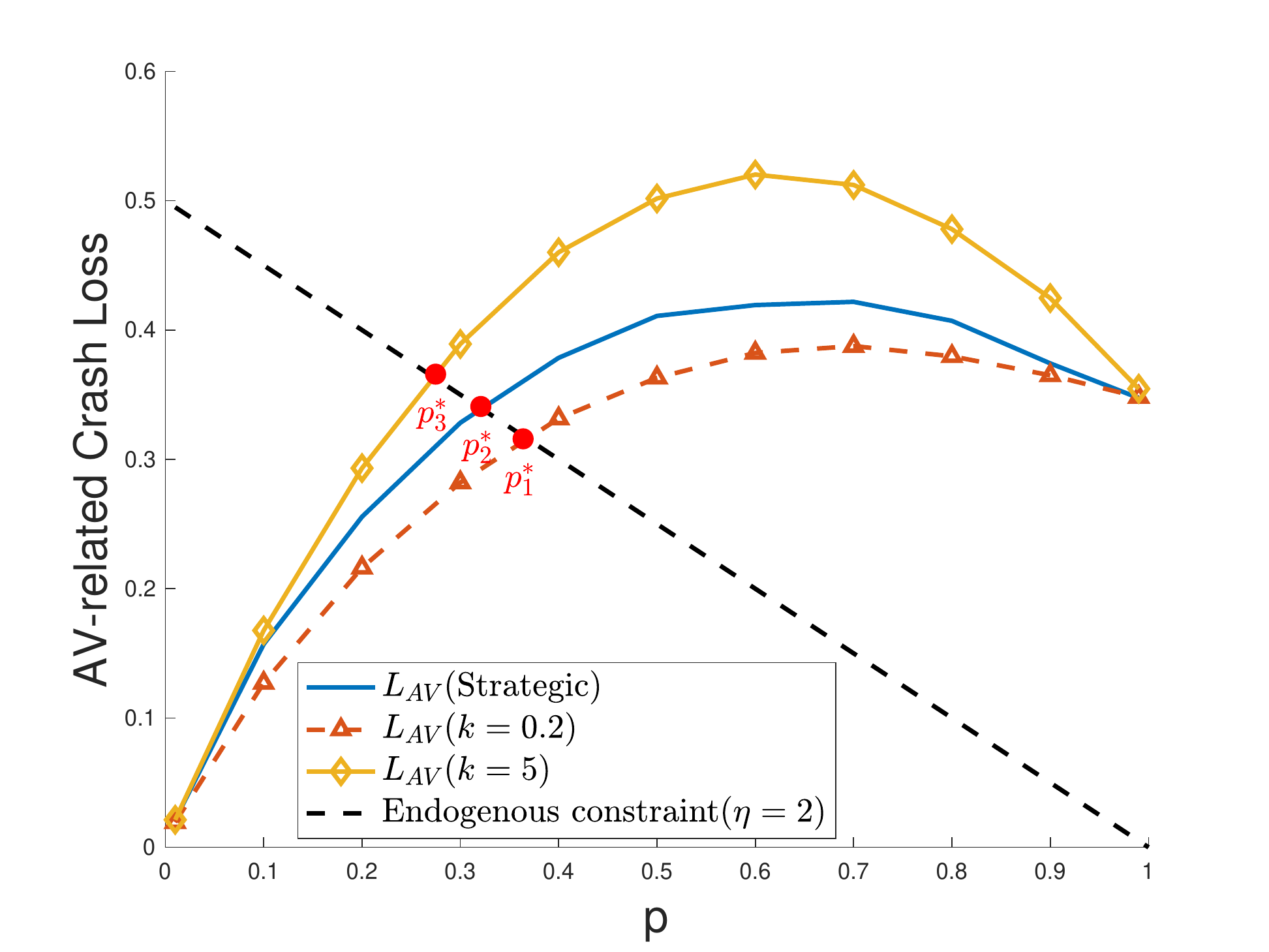}}
	
    \subfloat[Care level \label{fig:endogen_lawmaker_carelevel}]{\includegraphics[width=3in]{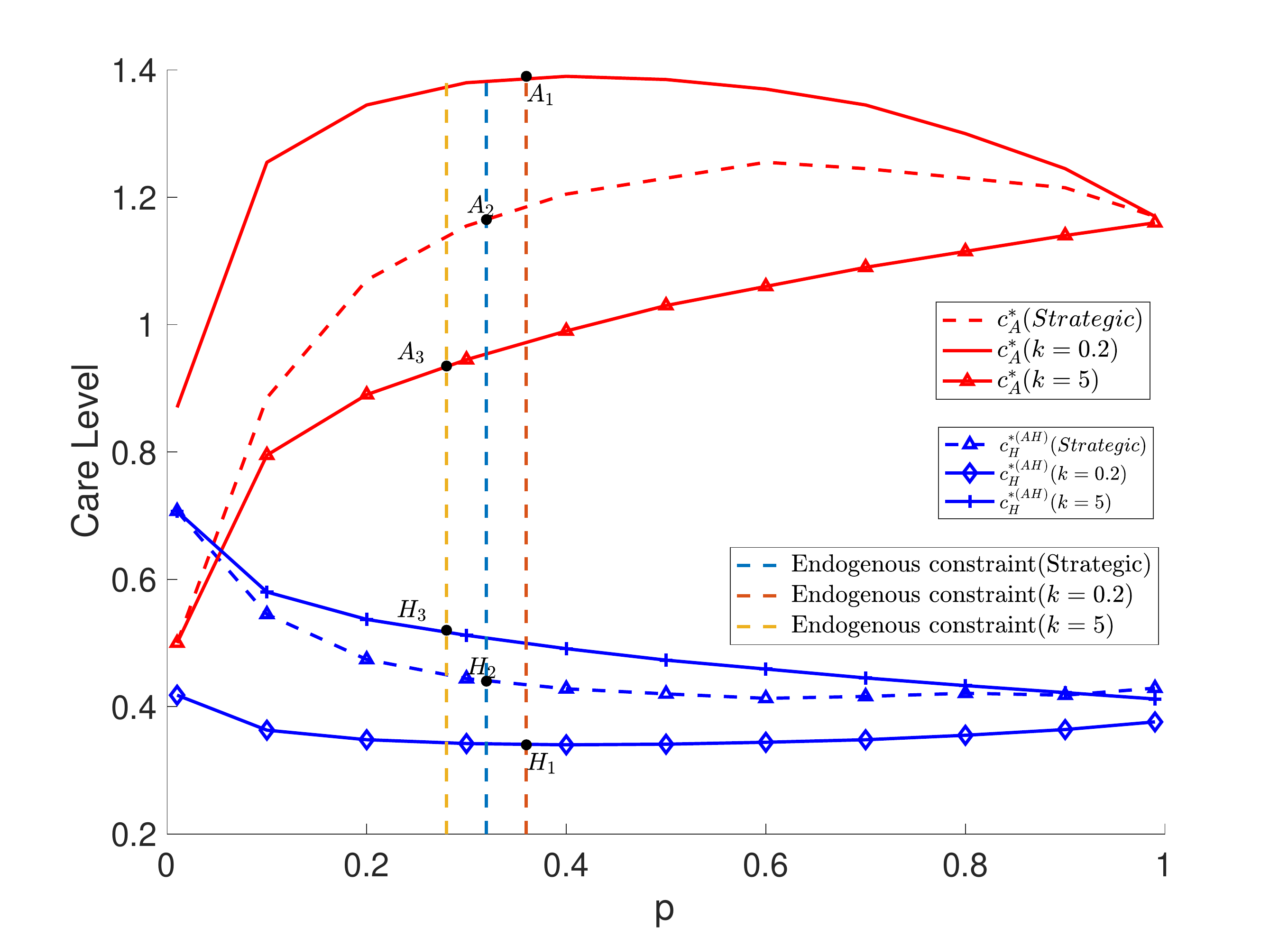}}
	\subfloat[Performance measures \label{fig:endogen_lawmaker_performance}]{\includegraphics[width=3in]{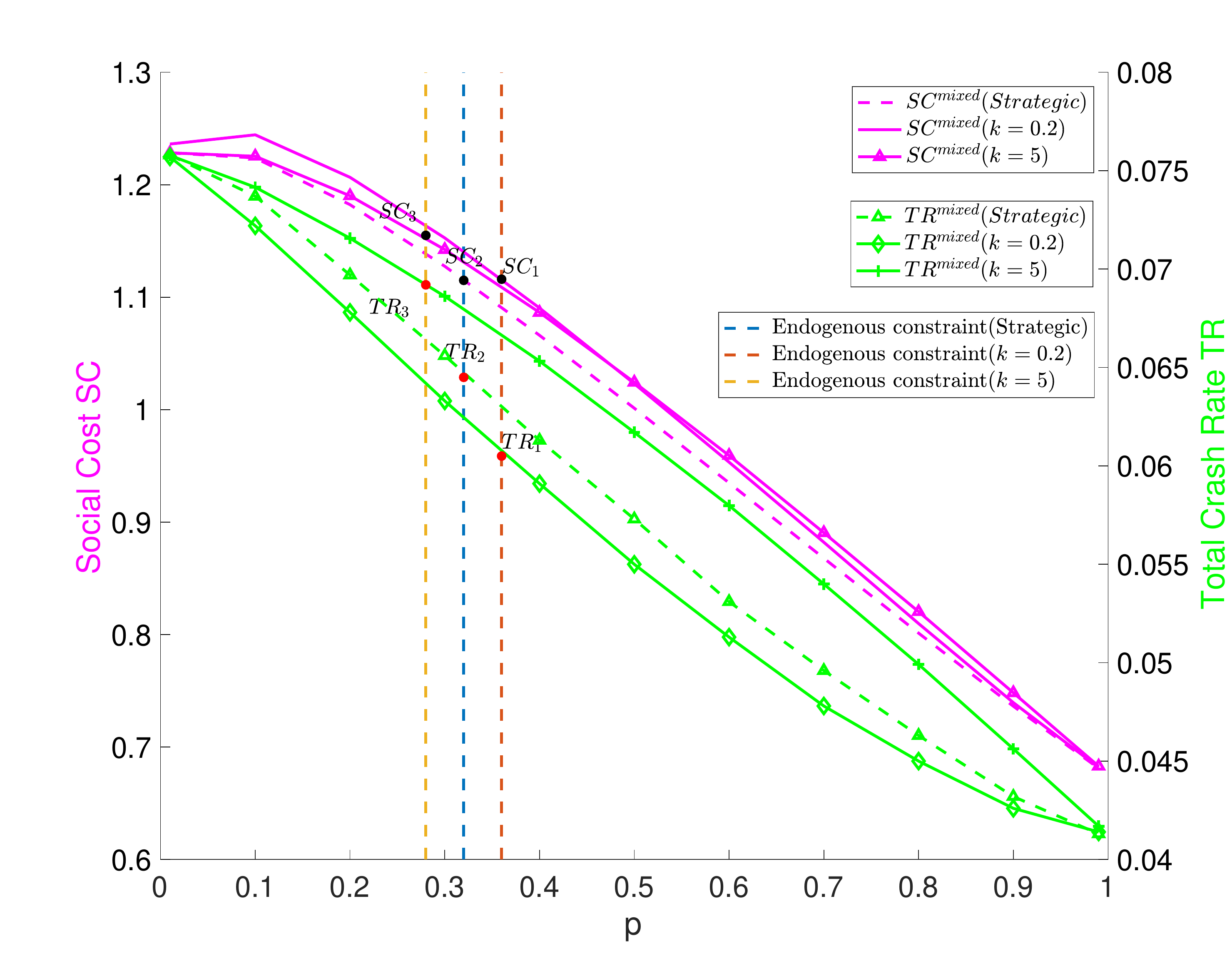}}
	\caption{Endogenous penetration rate with a strategic lawmaker}
	\label{fig:endogen_law}
\end{figure} 
How would an endogenous market share impact the optimal liability rule? 
\figref{fig:endogen_lawmaker} plots the AV-related crash loss curves under three different $k$'s. 
An endogenous constraint is added with the coefficient of $\eta=2$, indicated as the black dashed line. 
When $k = 0.2$ (the red dashed line with triangle markers), the equilibrium market share is $p^{*}_1 = 0.36$. When the lawmaker is strategic (the blue solid line), the equilibrium market share is $p^{*}_2 = 0.32$. When $k = 5$ (the yellow line with diamond markers), the equilibrium market share is $p^{*}_3 = 0.28$. 
When $k$ increases, i.e., the lawmaker punishes HVs more than AVs, the equilibrium penetration rate gradually decreases from the market equilibrium $p^{*}_1=0.36$ to $p^{*}_2=0.28$. 
The strategic lawmaker incurs a medium AV market share. 
This will be explained when we inspect care levels in \figref{fig:endogen_lawmaker_carelevel}.

\figref{fig:endogen_lawmaker_carelevel} and \figref{fig:endogen_lawmaker_performance}
are essentially \figref{fig:lawmaker_carelevel2} and \figref{fig:lawmaker_performance2}
but in an endogenous market. 
The three additional dashed lines correspond to three different $k$'s. 
In \figref{fig:endogen_lawmaker_carelevel}, the intersection points, marked as $A_1$ ($H_1$), $A_2$ ($H_2$), and $A_3$ ($H_3$) are the equilibrium care levels of AVs (HVs) with $k=0.2$, the strategic lawmaker, and $k=5$, respectively. 
When the lawmaker punishes AVs more than HVs ($k=0.2$), the AV manufacturer has to increase AVs' care level for a lower crash loss, leading to reduced public concerns and an increased market share. 
When the lawmaker punishes HVs more than AVs ($k=5$), the AV manufacturer lowers AVs' care level for a lower precaution cost, leading to a higher AV-related crash loss, which in turn increases public concerns and lowers the market share. 
In summary, to ease public concerns, the AV manufacture has to increase AVs' precaution cost in return.  
The strategic lawmaker needs to make a trade-off between the total precaution cost and the total crash loss, thus a medium market share is incurred with a medium crash loss.



In \figref{fig:endogen_lawmaker_performance}, the intersection points, i.e., $SC_1$ ($TR_1$), $SC_2$ ($TR_2$), and $SC_3$ ($TR_3$) are the equilibrium social welfare (total crash rate) with $k=0.2$, the strategic lawmaker, and $k=5$, respectively. 
Their relations are $SC_2<SC_1<SC_3$ and $TR_1<TR_2<TR_3$. 
Although the strategic lawmaker achieves the minimum social cost, 
the social cost with $k=0.2$ is not too much higher than that of the strategic lawmaker, but it maintains the smallest crash rate. This is because $k=0.2$ leads to the highest AV market share compared to other two cases, and more AVs improve social welfare in general.
From both \figref{fig:endogen_base_performance} and \figref{fig:endogen_lawmaker_performance} we can learn that the role of liability rules on social welfare improvement could be relatively more limited than that of easing public fear about this technology. Therefore, public education campaign could help the widespread adoption of AVs.


\section{Conclusion and Future Work}
\label{sec:conclusions} 

This paper investigates the strategic interactions between AVs and HVs using a hierarchical game and provides liability implications for the mixed traffic. 
The interactions between vehicles on roads can be categorized into three types of vehicular encounters: HV-HV, HV-AV, and AV-AV. 
Among them the HV-HV encounter is modeled as a game where each human driver selects her care level. 
AVs' selection of care levels while interacting with other AVs and HVs is determined by the AV manufacturer, once a one-time investment of sensors is fixed. 
To model the role the AV manufacturer plays in the design of autonomous driving, a Stackelberg game between the AV manufacturer and HVs is formulated. 
On the upper level, the lawmaker decides an optimal combination of driver liability and products liability rules 
to regulate both human drivers' care level selection in presence of AVs 
and the AV manufacturer's design of AVs' care level. 
The hierarchical game helps us to understand 
the human drivers' moral hazard, 
the AV manufacturer's impact on traffic safety, 
and the lawmaker's adaptation to the new transportation ecosystem. 
The game and its algorithm are tested on a set of numerical examples, offering insights into behavioral evolution of AVs and HVs as the AV penetration rate increases and as cost or environment parameters vary. 
The outcome of the developed game provides an analytical tool to identify optimal liability rules to ensure social welfare and road safety.  
We find that an optimally designed liability policy is crucial to help prevent human drivers from developing moral hazard and to help the AV manufacturer with a trade off between traffic safety and production costs.  

Albeit novel, this work can be extended in the follow ways:
(1) 
It is of great possibility that there will exist more than one AV manufacturers in the future, and any one of them causing a crash will damage its relative competitiveness in the market. 
Accordingly, the competition among different AV manufacturers should be modeled. 
We will consider more than one AV manufacturer and assume each AV manufacturer produces more than one AV type characterized by sensor specifications.

(2) 
This paper is primarily focused on the safety implication of AVs, not the economic nor planning aspect. 
Ideally we could model market segmentation and pricing of AV manufacturers and the purchasing behavior of consumers. 
On the one hand, AV manufacturers would consider profit as part of their utility and may sell AVs with low care levels at a lower price.  
On the other hand, risk-prone consumers may be willing to buy these AVs for a cheaper price.
Accordingly, consumers' car purchasing behavior could depend on AVs' care levels and eventually liability rules. 
Also, we could include city planners as another game player whose strategy is to plan a transportation system to accommodate AVs. Then we will have a multi-leader multi-follower game. 
The complex interaction between pricing and purchasing behaviors, as well as between city planners and AVs will complicate the analysis and weaken the main message of this paper. 
Thus we will leave it for future work.

(3) 
We will consider a wide spectrum of mixed automation other than the three discrete vehicle encounters. Modeling the in-between degree of vehicle automation, the role of HVs manufacturers in mixed automation, and its impact on liability will be left for future research. 



\section{Acknowledgements}
The authors would like to thank Data Science Institute from Columbia University for providing a seed grant for this research.

\bibliographystyle{elsarticle-harv} 
\biboptions{authoryear}
\bibliography{survey}

\end{document}